\documentclass[11pt,a4paper]{article}
\pdfoutput=1  
\usepackage{epsfig}
\usepackage{graphicx,psfrag}
\usepackage{multirow}
\usepackage{slashed}
\usepackage{pstricks}
\usepackage{caption}
\usepackage{subcaption}
\usepackage{url}
\usepackage{braket}
\usepackage{jheppub}
\usepackage{enumerate}
\usepackage{wasysym} 
\usepackage{mathrsfs} 
\usepackage{amsfonts} 
\usepackage{amsbsy} 
\usepackage{amscd}
\usepackage{color}

\makeatletter

\def\eeq{\end{equation}}
\def\beq{\begin{equation}}

\newcommand{\Rmnum}[1]{\expandafter\@slowromancap\romannumeral #1@}

\newcommand{\gsim}{\raisebox{-0.13cm}{~\shortstack{$>$ \\[-0.07cm]
      $\sim$}}~}
\newcommand{\lsim}{\raisebox{-0.13cm}{~\shortstack{$<$ \\[-0.07cm]
      $\sim$}}~}
\makeatother

\title{LHC signals of a heavy doublet Higgs as dark matter portal: cut-based approach and improvement with gradient boosting and neural networks}

\author[a]{Atri Dey,}
\author[a]{Jayita Lahiri,} 
\author[a]{Biswarup Mukhopadhyaya} 
  \affiliation[a]{Regional Centre for Accelerator-based Particle Physics,
Harish-Chandra Research Institute, HBNI,
Chhatnag Road, Jhunsi, Allahabad - 211 019, India} 
\emailAdd{atridey@hri.res.in}
\emailAdd{jayitalahiri@hri.res.in}
\emailAdd{biswarup@hri.res.in}

\abstract{Though the 125-GeV scalar, as the Higgs boson of the standard model,
is disfavoured as a dark matter portal by direct searches and the observations on
relic density, a heavier scalar in an extended electroweak sector can fit
into that role. We explore this possibility in the context of
two Higgs doublet models (2HDM). Taking Type I and Type II 2HDM as illustration,
and assuming a scalar gauge singlet dark matter particle, we show that the
heavy neutral CP-even scalar ($H$) can (a) serve as dark matter portal consistently 
with all data, and (b) have
a substantial invisible branching ratio, over a wide region of the parameter space.
Using this fact, we estimate rates of LHC signals where $H$ is produced via ({\it i}) gluon 
fusion, in association with a hard jet, and ({\it ii}) vector boson fusion. Invisible 
decays of the $H$ can then lead to monojet + $\slashed{E_T}$ in ({\it i}), and two forward jets 
with large rapidity gap + $\slashed{E_T}$ in ({\it ii}). The second kind of signal usually 
yields better significance for the high-luminosity run. We also supplement our
cut-based analyses with those based on gradient boosted decision trees (XGboost) 
and artificial neural network (ANN) techniques, where the statistical significance 
distinctly improves, especially for Type II 2HDM.

}

\preprint{HRI-RECAPP-2019-003\\$\textrm{}$}

\begin{document}

\maketitle

\newpage

\section{Introduction}

If we assume that some yet unknown particles constitute the dark matter (DM) content 
of our universe, how do they interact with the the known particles included in the standard
model (SM)? Also, can there be terrestrial signatures of such interactions? 
Numerous answers to such questions have  been offered in recent times.  
While it is by and large agreed that the interaction cannot exceed the weak coupling 
strength,  speculations abound on whether any particular sector among the SM matter
fields  has privileged interaction with a `dark sector'. 

The nature of interactions of the recently discovered 125-GeV scalar,
closely resembling the SM Higgs boson, has not been fully understood yet.
It has been speculated that this scalar could act as  portal to the dark sector~\cite{Djouadi:2012zc,Han:2016gyy}. 
However, recent results on direct search for dark matter, especially the data
from the XENON1T experiment, strongly disfavour that possibility unless the 
Higgs-DM coupling is extremely small ($\lsim 10^{-3}$ for an SU(2) singlet scalar DM)~\cite{LopezHonorez:2012kv,Greljo:2013wja,Fedderke:2014wda}. 
With such small coupling between Higgs and singlet scalar DM,
 the relic density exceeds the upper limit from Planck data~\cite{Ade:2013zuv}.

The possibility is, however, less constrained in an extended electroweak symmetry 
breaking (EWSB) sector. While the 125-GeV scalar may have too small an
interaction strength with a DM scalar to have any phenomenological consequence, 
other scalars that simultaneously participate in EWSB can have  appreciable 
interaction with it. This includes, for example, two Higgs doublet models (2HDM)
which are the simplest extensions of the minimal electroweak symmetry breaking (EWSB) sector of the SM~\cite{Gunion:1989we,Branco:2011iw}. Studies with various emphases have thus been carried out 
keeping 2HDM scenarios in view.  These include constraints  from relic density, direct searches and also from the Fermi-LAT results ~\cite{Han:2017etg}, the possibility
of new annihilation channels~\cite{Bandyopadhyay:2017tlq}, $\gamma$-ray signals from the 
galactic centre~\cite{Boucenna:2011hy}, implications of scalar spectra where the 125-GeV
particle is the heavier CP-even scalar~\cite{Han:2018bni,Arhrib:2018qmw}, or 
the viability of scenarios with pseudoscalar portals ~\cite{Berlin:2015wwa}. A recent 
summary of various extended EWSB sectors and constraints on their spectra from DM 
issues  as well as collider data can be found in~\cite{Arcadi:2019lka}. 

The present study confines itself to an SU(2) singlet  scalar DM candidate $\chi$.
It is prevented from mixing with fields occurring in the doublets if
it is odd under an imposed $Z_2$-symmetry and does not develop any
vacuum expectation value (vev). A crucial feature governing the
consistency requirement of a 2HDM with DM search data is the suppression of
the $hh\chi\chi$ quartic coupling, where $h$ is the 125-GeV physical state. 
What we emphasize here is a situation where the heavier scalar in 2HDM acting as the DM portal, 
something that is considerably less constrained than usual Higgs-portal scenarios.
As can be noticed, for example, in~\cite{Drozd:2014yla}, some heavier 
spin-zero state(s) can at the same time have sizeable interaction with a
$\chi$-pair. This state better be the heavier  CP-even state $H$ if
no CP-violation in the EWSB sector is postulated.  In such cases, one can have a 
non-negligible invisible branching ratio for $H$. Moreover, if the mass and the
gauge/Yukawa interaction strengths of H are such that it has substantial
rate of production at the high-luminosity run of the Large Hadron Collider (LHC),
new signals  can then be expected through  invisible decay of the  $H$, with 
appropriate tags alongside. We concentrate on situations where this indeed 
happens, and try to identify corresponding regions of the parameter space. 

Our benchmark scenarios are Type I and Type II models. We shall comment
on the Type X scenario later in the paper. An assortment of collider as well as 
flavour constraints exist on these models, and we make sure that the regions
where the signals are claimed to be noticeable are consistent with these constraints. 
In addition, all the usual constraints  apply on the DM candidate.  After ensuring that our 
benchmark points (BP) are restricted by  all these considerations (and some others 
like vacuum stability), we go on to study the production of the heavier scalar $H$
in both of the gluon fusion (ggF) and vector boson fusion (VBF) channels. While
in the first case we compute the $monojet~+~missing~energy$ signal rate, 
the cross section for $two~forward~jets~+~missing~energy$ is estimated in
the second, together with the backgrounds in each category. 

The plan of this work is as follows. In Section~\ref{sec2} we present a overview of our model. The constraints on the Higgs sector of the model have been discussed in Section~\ref{sec3}. The constraints on the dark sector will be discussed in Section~\ref{sec4}. In Section~\ref{sec5} we present the cut-based collider analysis for gluon fusion and vector boson fusion production of heavy Higgs. The scope of improvement over and above the cut-based analysis using the recently developed techniques has been explored in Section~\ref{sec6}. Finally, we summarize and conclude our work in Section~\ref{sec7}.

\section{Overview of the Model - 2HDM + DM}
\label{sec2}

We consider an extension of 2HDM along with a scalar dark matter candidate $\chi$. The dark matter candidate $\chi$ of our model interacts with the SM fermions and gauge-bosons via a portal through the 2HDM Higgs sector. The Lagrangian we consider is

\begin{equation}
{\cal L} = {\cal L}_{2HDM} + {\cal L}_{DM} + {\cal L}_{Int} 
\end{equation}
\begin{equation}
{\cal L}_{DM} + {\cal L}_{Int} = \frac{1}{2} \partial^{\mu} \chi  \partial_{\mu} \chi - \frac{1}{2} (M_{\chi}^2 + (\lambda_{1s} + \lambda_{2s}) v^2) \chi^2 + \lambda_S \chi^4 + \lambda_{1s} \chi^2 \Phi_1^{\dagger} \Phi_1 + \lambda_{2s} \chi^2 \Phi_2^{\dagger} \Phi_2
\label{lagdm}
\end{equation}

Here $\Phi_1$ and $\Phi_2$ are the Higgs doublets with $Y=1$ in the flavor basis and $\chi$ is a scalar of mass $M_{\chi}$ which is singlet under the SM gauge group. Furthermore, 
a $Z_2$ symmetry is postulated, under which $\chi$ is assumed to be odd, while
$\Phi_{1,2}$ are even. Thus $\chi$ does not have any vacuum expectation value(vev).

The above properties of $\chi$ do not allow it to mix with $\Phi_{1,2}$. Thus the
phenomenological constraints on such a scenario, other than those related to
dark matter issues, mostly arise in the sector spanned by the two scalar  doublets.
Since these constraints need to be respected in what follows, we start with a
recapitulation of the different components of ${\cal L}_{2HDM}$. We first consider
those components which are identical in the various popular versions of 2HDM.

\newpage

{\bf The scalar potential:}

The most general 2HDM scalar potential consistent with SU(2)$_L \times$ U(1)$_Y$ gauge invariance is
\begin{eqnarray}
\label{lambdapotential}
\mathcal{V} &=& m_{11}^2\Phi_1^\dagger\Phi_1+m_{22}^2\Phi_2^\dagger\Phi_2
-[m_{12}^2\Phi_1^\dagger\Phi_2+{\rm h.c.}]
 +\frac{1}{2}\lambda_1(\Phi_1^\dagger\Phi_1)^2
+\frac{1}{2}\lambda_2(\Phi_2^\dagger\Phi_2)^2
+\lambda_3(\Phi_1^\dagger\Phi_1)(\Phi_2^\dagger\Phi_2) \nonumber \\
&+&\lambda_4(\Phi_1^\dagger\Phi_2)(\Phi_2^\dagger\Phi_1)
+\left\{\frac{1}{2}\lambda_5(\Phi_1^\dagger\Phi_2)^2
+\big[\lambda_6(\Phi_1^\dagger\Phi_1)
+\lambda_7(\Phi_2^\dagger\Phi_2)\big]
\Phi_1^\dagger\Phi_2+{\rm h.c.}\right\}\,. 
\end{eqnarray}

where $m_{12}^2$, $\lambda_5, \lambda_6$ and $\lambda_7$ can be complex in
general. However, we neglect the possibility of CP-violation
in the scalar sector.  

As we shall discuss below, one is faced with the task of avoiding flavour-changing
neutral current (FCNC) enhancement.  A popular way  of suppressing tree-level FCNC
is to impose a softly-broken $Z_2$ symmetry on the Higgs potential, which implies 
$ \lambda_6=\lambda_7 = 0$. Moreover, CP-conservation
in the tree-level potential  renders both  $\lambda_5$ and $m_{12}^2$ real. Upon
electroweak symmetry breaking, the two Higgs doublets acquire non-zero vev. Under the assumption of CP-conserved and charge-neutral vacuum, the vev of the Higgs fields 
can be expressed as

\begin{equation}
\langle \Phi_1 \rangle={1\over\sqrt{2}} \left(
\begin{array}{c} 0\\ v_1\end{array}\right), \qquad \langle
\Phi_2\rangle=
{1\over\sqrt{2}}\left(\begin{array}{c}0\\ v_2
\end{array}\right)\,.\label{potmin}
\end{equation}
\noindent
where $\tan \beta = \frac{v_2}{v_1}$.

Diagonalisation of the CP-odd neutral scalar mass-matrix yields the
physical states 

\begin{eqnarray*}
H &=&(\sqrt{2}{\rm Re\,}\Phi_1^0-v_1)\cos \alpha+
(\sqrt{2}{\rm Re\,}\Phi_2^0-v_2)\sin \alpha\,, \\
h &=&-(\sqrt{2}{\rm Re\,}\Phi_1^0-v_1)\sin \alpha+
(\sqrt{2}{\rm Re\,}\Phi_2^0-v_2)\cos \alpha\,,
\label{hZ2scalareigenstates}
\end{eqnarray*}

\noindent
where $\alpha$ is the mixing angle between the  CP-even parts of the two fields. We should mention here that  only the CP-even scalars ($h$ and $H$) can act as a portal for dark matter so long as CP is conserved. Here $h$ stands for the SM-like Higgs,
which is the lighter state in most models.

\medskip
{\bf Gauge interactions:}

Gauge interactions not only decide the scalar decay  branching ratios but also
rates in the various scalar production channels; they are thus
worth specifying.  They arise from the covariant kinetic energy terms

\begin{equation}
{\cal L}_{gauge} = (D_{\mu} \Phi_1)^{\dagger} D^{\mu} \Phi_1 + (D_{\mu} \Phi_2)^{\dagger} D^{\mu} \Phi_2
\end{equation}

Where $D_{\mu} = \partial_{\mu} -   \frac{i}{2} g W_{\mu}^a \tau^a -  \frac{i}{2} g' B_{\mu}$. The gauge couplings, being the same for all variants of CP-conserving 2HDM,
can be found, for example, in ~\cite{Branco:2011iw}. We just remind the reader that the two interactions strengths relevant to the current work, namely,  those of the $hVV$ ($HVV$) interactions
(with $V = W,Z$)  are given by 

\begin{eqnarray*}
g_{hVV} = g_{SM} \times \sin(\beta - \alpha)  \\ 
h_{HVV} = g_{SM} \times \cos(\beta - \alpha)
\end{eqnarray*}
where $g_{SM}$ is the corresponding coupling strength of the SM Higgs.

\medskip

{\bf Yukawa interactions:}

The Yukawa sector is what essentially distinguishes different variants of 2HDM from each other. The variants of 2HDM have been classified on the basis of the couplings of the up and down-type quarks and leptons with the two Higgs doublets. Different discrete symmetries are postulated in different types of 2HDM to ensure the absence of tree-level FCNC in the model. We  concentrate here on two types of 2HDMs namely Type I and Type II 2HDM. We mention here that the other two types of 2HDM, namely, the  Type X (lepton-specific) and flipped ones, differ from Type I and Type II respectively in terms of their lepton coupling. All that can be mentioned at this stage is that the decay branching ratio of the H to a pair of DM
particles  can {\it prima facie} be larger in the Type X scenario, as compared to the two
cases considered here. We shall comment more on this in section 4.

In case of Type I 2HDM, up and down type quarks and leptons couple to the same doublet. Then the Yukawa Lagrangian takes the form as follows

\begin{equation}
\label{type1}
{\cal L}_{Yukawa} = y^1_{ij}\bar Q_{iL} \Phi_2 d_{jR} + y^2_{ij}\bar
Q_{iL} {\bar{\Phi}}_2 u_{jR} + y^5_{ij}\bar L_{iL} \Phi_2 e_{jR}
\end{equation}
This can be achieved by imposing the discrete symmetry on the 
${\cal L}_{Yukawa}$, $\Phi_1 \rightarrow -\Phi_1$.

In Type II 2HDM, up-type
quarks couples to one doublet, and down-type quarks and leptons to to another. Under this assumption, the ${\cal L}_{Yukawa}$
becomes

\begin{equation}
\label{type2}
{\cal L}_{Yukawa} = y^1_{ij}\bar Q_{iL} \Phi_1 d_{jR} + y^2_{ij}\bar
Q_{iL} {\bar{\Phi}}_2 u_{jR} + y^5_{ij}\bar L_{iL} \Phi_1 e_{jR}
\end{equation}
 This can be enforced by demanding that the ${\cal L}_{Yukawa}$
 remains invariant under $\Phi_1 \rightarrow -\Phi_1$ and $d_R
 \rightarrow -d_R$ and $e_R \rightarrow -e_R$.  

Having thus outlined the basic features, we now discuss in turn the various
constraints applicable on such a scenario, arising (a) from the usual 
phenomenology of a 2HDM, and (b) from considerations related to dark matter.

\section{Constraints on relevant parameters of ${\cal L}_{2HDM}$}\label{sec3}

When there is no mixing between the two doublets and the DM particle $\chi$, 
the constraints on the two-doublet sector does not differ appreciably from
those on a pure 2HDM. We discuss such constraints below, while those 
applicable to the sector involving $\chi$ will be taken up in the next section.

The theoretical constraints come from the consideration of vacuum stability, perturbativity. 
As regards vacuum stability, it is sufficient at this stage to ensure it around
the electroweak scale only, since we are not concerned with ultraviolet completion. 
Thus the low-energy phenomenological model should retain positivity of 
the potential for sufficiently large values of the field, in order to have it  bounded 
from below. Since at large field values the potential is dominated by the quartic terms, the stability condition thus implies the following conditions on the quartic couplings  ~\cite{Deshpande:1977rw,Nie:1998yn}:

\begin{eqnarray}
\lambda_1 > 0,~~~ \lambda_2 > 0,~~~ \lambda_3 > -(\lambda_1 \lambda_2)^{1/2} \nonumber \\
\lambda_3 + \lambda_4 - |\lambda_5| > (\lambda_1 \lambda_2)^{1/2} (\text{when}~~ \lambda_6 = \lambda_7 = 0)
\end{eqnarray}

For perturbativity of quartic interactions at the electroweak scale, one should demand that the quartic couplings  at the EWSB scale obey  

\begin{equation}
C_{H_i H_j H_k H_l} < 4\pi
\end{equation}

The further requirement of perturbativity in the Yukawa sector disallow
very large ($\gsim 50$) values of $\tan \beta$. 
In addition, tree-level unitarity in the scattering of Higgs bosons and the longitudinal components of the EW gauge bosons ~\cite{Arhrib:2000is,Kanemura:1993hm} requires that eigen-values of the scattering matrices have to be less that $16 \pi$~\cite{Lee:1977yc,Lee:1977eg,Ginzburg:2005dt}.

Next come phenomenological constraints.  They do  arise from electroweak
precision measurements, especially from the oblique parameters~\cite{Peskin:1991sw}.
The addition of extra Higgs doublets (and also singlets) in general do not affect
them, especially the $T$-parameter,  as the custodial SU(2) remains unbroken.  It
can however be broken at the loop level if the masses of the additional scalar physical states are very different from each other. It has been ensured in this study that the mass-gap among such scalars  (and the potential in general) are consistent with such 
constraints~\cite{Erler:2019hds}. 

The extended Higgs sector of 2HDM also contributes to the anomalous magnetic moment of the muon, which continues to show a 3$\sigma$ deviation from the SM prediction
\cite{Bennett:2006fi,Aoyama:2012wk,Czarnecki:2002nt}. All the additional spin-zero particles in our scenario (except $\chi$)
contribute to the anomalous magnetic moment;  particularly important, however,
are the two loop contributions via the Bar-Zee diagrams mediated by a light
pseudoscalar~\cite{Cherchiglia:2017uwv}. While we have not attempted to explain the $(g-2)_\mu$ excess,
it has been made sure that the region of the parameter space used here has not exceeded
the SM contribution by more than 3$\sigma$.

 Constraints also come form rare B-decays such as $b \rightarrow s \gamma$, $B_s \rightarrow \mu^+ \mu^-$, $B^+ \rightarrow \tau^+ \nu$. These constraints are much stronger on Type II than on Type I, largely because of  charged Higgs contributions. 
 The available results for $b \rightarrow s \gamma$ imply  $M_H^{\pm} \gsim 600$ GeV~\cite{Czakon:2015exa,Misiak:2006zs,Amhis:2016xyh}, while
 for Type I scenarios the lower bound is about 80 GeV only. Side by side,
 high $\tan \beta$ regions in Type II are restricted by the observed rate of
 $B_s \rightarrow \mu^+ \mu^-$ \cite{CMS:2014xfa,Adachi:2012mm,Kronenbitter:2015kls}.
 
CMS and ATLAS data from  runs I and II on the 125-GeV scalar have been 
limiting its signal strengths in various channels with increasing precision~\cite{Khachatryan:2016vau,CMS:2017rli,CMS:2017jkd,CMS:2017pzi,Sirunyan:2017khh,ATLAS-CONF-2017-045,ATLAS-CONF-2017-043}.  The net outcome is a gradual convergence
towards the so-called alignment limit, namely, $(\beta - \alpha) = \pi/2$. Consistency with
the current constraints from this consideration has been ensured for our extended Higgs 
sector here, together with the restriction that the invisible branching ratio
of the already observed scalar is not more than 15\%~\cite{Sirunyan:2018owy}.

As regards direct search for the additional scalars, no positive results are
available from the LHC so far. 
The experiments, on the  other hand,  have put upper limits on 
the value  $\sigma \times \text{Br}$ for each of them, which in turn can be translated to a limit on the parameter space. The most stringent constraint so far has been obtained from the decay mode $H \rightarrow \tau \tau$. In case of Type II, the Higgs coupling of the heavier neutral Higgs with a $\tau$-pair is proportional to $\tan \beta$.
on which a strong upper bound on therefore exists. Higher $\tan \beta$ values will be allowed only for higher mass of $H$. Regions in the $m_H - \tan\beta$ space are
thus restricted  by the non-observation of such signals ~\cite{Chowdhury:2017aav}. Type I scenarios
are free from such a constraint.

\section{Constraints on the dark matter sector}\label{sec4}

For $\chi$ to be considered as a thermal dark matter candidate, it has been required
here to satisfy the following constraints:

\begin{itemize}
\item The thermal relic density of $\chi$ should not exceed the latest Planck data~\cite{Ade:2013zuv} at the 2$\sigma$ level.

\item The $\chi$-nucleon cross section  should be below 
the current upper bound from XENON1T~\cite{Aprile:2018dbl}.

\item Constraints from indirect detection experiments should be satisfied. Therefore the annihilation rate of $\chi$ has been consistent at
 the 95\% confidence level with both isotropic gamma-ray distribution data
 and the gamma ray observations from dwarf spheroidal galaxies~\cite{Ackermann:2015zua}.

\item The invisible decay of the 125-GeV scalar $h$ has been limited to
15\%~\cite{Sirunyan:2018owy}.
\end{itemize}

The vacuum stability and perturbativity conditions discussed in the previous section, will be slightly modified in presence of an extra scalar singlet field $\chi$. The stability criteria in this case imply~\cite{Drozd:2014yla}

\begin{eqnarray}
& \lambda_1, \lambda_2, \lambda_S >0, \quad\lambda_3+\lambda_4-|\lambda_5|>-\sqrt{\lambda_1 \lambda_2}, \quad
\lambda_3>-\sqrt{\lambda_1 \lambda_2} & \label{first3}\\
&  \lambda_{1s} > - \sqrt{\frac{1}{12}\lambda_{S}\lambda_1}, \quad 
\lambda_{2s} > - \sqrt{\frac{1}{12}\lambda_{S}\lambda_2}\,. &
\label{kapsimp}
\end{eqnarray}

If $\lambda_{1s}~{\rm or}~ \lambda_{2s}<0$, then we also have to satisfy
\begin{eqnarray}
& - 2\lambda_{1s}\lambda_{2s}+\frac{1}{6}\lambda_{S}\lambda_3>-\sqrt{ 4\left(\frac{1}{12}\lambda_{S}\lambda_1 - \lambda_{1s}^2 \right)\left(\frac{1}{12}\lambda_{S}\lambda_2-\lambda_{2s}^2\right)}& \label{kapcomp1}\\ 
& - 2\lambda_{1s}\lambda_{2s}+\frac{1}{6}\lambda_{S}(\lambda_3 +\lambda_4-|\lambda_5| )>-\sqrt{ 4\left(\frac{1}{12}\lambda_{S}\lambda_1 - \lambda_{1s}^2 \right)\left(\frac{1}{12}\lambda_{S}\lambda_2-\lambda_{2s}^2\right)} \,.&
\label{kapcomp2}
\end{eqnarray}

The perturbativity condition is determined by looking at quartic terms involving the singlet scalar field $\chi$. They turn out to be $0 < \lambda_S < 4 \pi$, $|\lambda_{1s}| , |\lambda_{2s}| < 4\pi$~\cite{Drozd:2014yla}.

We perform a scan of the  parameter space and choose a few benchmark points which satisfy all the aforementioned constraints. The ranges of scan for the two Types of models are as follows(Table.~\ref{scan}):

 \begin{table}[!hptb]
\begin{center}
\begin{footnotesize}
\begin{tabular}{| c | c |}
\hline
 Type I & Type II  \\
\hline
 $80 \text{GeV} < m_H < 900 \text{GeV}$ & $500 \text{GeV} < m_H < 900 \text{GeV}$  \\
$100 \text{GeV} < m_{\chi} < 400 \text{GeV}$ & $60 \text{GeV} < m_{\chi} < 400 \text{GeV}$  \\
$-12 < \lambda_{1s} < 12$ & $-12 < \lambda_{1s} < 12$ \\
$-12 < \lambda_{2s} < 12$ & $-12 < \lambda_{2s} < 12$ \\
$2 < \tan \beta < 20$   &  $2 < \tan \beta < 20$ \\
$0.8 \lsim \sin (\beta - \alpha) < 1.0$ & $0.8 \lsim \sin (\beta - \alpha) < 1.0$ \\
\hline
\end{tabular}
\end{footnotesize}
\caption{The range of scan for relevant parameters for Type I and Type II 2HDM.}
\label{scan}
\end{center}
\end{table}

\begin{figure}[!hptb]
\includegraphics[width=6.1cm, height=7.5cm,angle=-90.0]{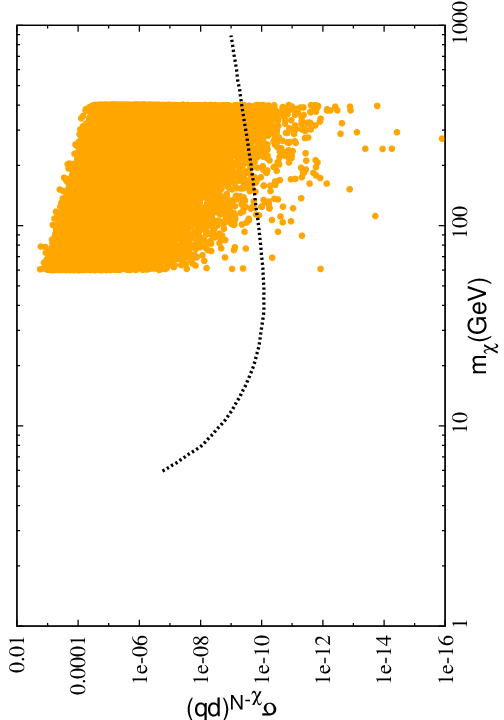} 
\hspace{0.05cm}
\includegraphics[width=6.1cm, height=7.5cm,angle=-90.0]{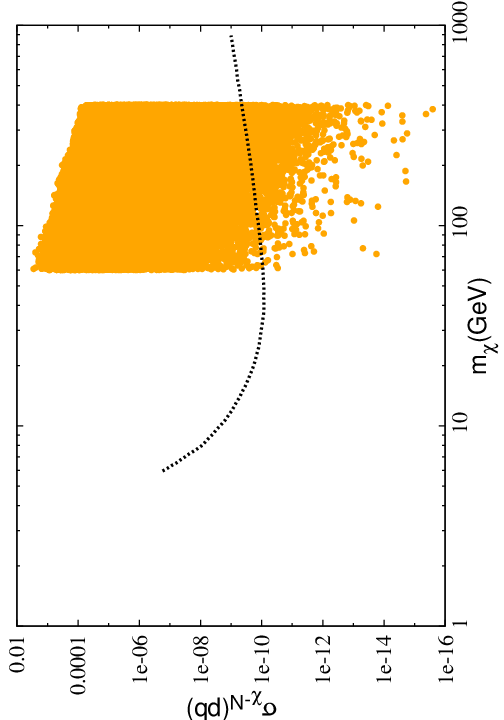} 
\caption{Parameter space allowed by the relic density observation for Type I (left) and Type II (right) 2HDM. The black dashed line is the upper limit on the $\chi-N$ scattering cross section from XENON1T experiment. }
\label{omega_dd}
\end{figure}

We  show in Figure.~\ref{omega_dd} scatter plots generated from the scan, compared 
with the allowed
 region in the $m_\chi - \sigma_{(\chi - N)}$ space obtained from the XENON1T data~\cite{Aprile:2018dbl}. 
The plots contains only those  points which  satisfy  all constraints including those from relic density. The black curve in each case shows the upper limit on cross section for nucleon-DM scattering which is relevant in direct detection.  All our benchmark   points used
for LHC predictions are chosen from regions beneath the curves.

\begin{figure}[!hptb]
\includegraphics[width=6.1cm, height=7.5cm,angle=-90.0]{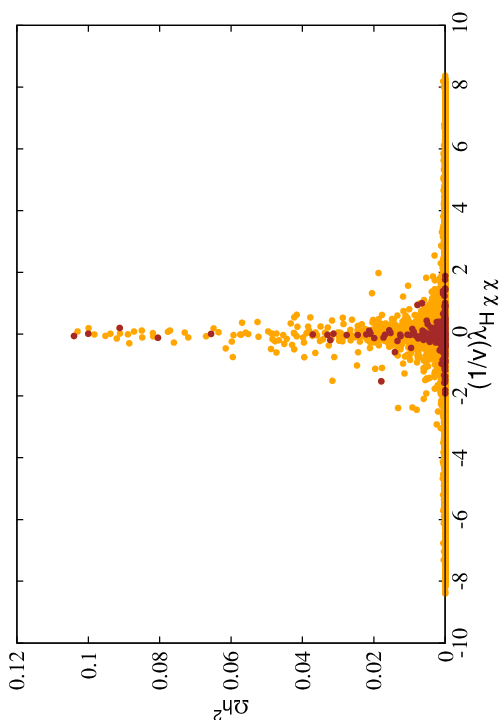} 
\hspace{0.05cm}
\includegraphics[width=6.1cm, height=7.5cm,angle=-90.0]{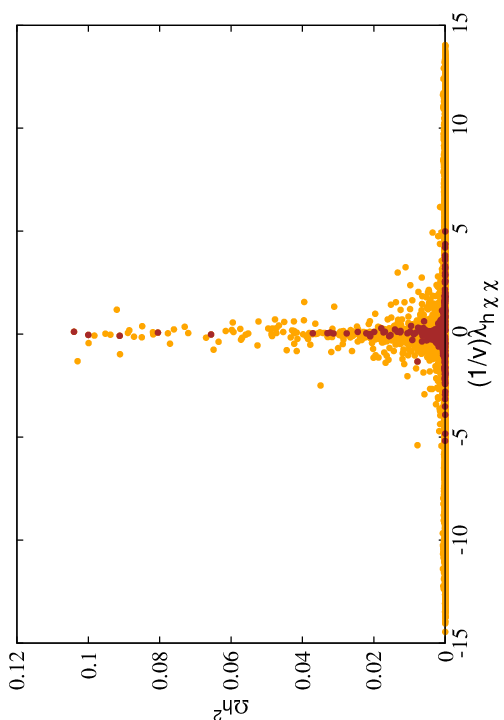} \\
\vspace*{0.1cm}
\includegraphics[width=6.1cm, height=7.5cm,angle=-90.0]{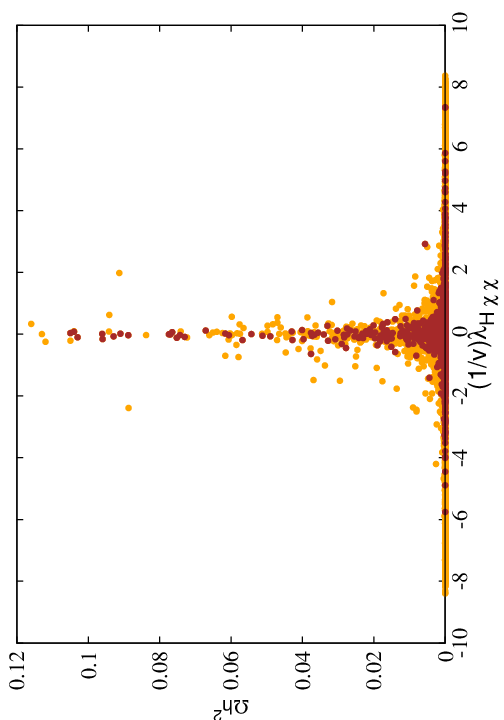} 
\hspace{0.05cm}
\includegraphics[width=6.1cm, height=7.5cm,angle=-90.0]{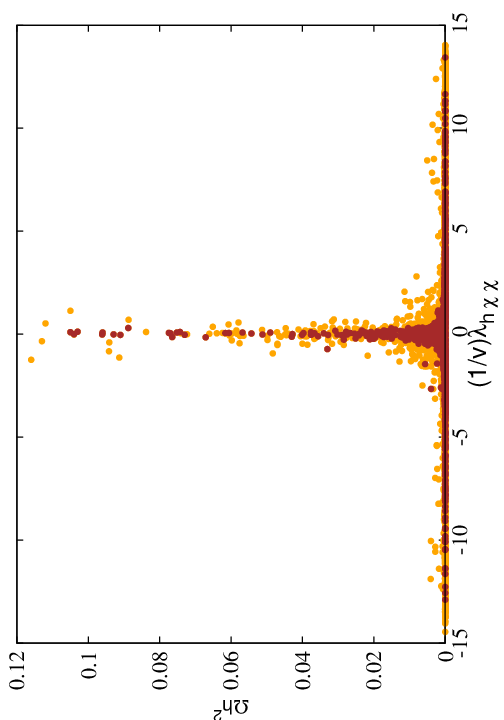} \\
\caption{Dependence of relic density on $\lambda_{eff}$ for Type I (top) and Type II (bottom) 2HDM. The orange points satisfy relic density constraints. Maroon points in addition satisfy direct detection constraints as well. }
\label{omega_lam1}
\end{figure}

We proceed further to explore the dependence of relic density and direct detection cross section on the coupling of dark matter with Higgs fields. In order to do that we have first calculated the coupling of dark matter candidate $\chi$ with two physical CP-even Higgs fields ($h$ and $H$) in terms of the coupling parameters $\lambda_{1s} $ and $\lambda_{2s}$ of the Lagrangian~\ref{lagdm}:

\begin{eqnarray}
\lambda_{H \chi \chi} = \lambda_{1s} v \cos \alpha \cos \beta + \lambda_{2s} v \sin \alpha \sin \beta \\
\lambda_{h \chi \chi} = - \lambda_{1s} v \sin \alpha \cos \beta + \lambda_{2s} v \cos \alpha \sin \beta 
\end{eqnarray}
The dependence of relic density on $\lambda_{H \chi \chi}$ and $\lambda_{h \chi \chi}$ has been shown in Figure.~\ref{omega_lam1}, with the scatter plots containing orange points which satisfy the upper limit on relic density and maroon points which satisfy both relic density and direct detection constraints. For larger magnitudes of $\lambda_{H \chi \chi}$ or $\lambda_{h \chi \chi}$, the annihilation cross section increases and consequently the relic density goes down, as is evident from the figure.

\begin{figure}[!hptb]
\includegraphics[width=6.1cm, height=7.5cm,angle=-90.0]{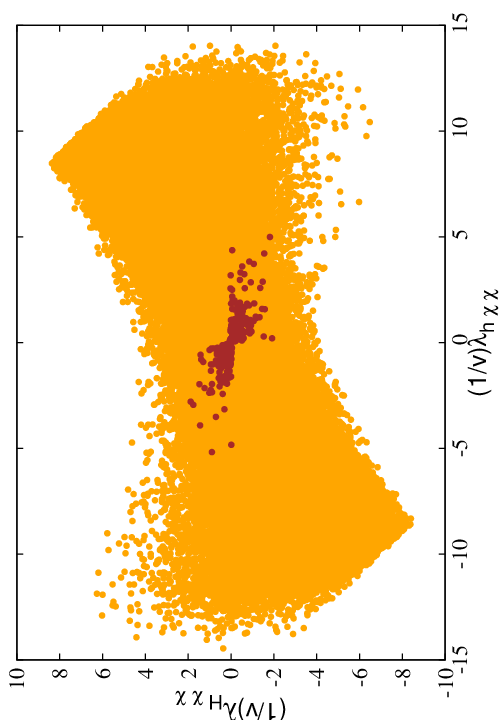} 
\hspace{0.05cm}
\includegraphics[width=6.1cm, height=7.5cm,angle=-90.0]{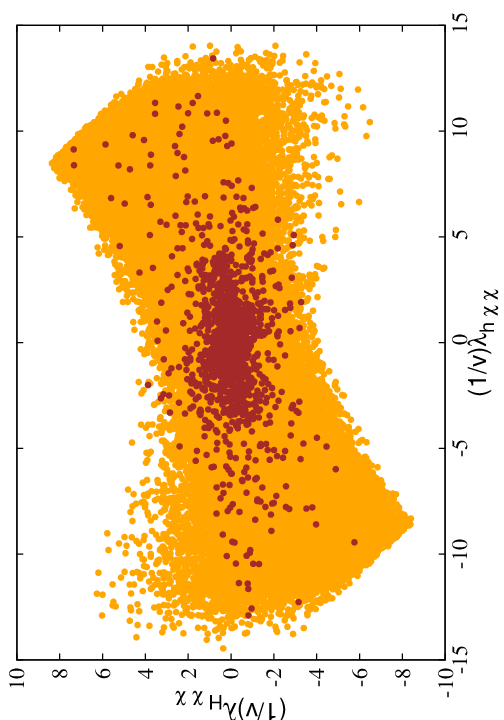} 
\caption{Scatter plot in the $\lambda_{h \chi \chi}-\lambda_{H \chi \chi}$ plane for Type I (top) and Type II (bottom) 2HDM. Color code is same as in Figure.~\ref{omega_lam1}. }
\label{lamh_lamH}
\end{figure}

\begin{figure}[!hptb]
\begin{center}
\includegraphics[width=9.0cm, height=7cm]{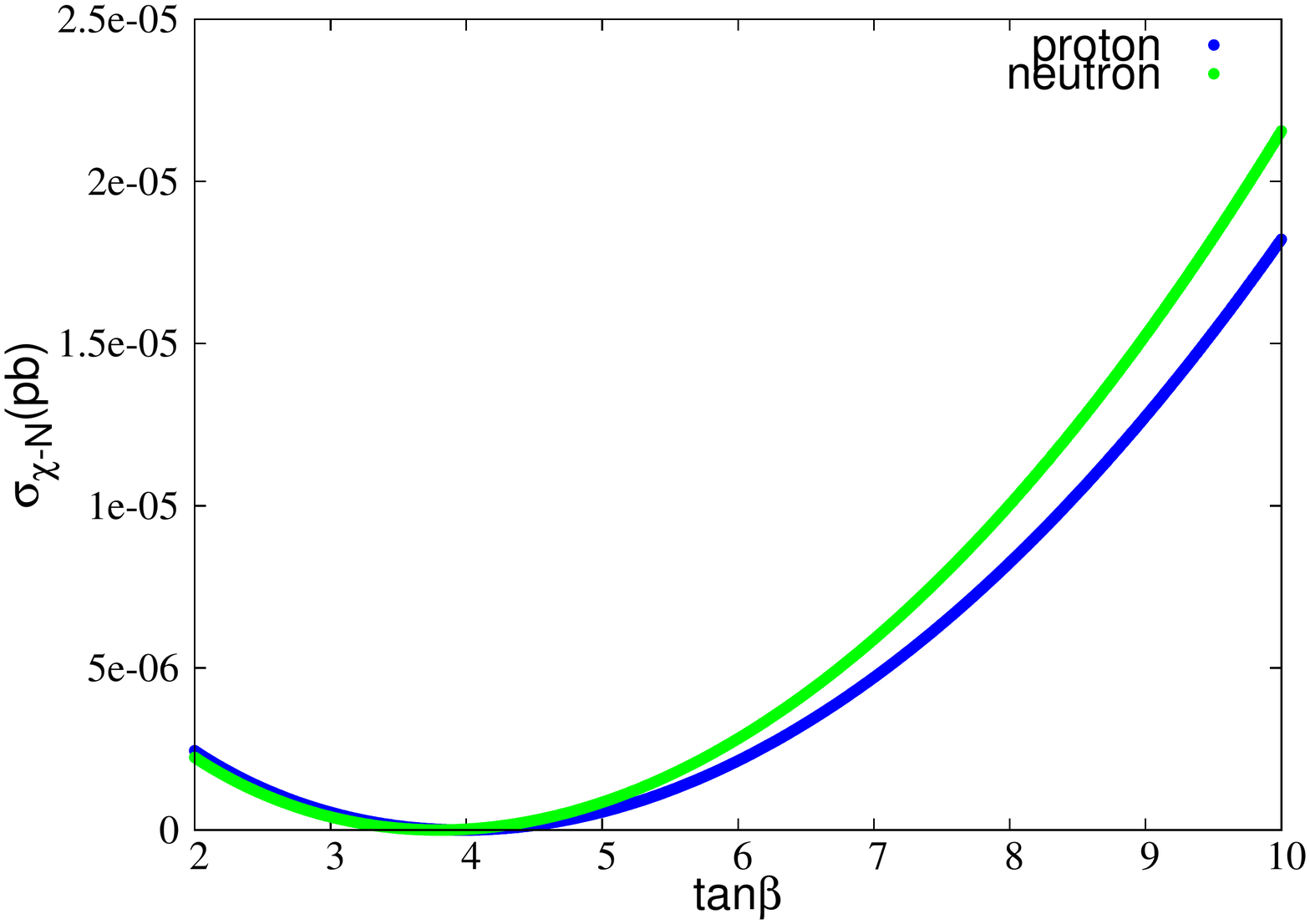}
\caption{$\chi-N$ scattering cross section as a function of $\tan \beta$ with all other parameters fixed as Type II BP I(Table~\ref{benchmark}). Here $N$ stands for nucleon type, namely, proton(blue curve) and neutron(green curve).}
\label{tanb}
\end{center}
\end{figure}

In Figure.~\ref{lamh_lamH} we present a scatter plot in the $\lambda_{h \chi \chi}$-$\lambda_{H \chi \chi}$ plane, with a factor of $v = \sqrt{v_1^2 + v_2^2}$ taken out of each tri-linear coupling in order to make it dimensionless. While all the points shown in the scatter-plots  satisfy relic density constraints,
the XENON1T limits are respected only by the maroon points. We can see that the direct detection limit apparently looks more restrictive in case of Type I than Type II 2HDM. In fact, a cancellation takes place between the contribution via the light ($h$) heavy Higgs ($H$) bosons to the $\chi - N$ t-channel scattering amplitude for a specific range of $\tan \beta$ (See Figure.~\ref{tanb}). Because of this cancellation the direct detection limit becomes weaker in case of Type II. We have checked that in Type I, this type of cancellation takes place only in low $\tan \beta$ region. In our study we have not scanned very low $\tan \beta$(0.5 - 3.0) region. Therefore in Figure.~\ref{lamh_lamH}, the parameter space allowed by the direct search appears to be smaller in the coupling plane in case of Type I as compared to Type II.

The other constraining factor is the maximum allowed invisible decay branching ratio
of the Higgs boson. The upper limit on Br($h_{SM} \rightarrow \text{invisible}$) is 15\%~\cite{Khachatryan:2016whc,Aaboud:2019rtt,Sirunyan:2018owy}. 
This in principle should restrict the parameter space comprising of $\lambda_{h\chi\chi}, \tan \beta$ and $\sin(\beta - \alpha$) of each scenario under consideration.
However, our purpose in this work is to identify regions where one of the heavy scalars
(the H in particular) can serve as the dark matter portal. Our scan reveals that this
happens, though appropriate annihilation rates ensured by the $\chi \chi H$ interaction,
for regions for which $m_{\chi}$ is well above $\frac{m_h}{2}$. Therefore, the invisible
branching ratio for the 125-GeV scalar is not something that restricts  us in practice.

\begin{figure}[!hptb]
\includegraphics[width=6.1cm, height=7.5cm,angle=-90.0]{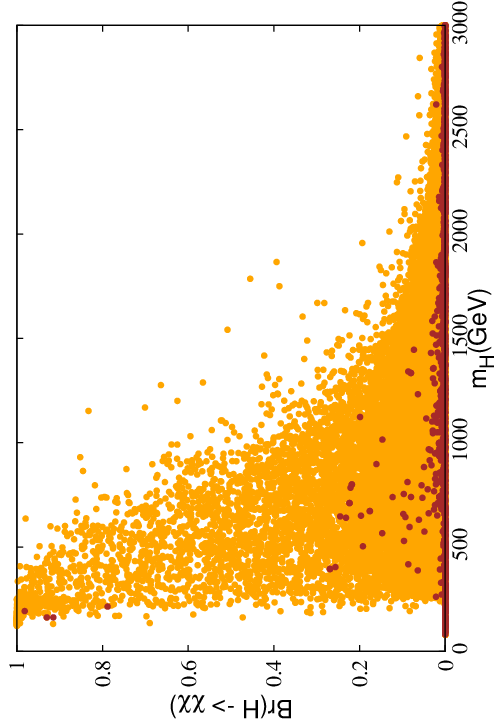} 
\hspace{0.05cm}
\includegraphics[width=6.1cm, height=7.5cm,angle=-90.0]{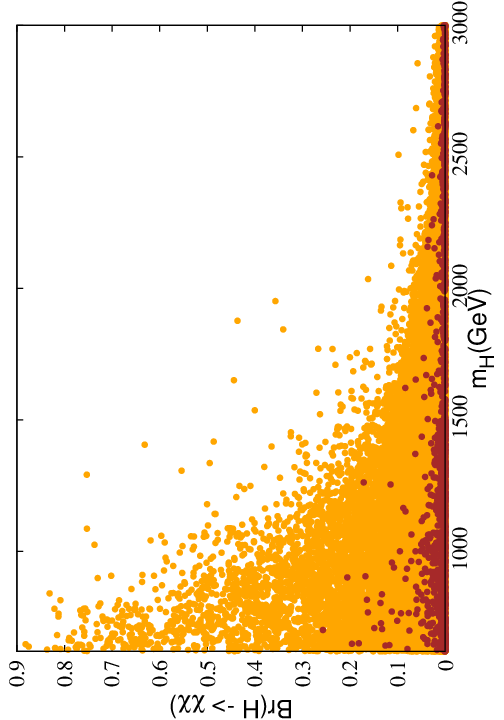} 
\caption{Br$(H \rightarrow \chi \chi$) as a function of $m_H$ for Type I (left) and Type II (right) 2HDM. Color code is same as in Figure.~\ref{omega_lam1}.}
\label{br_mH}
\end{figure}

Having identified regions in which the heavy neutral scalar H is the DM portal,
we would finally like to see if the invisible decays of the H produced at the LHC
can lead to some characteristic signal at observable rates. One
crucial  deciding factor in this is the heavy Higgs invisible branching ratio.  
With this in view, we plot   Br($H \rightarrow \chi \chi$) as a function $m_H$ in Figure.~\ref{br_mH}. The orange points satisfy only relic density constraints and maroon points satisfy both relic density and direct detection constraints. We can see from this figure that it is possible to achieve typically 20-30\% invisible branching fraction for the heavy Higgs. 

It should be mentioned here that for type II 2HDM, the invisible branching fraction of the heavier Higgs depends on $\tan \beta$ as well. In the low $\tan \beta$ region, the $Ht \bar t$ interaction strength gets enhanced for $m_H > 2 m_t$, the decay of heavy Higgs occurs predominantly in the $t \bar t$ channel. Another dominant decay channel for $H$ in the low $\tan \beta$ region is $H \rightarrow h h$ when it is kinematically allowed. At large $\tan \beta$ the $b \bar b$ decay mode of heavy Higgs becomes most important. In the intermediate region the invisible decay modes become important along with decays into weak gauge boson pairs.

In case of Type I, the branching ratio $H \rightarrow \chi \chi$ decreases monotonically with $\tan \beta$, because the coupling $\lambda_{H \chi \chi}$ decreases with $\tan \beta$. We see that it is possible to obtain $\sim 20-30$\% invisible branching fraction only in the small and intermediate range of $\tan \beta$ in Type I 2HDM.

\section{Collider Analysis (Cut based)}\label{sec5}

The discussion in the foregoing sections convince us that a heavy neutral scalar
in a two-Higgs doublet scenario may serve as portal to the dark sector,
consistently with all constraints   related to dark matter as well as the Higgs sector itself.
 We concentrate next on the strategies to look for any signal such a scenario at the 
high luminosity LHC. We have considered the case when only the heavier CP-even neutral Higgs can go to invisible decay modes. As the process will always involve missing energy because of the presence of the singlet scalar stable DM candidate ($\chi$) in the final state, one has to consider some visible final state which recoils against $\chi$. The most promising channel in this regards seems to be monojet + $\slashed{E_T}$ which is
already well-studied in the context of collider searches for DM particles.

Of course, the most obvious production channel for the heavy Higgs is  gluon fusion. 
One has to keep in mind at the same time the copious QCD background
to monojets, which may be difficult to manage in such regions of the parameter space
where the production is on the lower side. Therefore, we also investigate
the other  option, namely,  to look for the final state with two forward jets + $\slashed{E_T}$, when the heavy Higgs is produced through vector boson fusion (VBF) and decays  invisibly. The missing $p_T$ recoils in the azimuthal plane against the two forward jets. 
This channel has the promise of better background reduction despite its relatively
lower signal rates. Keeping all this mind, we present our results on both of the
above channels.

All signals and the corresponding backgrounds have been calculated at the
next-to-leading order, using Madgraph@MCNLO~\cite{Alwall:2014hca}.
MLM matching has been performed using using appropriate XQCUT variables.
nn23lo1 parton distribution functions have been used, with the renormalisation and
factorisation scales set at the $p_T$ of the hardest jet. We  have checked that 
such scale choice does not cause the rates to differ by more than 10\%, compared
to other choices such as the heavy Higgs mass. The showering and hadronization have been performed by PYTHIA8~\cite{Sjostrand:2006za}.
The detector simulation has been done by Delphes-3.4.1~\cite{deFavereau:2013fsa}.

We will discuss the results of our cut-based analysis for a few benchmarks from both Type I and Type II 2HDM. The benchmark points (BP) are so chosen that they obey all the theoretical and experimental constraints on Higgs and dark sector. Moreover, we 
highlight those regions of the parameter space where the invisible branching fraction of the heavy Higgs is non-negligible (4 - 22 \% for the various  BP's).  Based on such  considerations, we have chosen two benchmarks each for the Type I and  Type II 
scenarios. These benchmark points are presented in Table.~\ref{benchmark}.

Each benchmark point listed in Table.~\ref{benchmark} is representative of a substantial region in the parameter space, where the invisible branching ratio is shown in the last column. In addition, there are some combinations of parameters, which yield consistently with all constraints, large $H\chi\chi$ coupling (with $\frac{H\chi\chi}{v}$ close to the perturbative limit), and invisible branching ratios as large as 80\%. We have not used such points in our analysis, since they correspond to rather small and isolated regions of the parameter space.

\begin{table}[!hptb]
\begin{flushleft}
\begin{footnotesize}
\begin{tabular}{| c | c | c | c | c | c | c | c | c | c |}
\hline
& $m_H$(GeV) & $m_H^{\pm}$(GeV) & $m_A$(GeV) & $m_{\chi}$(GeV) & $\lambda_1$ & $\lambda_2$ & $\tan \beta$ & $\sin(\beta - \alpha)$ & Br($H \rightarrow \chi \chi$)  \\
\hline
Type I BP I & 236.0 & 279.8  & 277.6 & 113.4 & -0.86 & 0.14 & 4.42 & 0.88 & 17\% \\
\hline
Type I BP II & 146.4 & 146.0 & 141.5 & 71.3 & 0.011 & 0.011 & 10.0 & 0.88 & 4\% \\
\hline
Type II BP I & 629.4 & 668.6 & 654.4 & 252.1 & 4.99 & 4.99 & 4.5 & 0.88 & 21\% \\
\hline
Type II BP II & 644.0 & 661.1 & 671.1 & 280.3 & 2.98 & -3.28 & 5.74 & 0.92 & 30\% \\
\hline
\end{tabular}
\end{footnotesize}
\caption{The Benchmark points for Type I and Type II}
\label{benchmark}
\end{flushleft}
\end{table}

It is in order to also mention the Type X 2HDM scenario in the context of invisible decay of $H$ as was promised in Section~\ref{sec2}. The Type X 2HDM is similar to Type I case with the only difference in the coupling of the Higgs bosons with the lepton sector. The coupling of $H$ with the leptons is proportional to $\tan \beta$ near the alignment limit in Type X case. Therefore at large $\tan \beta$, the $H \tau \tau$ interaction strength dominates over all other interactions of $H$. For low enough $m_H$, when $m_H < 2m_t$ or $m_H < 2m_h$, the $t \bar t$, $h h$ decay modes of $H$ are kinematically forbidden. Therefore in this region the invisible branching fraction of $H$ can be comparable with $\text{Br} (H \rightarrow \tau \tau$), when $\tan \beta$ is not so large. For large $m_H$, many other decay modes open up and therefore it is possible to get moderate invisible branching fraction for heavy Higgs only with low enough $\tan \beta$. An interesting scenario with a scalar
  dark matter in the Type X 2HDM has been considered in~\cite{Bandyopadhyay:2017tlq}.

\subsection{Gluon fusion}

The production of $H$ in gluon fusion in association with a hard jet leads to monojet + $\slashed{E_T}$, a much advertised tell-tale signature of WIMP DM at the LHC, when the $H$ decays invisibly. While this final state has been looked for in the recent experiments~\cite{Aaboud:2017phn}, we present a projection for the high luminosity run which is essential in unveiling the scenario under consideration.

\medskip

\noindent
{\bf Signal:}  The signal here is one hard  jet + $\slashed{E_T}$. \\
The jet can originate in the hard scattering as well as from the radiation from the initial gluon legs. But as we demand that this jet should recoil largely against the massive heavy Higgs decaying to invisible final states, it must be characterised by the absence of
any other jet with comparable $p_T$ whose threshold is set to be
 sufficiently large.

\medskip
\noindent
{\bf Backgrounds:} The major backgrounds come from~\cite{Aaboud:2017phn}

\begin{itemize}
\item $Z(\rightarrow \nu \bar{\nu})$ + jets.
\item $W(\rightarrow l \nu)$ + jets.
\item QCD multijet events where mismeasurement of jet energy can give rise to $\slashed{E_T}$ .
\item $t \bar t$ production with one or both tops decaying hadronically.
\end{itemize}

\medskip
\noindent
{\bf Distributions:}

\medskip
\noindent
Before we present our results of the cut based analysis we present here the distributions of relevant observables for signal and backgrounds. These distributions have led us to suitable cuts to enhance the signal significance.

\begin{figure}[!hptb]

\includegraphics[width=9.0cm, height=7cm]{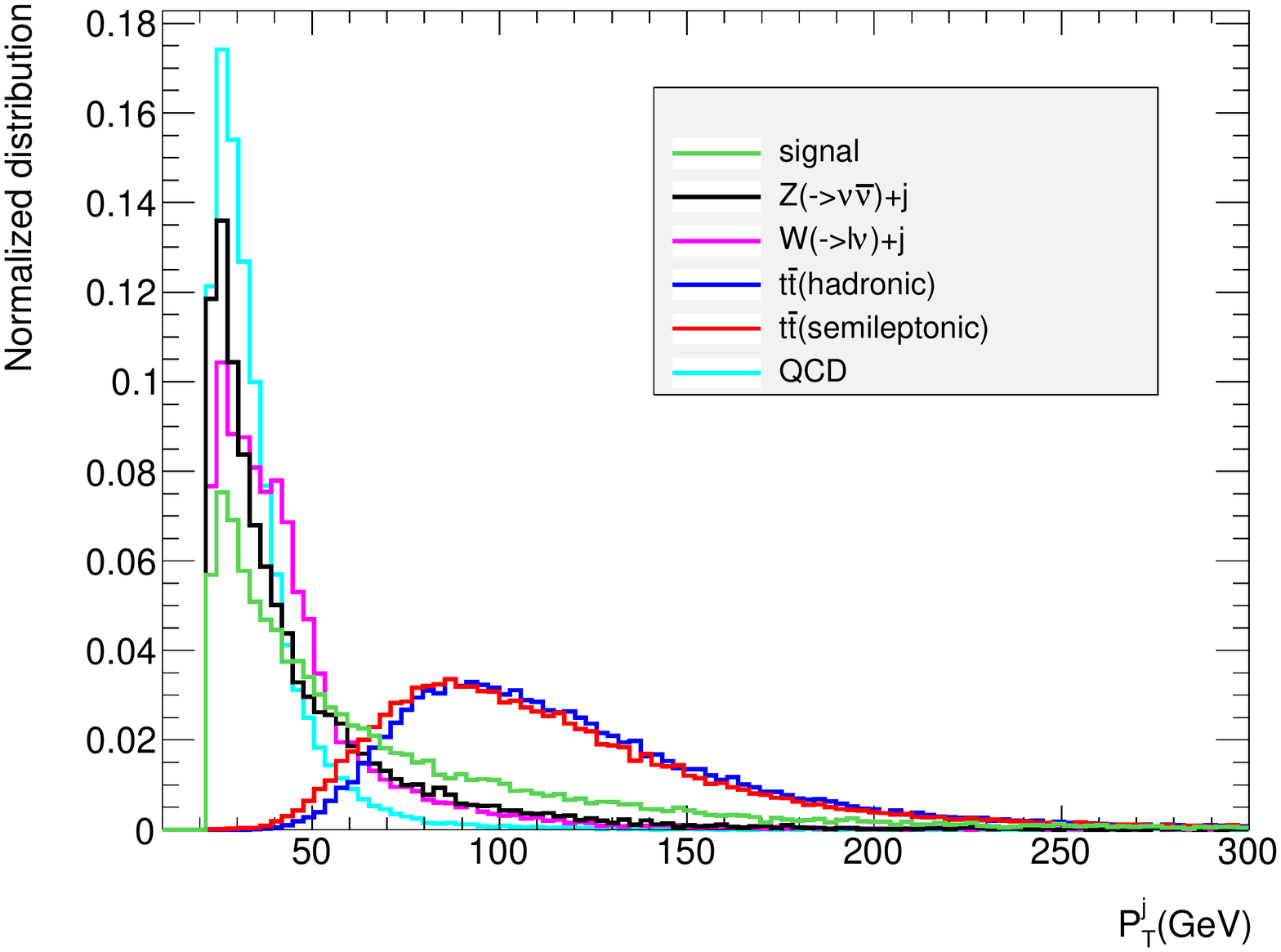}
\hspace{0.00001cm}
\includegraphics[width=9.0cm, height=7cm]{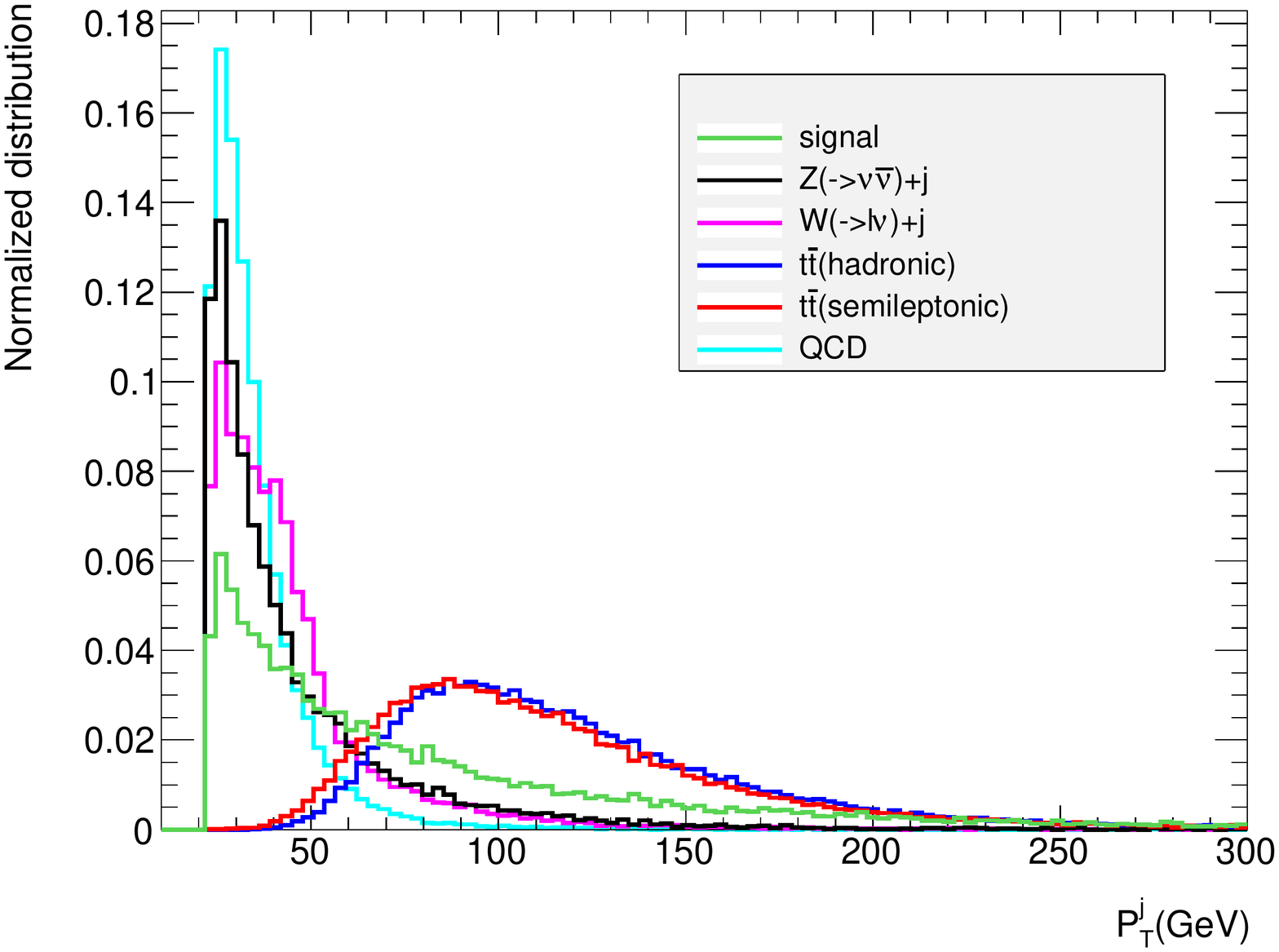} \\
\vspace*{0.0001cm}
\includegraphics[width=9.0cm, height=7cm]{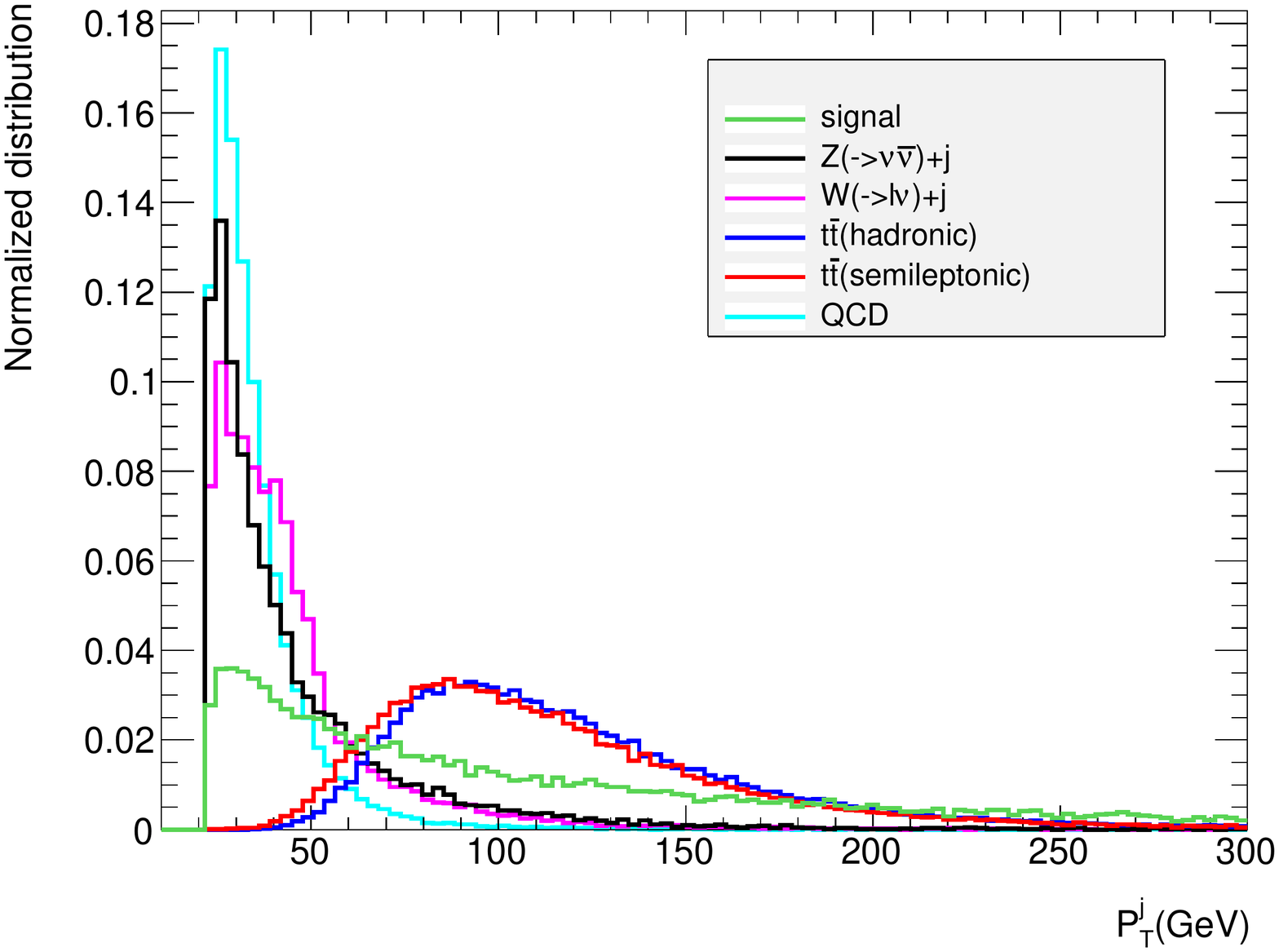}
\hspace{0.02cm}
\includegraphics[width=9.0cm, height=7cm]{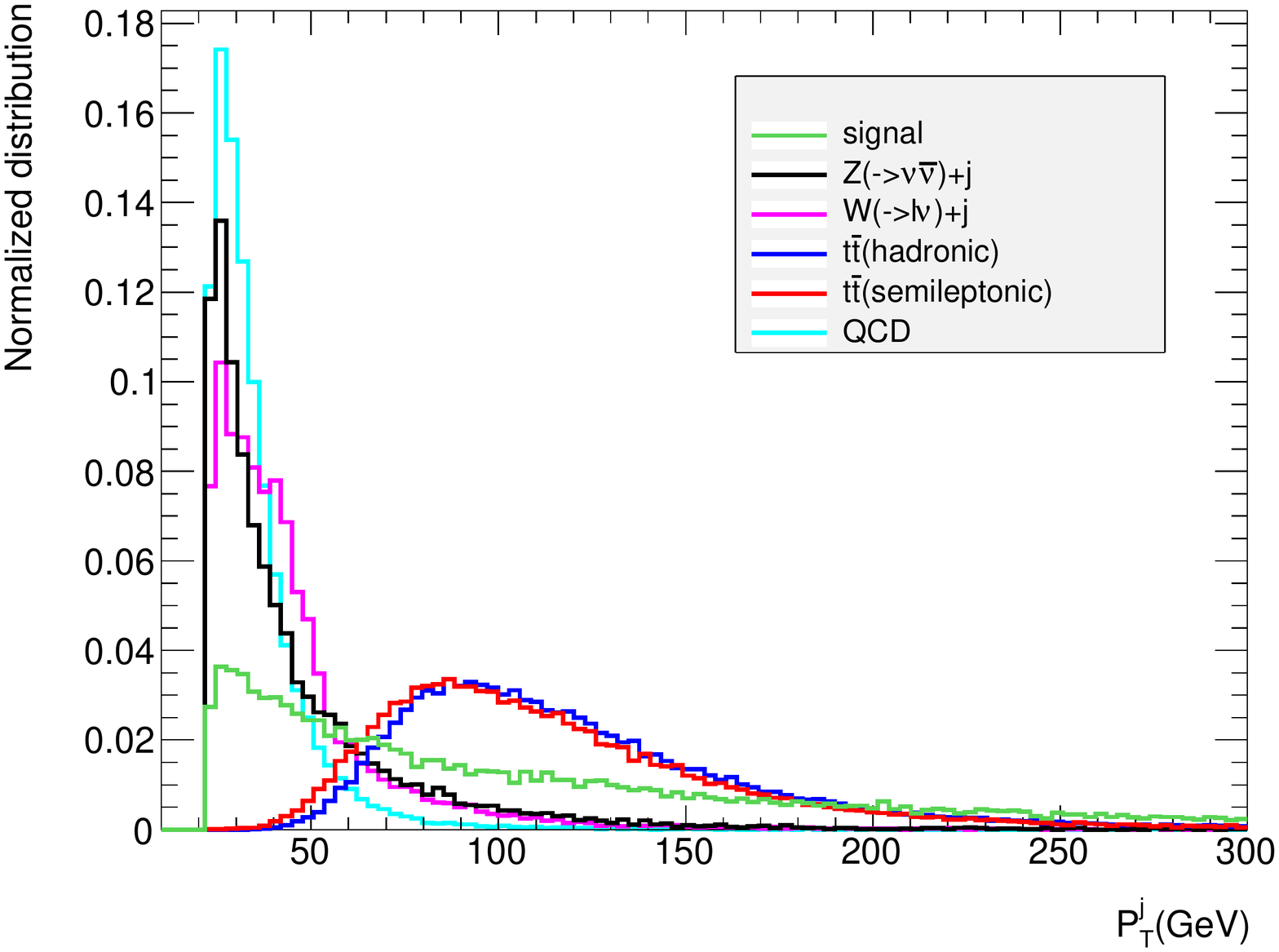}

\centering

\caption{$P_T$ distribution of the leading jet for gluon fusion signal and background processes, Type I BP I(top left), Type I BP II(top right), Type II BP I(bottom left) and Type II BP II(bottom right).}
\label{ggfjetpt}
\end{figure}

\begin{figure}[!hptb]
\includegraphics[width=9.0cm, height=7cm]{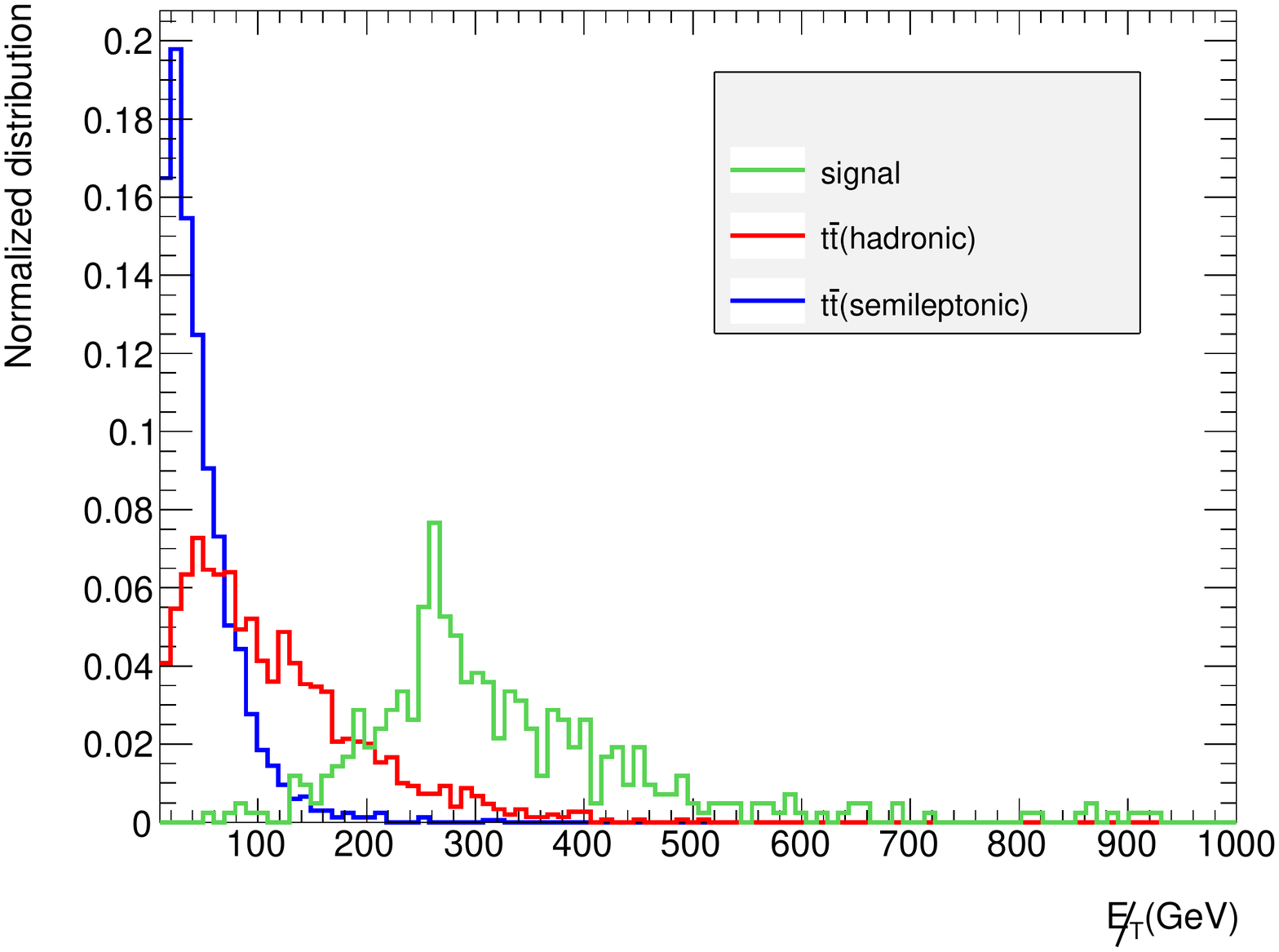}
\hspace{0.02cm}
\includegraphics[width=9.0cm, height=7cm]{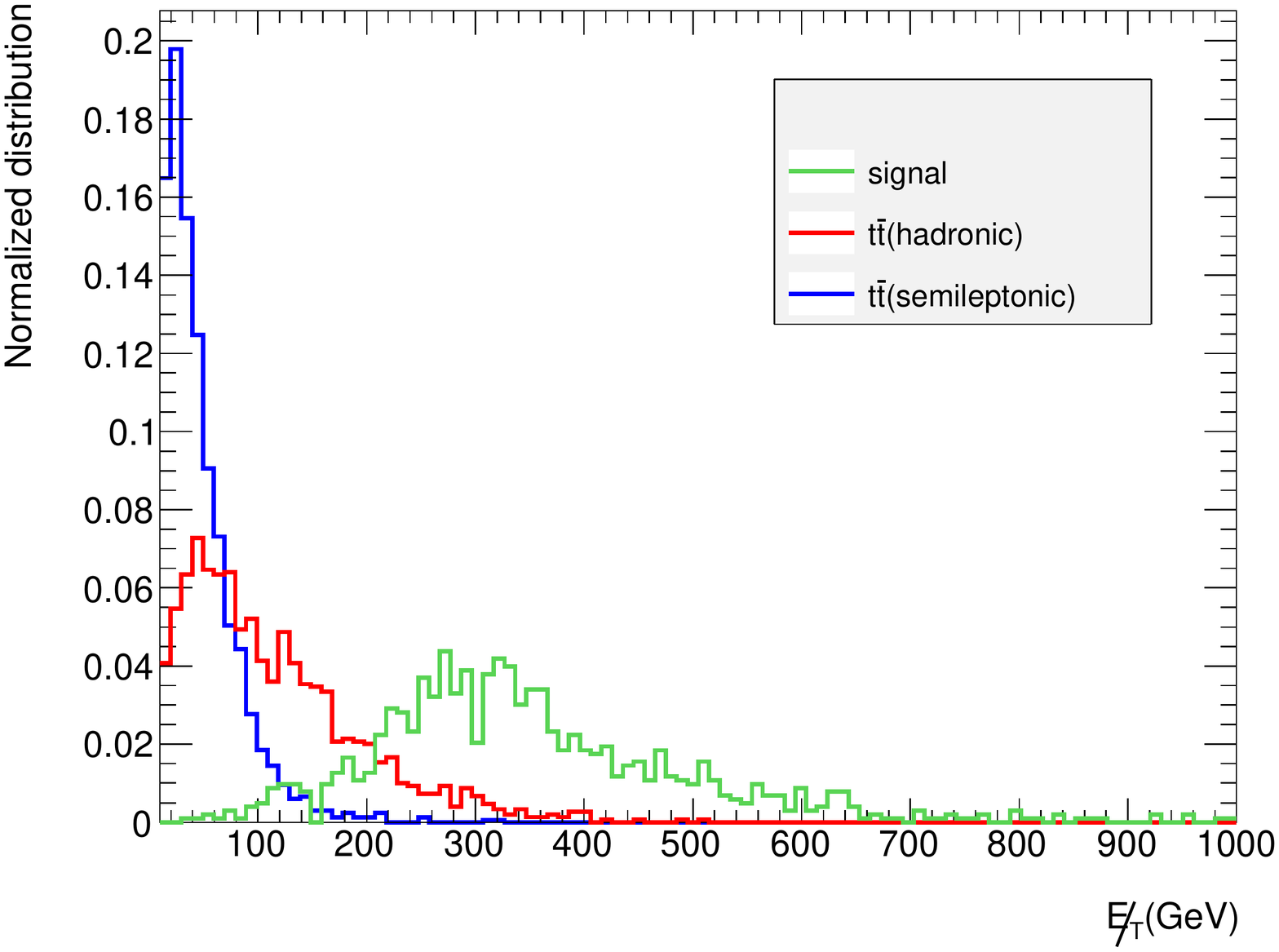} \\
\vspace*{0.0001cm}

\includegraphics[width=9.0cm, height=7cm]{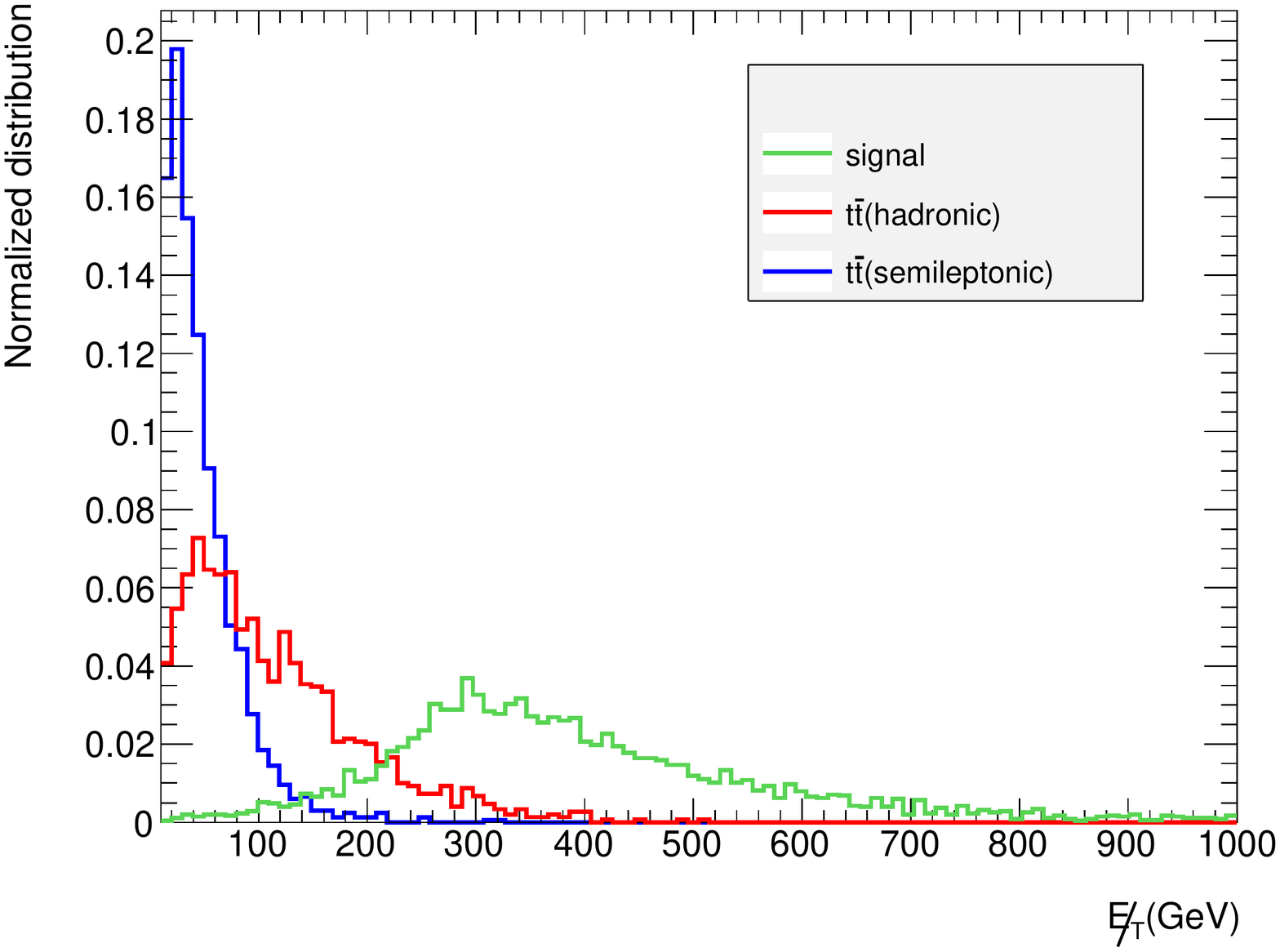}
\hspace{0.02cm}
\includegraphics[width=9.0cm, height=7cm]{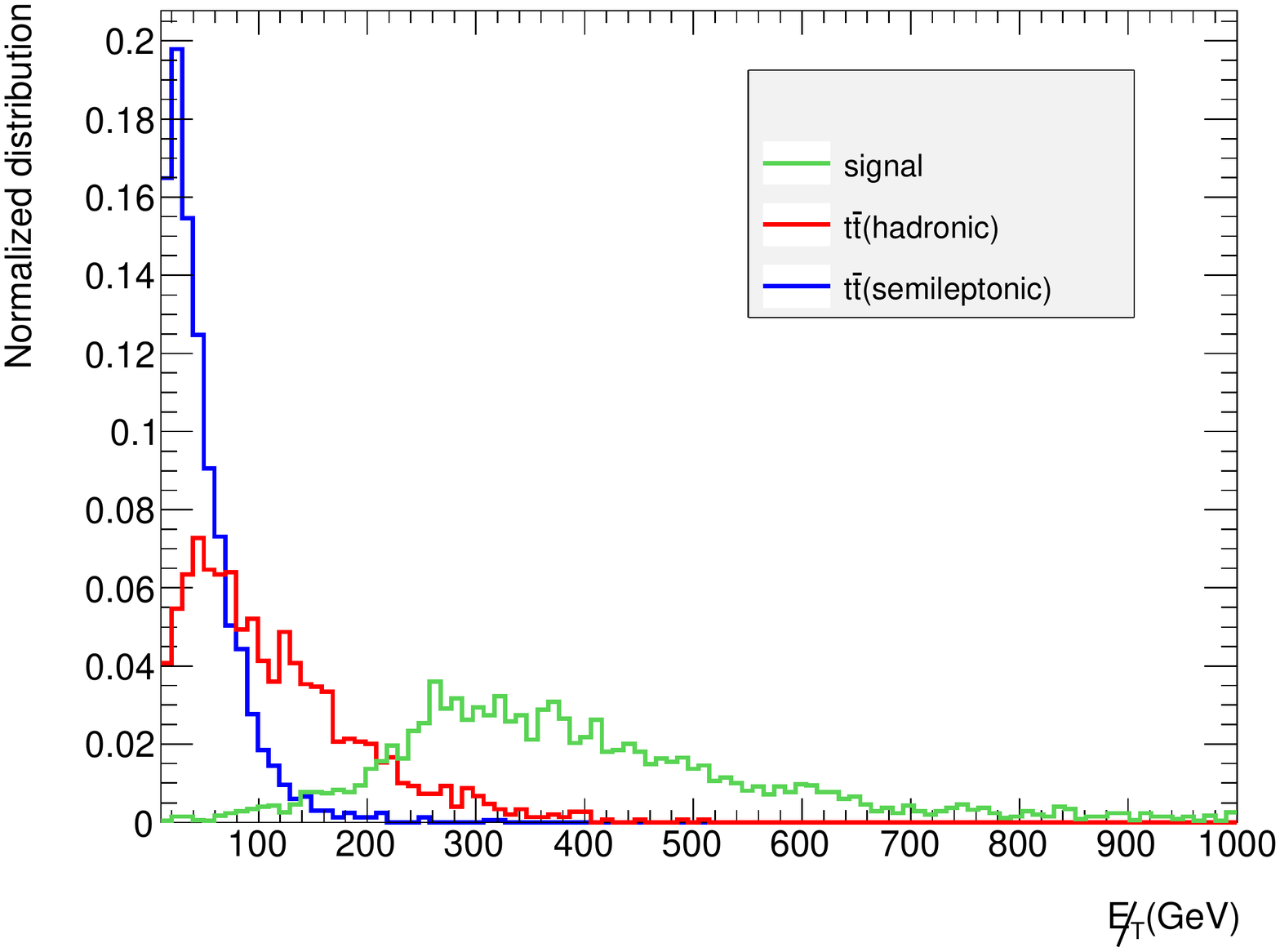}

\centering

\caption{$\slashed{E_T}$ distribution for gluon fusion signal and background processes, Type I BP I(top left), Type I BP II(top right), Type II BP I(bottom left) and Type II BP II(bottom right).}
\label{ggfmisspt}
\end{figure}

\begin{figure}[!hptb]
\includegraphics[width=9.0cm, height=7cm]{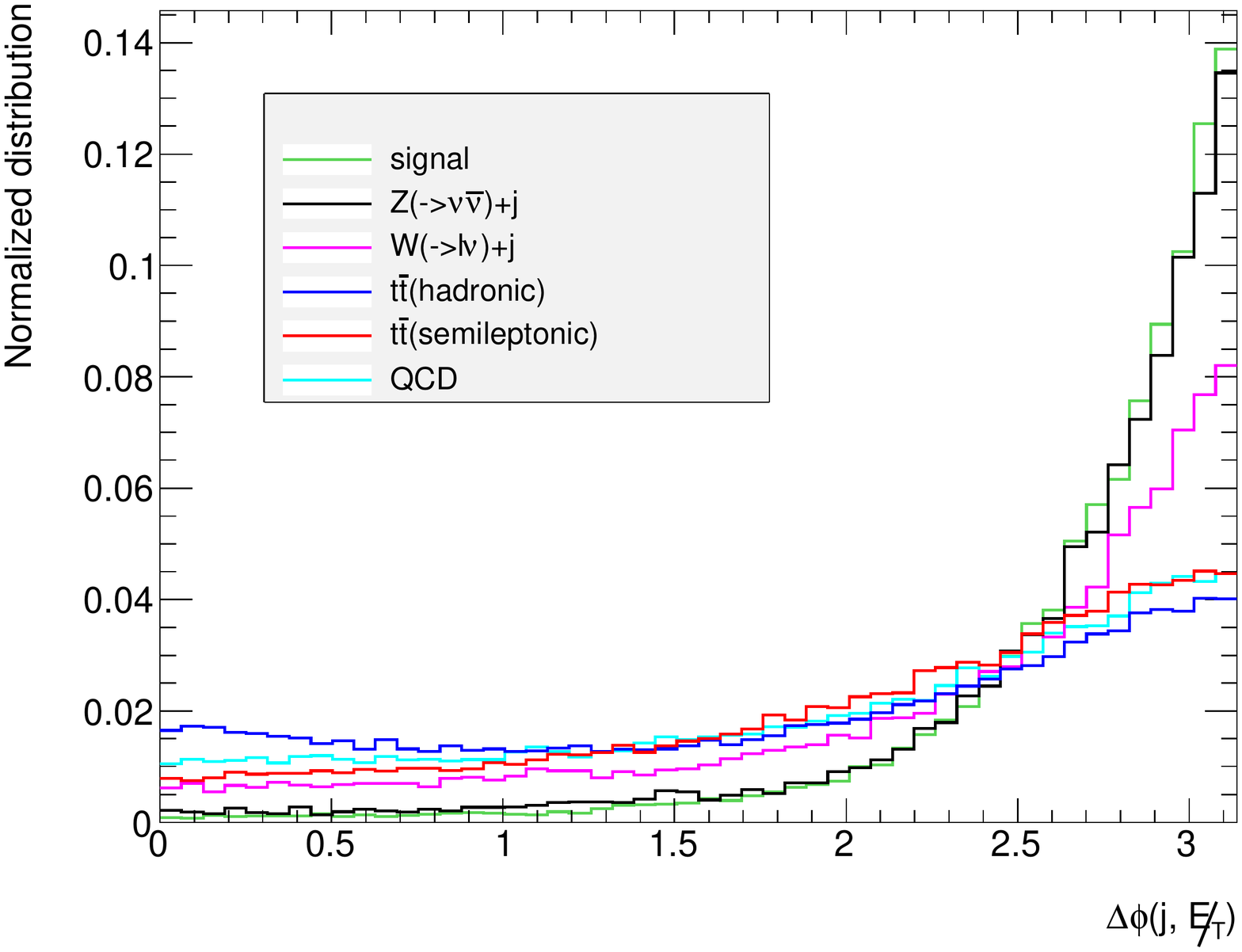}
\hspace{0.02cm}
\includegraphics[width=9.0cm, height=7cm]{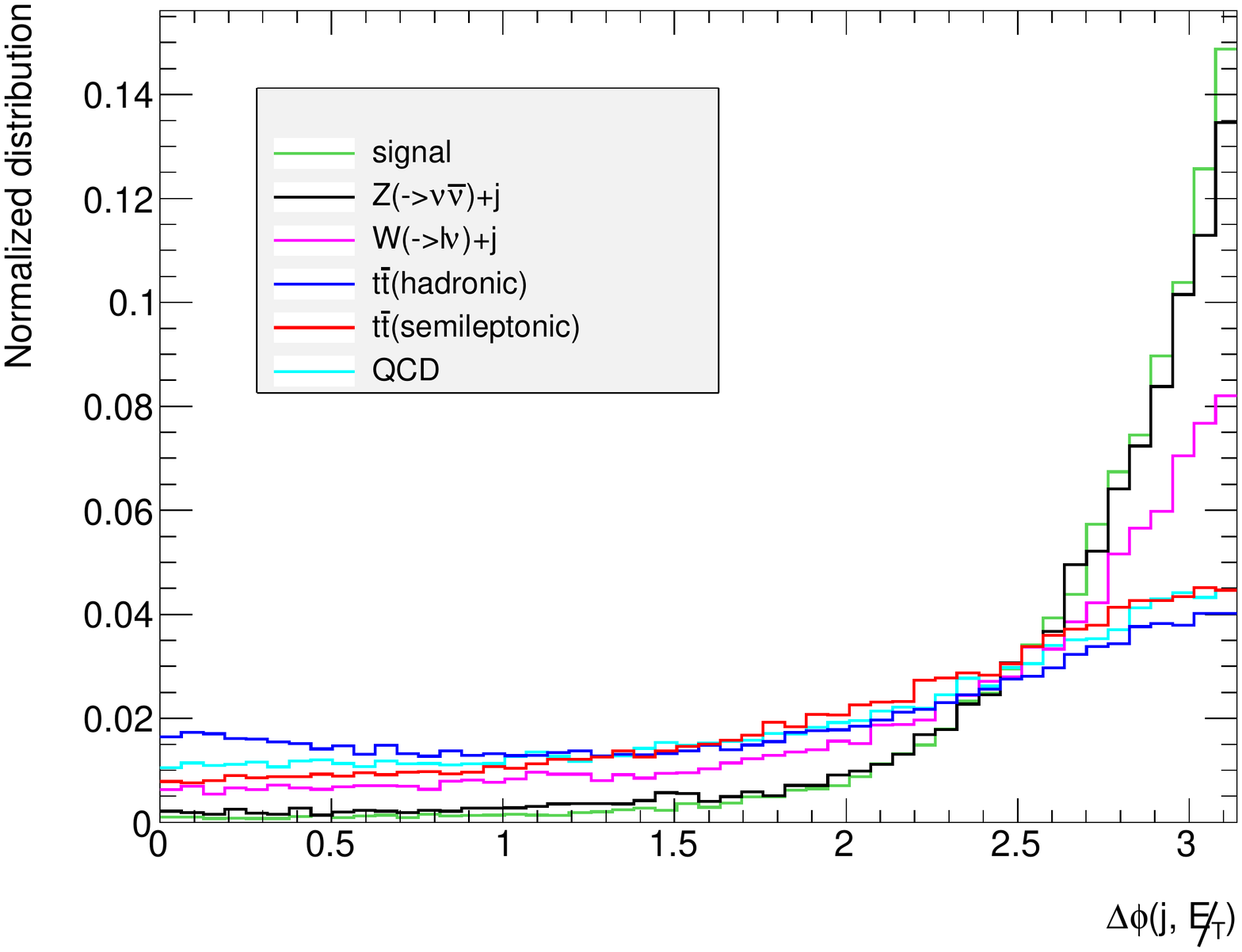} \\
\vspace*{0.0001cm}

\includegraphics[width=9.0cm, height=7cm]{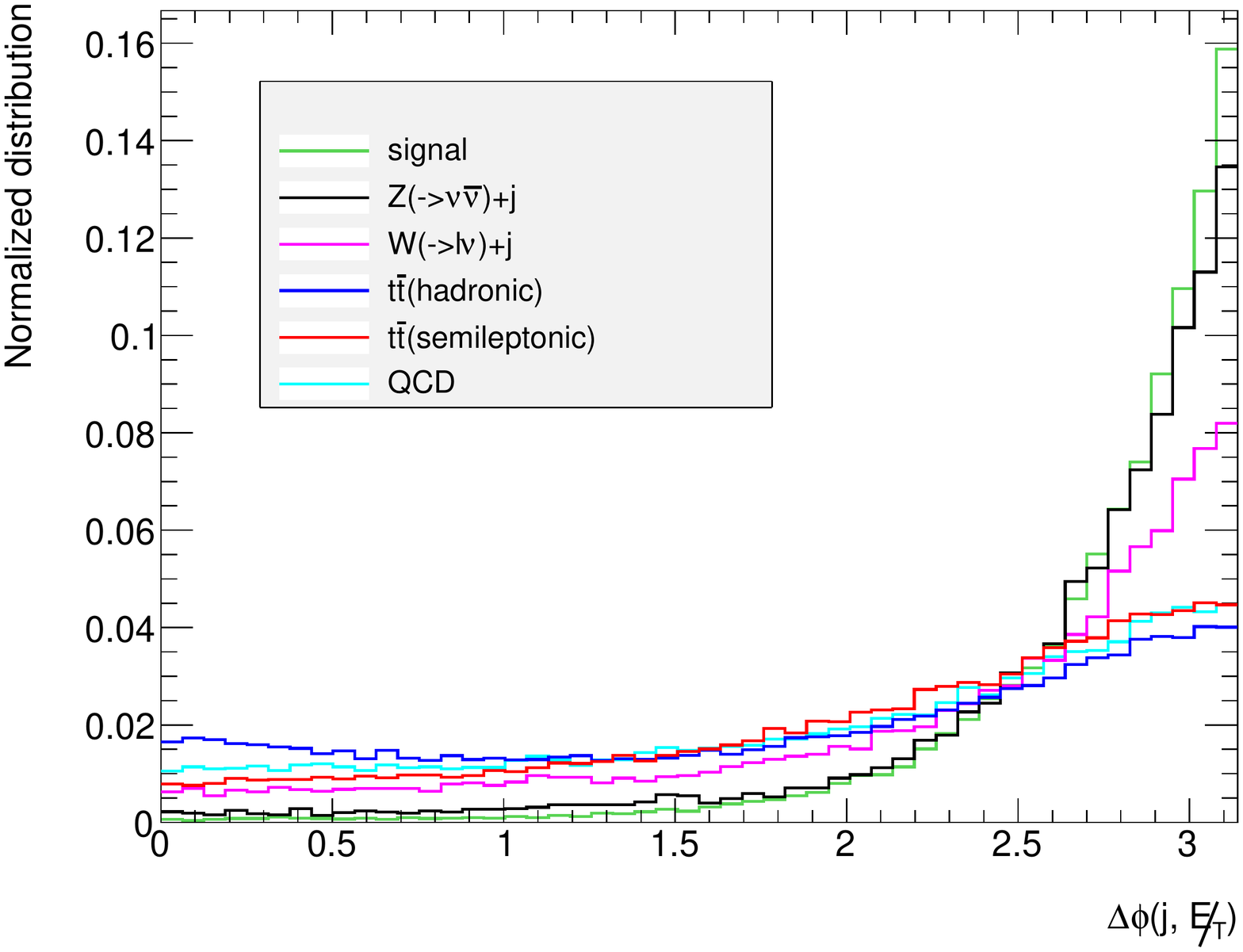}
\hspace{0.02cm}
\includegraphics[width=9.0cm, height=7cm]{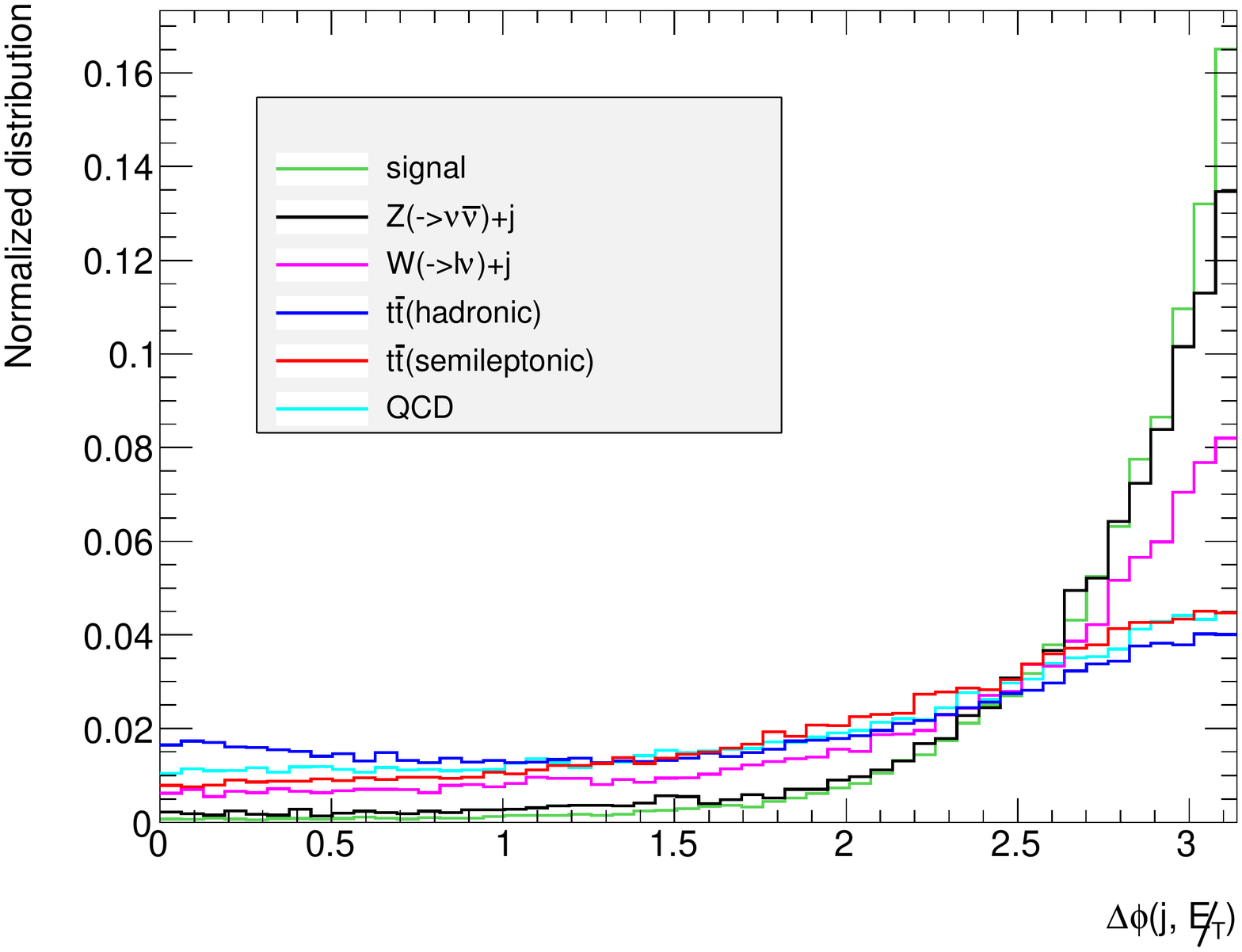}

\centering

\caption{$\Delta \phi$(jet, $\slashed{E_T}$) distribution for gluon fusion signal and background processes, Type I BP I(top left), Type I BP II(top right), Type II BP I(bottom left) and Type II BP II(bottom right).}
\label{ggfjetmiss}
\end{figure}

\begin{figure}[!hptb]
\includegraphics[width=9.0cm, height=7cm]{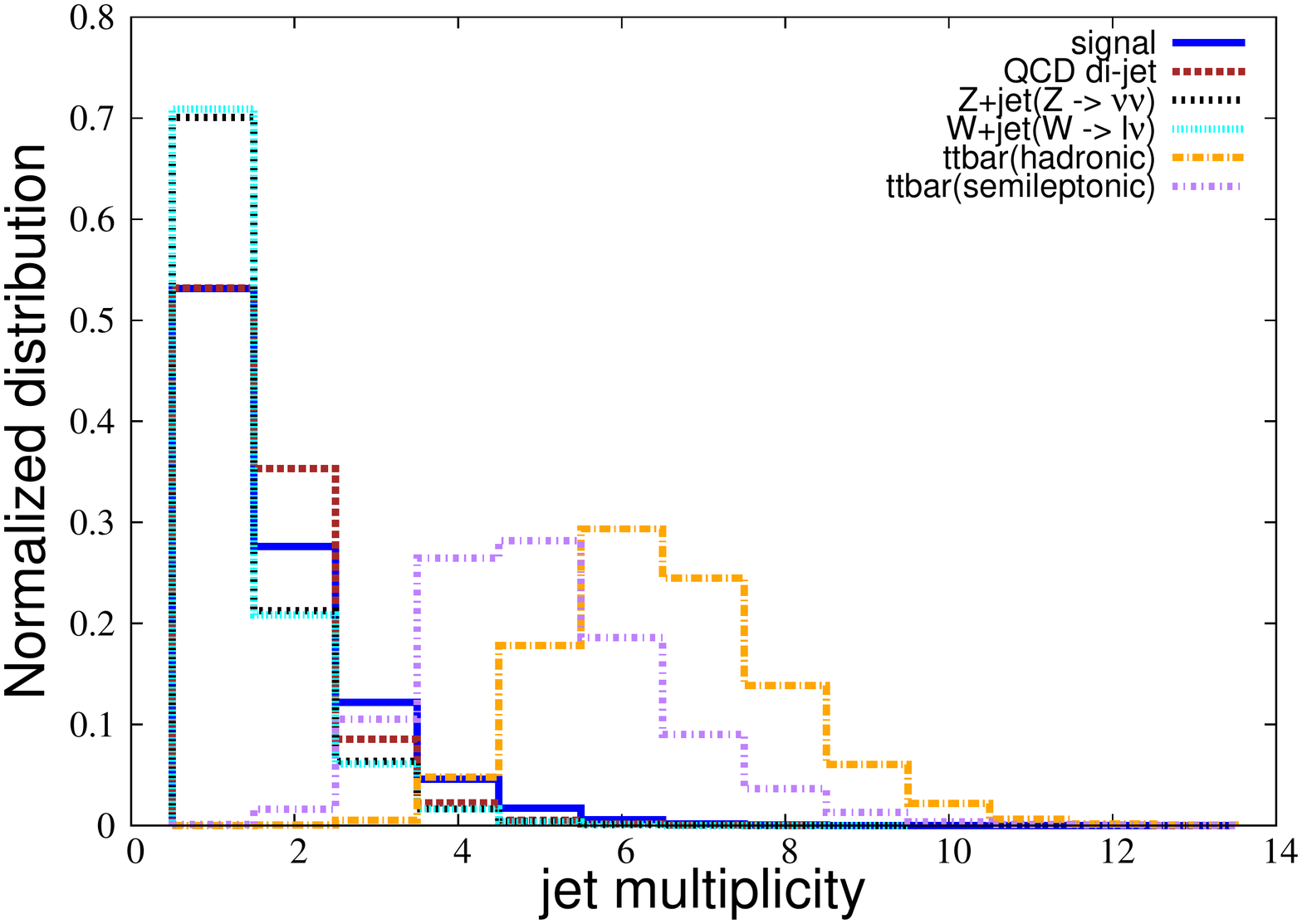}
\hspace{0.02cm}
\includegraphics[width=9.0cm, height=7cm]{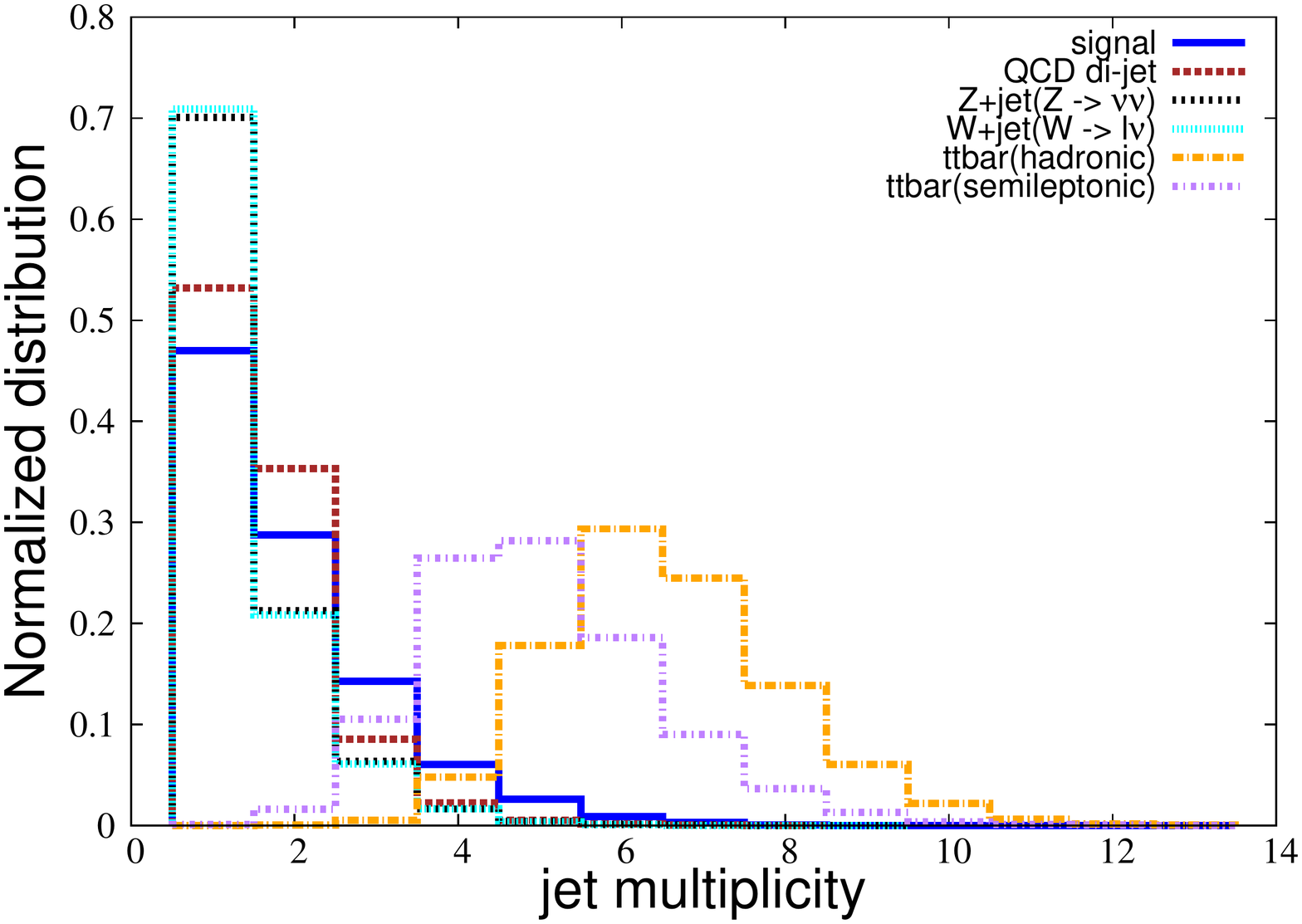} \\
\vspace*{0.0001cm}

\includegraphics[width=9.0cm, height=7cm]{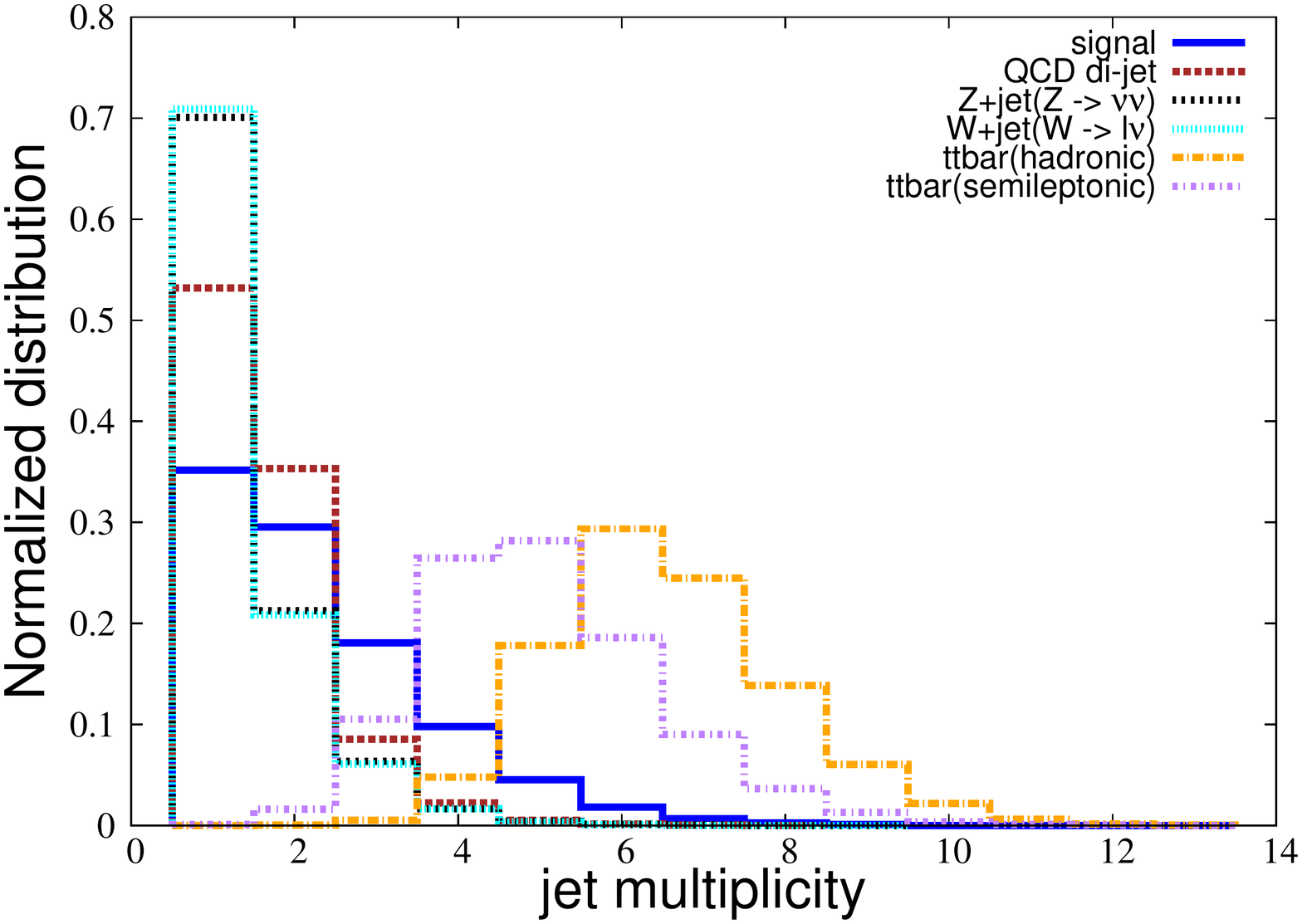}
\hspace{0.02cm}
\includegraphics[width=9.0cm, height=7cm]{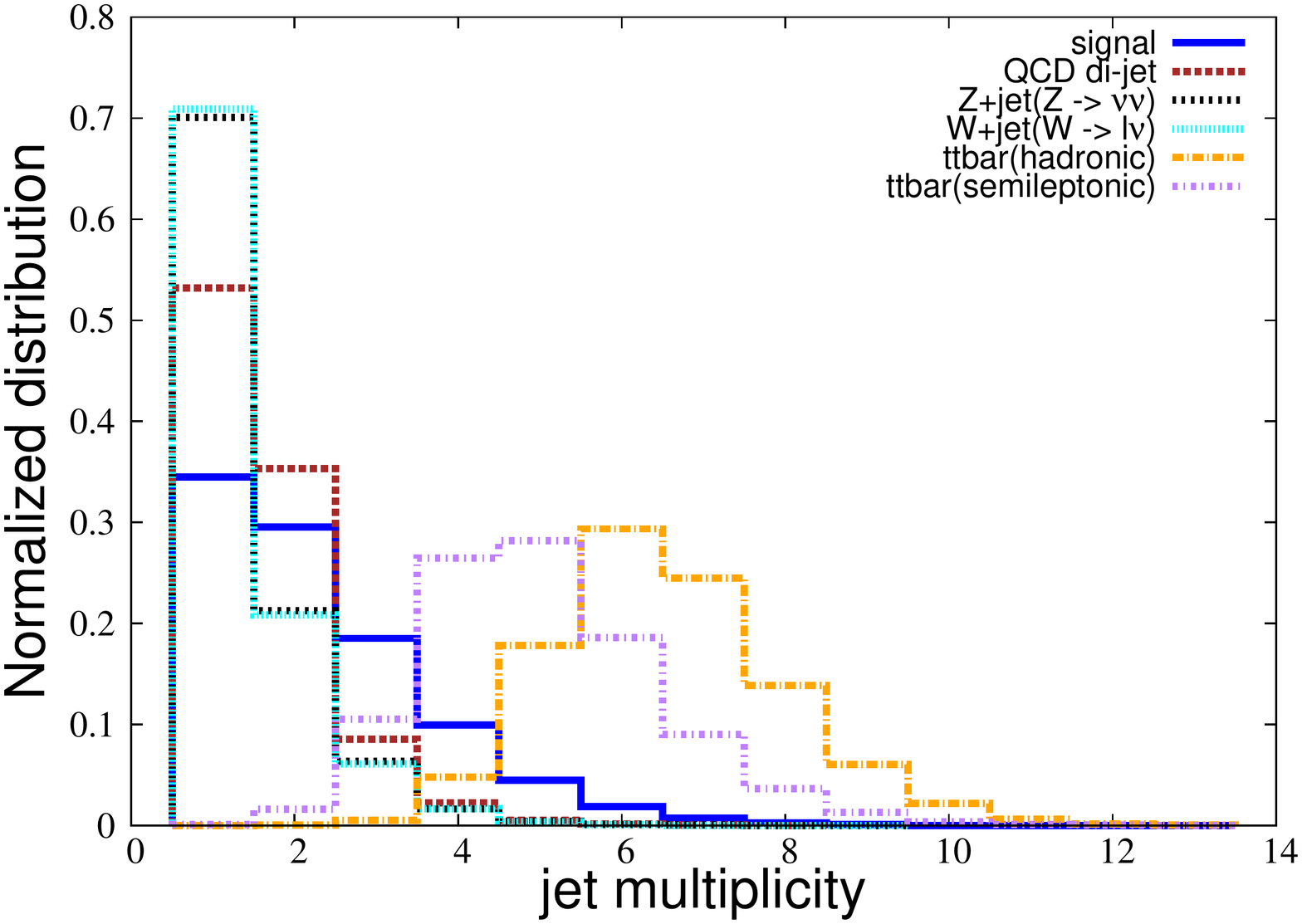}

\centering

\caption{jet multiplicity distribution for gluon fusion signal and background processes, Type I BP I (top left), Type I BP II (top right), Type II BP I (bottom left) and Type II BP II (bottom right).}
\label{jet_mult}
\end{figure}

In Figure.~\ref{ggfjetpt}, we show the normalised $p_T$ distribution of the leading jet for the four benchmark points, together with those for all the aforesaid backgrounds. We find the distribution of $p_T$ of the leading jet fall slower than that for the backgrounds coming from $Z+$ jets, $W$ + jets and QCD multijet. A hard $p_T$ cut thus helps us reduce these backgrounds. The jet $p_T$ distribution for $t \bar t$ channel peaks at a larger value. Therefore considerable $t \bar t$ background will remain after applying this cut which is also evident from the figure. 

We next plot the $\slashed{E_T}$ distribution of signal and the $t \bar t$ backgrounds in Figure.~\ref{ggfmisspt} after applying a hard $p_T$ cut of 250 GeV on the leading jet. The $m_H$ going entirely invisible causes the $\slashed{E_T}$-distribution to peak at higher values for all benchmark points. Thus a suitable $\slashed{E_T}$ cut, too is of considerable help.

 In Figure.~\ref{ggfjetmiss} we plot the $\Delta \phi$(jet, $\slashed{E_T})$ distribution for the signal and the backgrounds. In case of QCD multijet backgrounds, $\slashed{E_T}$ is expected to arise mainly from the mismeasurement of jet energy, from unclustered particles or from invisible decay of jet components. On the other hand, in case of signal the $\slashed{E_T})$ is most likely to recoil against the leading jet in the azimuthal plane. Therefore the $\Delta \phi$(jet, $\slashed{E_T})$ distribution sharply peaks close to $\sim \pi$ in case of signal as well as $Z$ + jets or $W$+ jets background. A cut on this variable will help us reduce the QCD background.Moreover, the veto on hard leptons largely reduce the $W(\rightarrow l \nu)$ and $t \bar t$(semileptonic) backgrounds. 

We mention here that it is possible to reduce the background systematic uncertainty by data driven background estimation method adopted in~\cite{Aaboud:2017phn}. The extrapolation of background events from the control region to the signal region does reduce the systematic uncertainty and enhance signal significance.

We also show the jet-multiplicity distribution for signal and all the backgrounds in Figure.~\ref{jet_mult}. The jet multiplicity peaks at larger value for the $t \bar t$ background. We do not put any hard cut on this variable, but rather use it as an input in the neural network and boosted decision tree approach discussed in the Section.~\ref{sec6}. Before proceeding in that direction, however, we go ahead and present our results based on various rectangular cuts.

\subsubsection{Results}

\medskip
{\bf Event selection criteria:}
Over and above the basic acceptance cuts listed, for example in \cite{Aaboud:2017phn}, 
the following selection criteria are imposed, based on the foregoing discussion : 

\begin{itemize}
\item Cut 1: $p_T$ of the singled-out jet $>$ 250 GeV. 
\item Cut 2: $\slashed{E_T} >$ 250 GeV. 
\item Cut 3: Lepton veto: Events with electrons with $p_T > 20$ GeV or muons with $p_T > 10$ GeV are not selected.
\end{itemize}

\noindent
Table.~\ref{cuts} contains the results for the gluon fusion channel after applying these cuts in succession in the signal and background processes, thus revealing the response of each cut.
In Table.~\ref{significance_ggf} we calculated the projected significance (${\cal S}$) in the gluon fusion channel for each benchmark point, for 14 TeV LHC with 3000 fb$^{-1}$. The significance ${\cal S}$ is defined as follows:

\begin{equation}
{\cal S} = \sqrt{2 [(S+B) \text{Log}(1+\frac{S}{B}) - S]}
\label{significance}
\end{equation}

Where $S$ and $B$ are the number of signal and background events surviving after applying all the cuts respectively.

\begin{table}[!hptb]
\begin{center}
\begin{footnotesize}
\begin{tabular}{| c | c | c | c | c | c | c | c | c |}
\hline
 & TypeI BP I & TypeI BP II & TypeII BP I & TypeII BP II & $Z(\nu \bar{\nu})+j$ & $W(l \nu) + j$ & $t \bar t$ hadronic & $t \bar t$ semileptonic  \\
\hline
$\sigma$(pb) & 0.25 & 0.22 & 0.042 & 0.035 & $1.3\times 10^4$  & $2.8 \times 10^4$  & 605.7  & 302.9  \\
\hline
Cut 1 & 3.9\% & 1.8\% & 12.1$\%$ & 12.1\% & 0.08\% & 0.05\% & 3.6\% & 3.0\% \\
\hline
Cut 2 & 3.0\% & 1.3\% & 10.0\% & 10.0\% & 0.05\% & 0.002\% & 0.008\% & 0.2\% \\
\hline
Cut 3 & 3.0\% & 1.3\% & 10.0\% & 10.0\% & 0.05\% & 0.002\% & 0.002\% & 0.002\% \\
\hline
\end{tabular}
\end{footnotesize}
\caption{Signal and background efficiencies after applying various cuts for the gluon fusion production channel at 14 TeV with ${\cal L}$ = 3000 fb$^{-1}$. The cross sections are calculated at NLO. }
\label{cuts}
\end{center}
\end{table}


\begin{table}[!hptb]
\begin{center}
\begin{footnotesize}
\begin{tabular}{| c | c |}
\hline
BP & $
{\cal S}$   \\
\hline
Type I BP I  &  5.1 $\sigma$  \\
\hline
Type I BP II  & 1.9 $\sigma$  \\
\hline
Type II BP I  & 2.7 $\sigma$  \\
\hline
Type II BP II  & 2.3 $\sigma$  \\
\hline
\end{tabular}
\end{footnotesize}
\caption{Signal significance for the benchmark points at 14 TeV with ${\cal L}$ = 3000 fb$^{-1}$ in the gluon fusion channel. }
\label{significance_ggf}
\end{center}
\end{table}

In Table.~\ref{significance_ggf}, we see that for Type I BP I the significance is largest. The reason is, for this benchmark point both production of $H$ and the branching fraction ($H \rightarrow \chi \chi$) are considerable. Although the production cross section for Type I BP II is also large, the low invisible branching fraction affects the signal significance. In case of Type II, the production cross section is smaller than that in case of Type I, because of large $m_H$. However, because of considerable invisible branching fraction and better separation between signal and background distributions Type II benchmark points result in moderate signal significance.

\subsection{Vector boson fusion}
Next, we explore the vector boson fusion (VBF) channel which is characterized by two energetic forward jets with negligible hadronic activity in the intervening rapidity gap. It is evident that in the 2HDM, the 125-GeV Higgs data pushes us to the `alignment limit' where $\sin (\beta-\alpha)$ is close to 1. This constrain in turn decreases the  heavy Higgs coupling to gauge bosons which is proportional to $\cos (\beta - \alpha)$. Therefore the cross section for VBF production of heavy Higgs is much suppressed as compared to that of gluon fusion process. However, the VBF channel has an advantage over the gluon fusion process, a much better discriminating power between the signal and the backgrounds. We will elaborate on this in the subsequent discussions. 

\noindent
{\bf Signal:}  The signal we consider here is two hard forward jets + $\slashed{E_T}$ \\

\noindent
{\bf Backgrounds:}

\begin{itemize}
\item $Z(\rightarrow \nu \nu)$ + jets (QCD)
\item $Z(\rightarrow \nu \nu)$ + jets (EW)
\item $W(\rightarrow l \nu), l= e, \mu, \tau)$ + jets (QCD)
\item $W(\rightarrow l \nu), l= e, \mu, \tau)$ + jets (EW)
\item QCD multijet with $\slashed{E_T}$ caused by jet-energy mismeasurement 
\item $p p \rightarrow h $+ 2 jets ($h \rightarrow Z Z, Z \rightarrow \nu \bar{\nu}$)
\end{itemize}
 
\noindent
{\bf Distributions:}

VBF is a pure electroweak process without color flow in the central region. The process
naturally leads to high $p_T$ jets in the forward region and allows for
no hadronic activity in the central region except from the decay of the Higgs itself. This feature of VBF is in contrast with most background processes, which typically involve t-channel color exchange and therefore lead to hadronic activity in the central region~\cite{Rainwater:1998kj}. Therefore it is possible to observe small signal rates in a region of phase
space which is not very populated by QCD background events. This advantage of VBF has
been used for the production of intermediate and large masses of Higgs
bosons in the literature~\cite{Hankele:2006ma,Sirunyan:2018owy,Khachatryan:2016whc,Aaboud:2019rtt}. In addition, the production of `invisible' objects for other scenarios have also been considered in the literature~\cite{Datta:2001hv}.

\begin{figure}[!hptb]
\includegraphics[width=9.0cm, height=7cm]{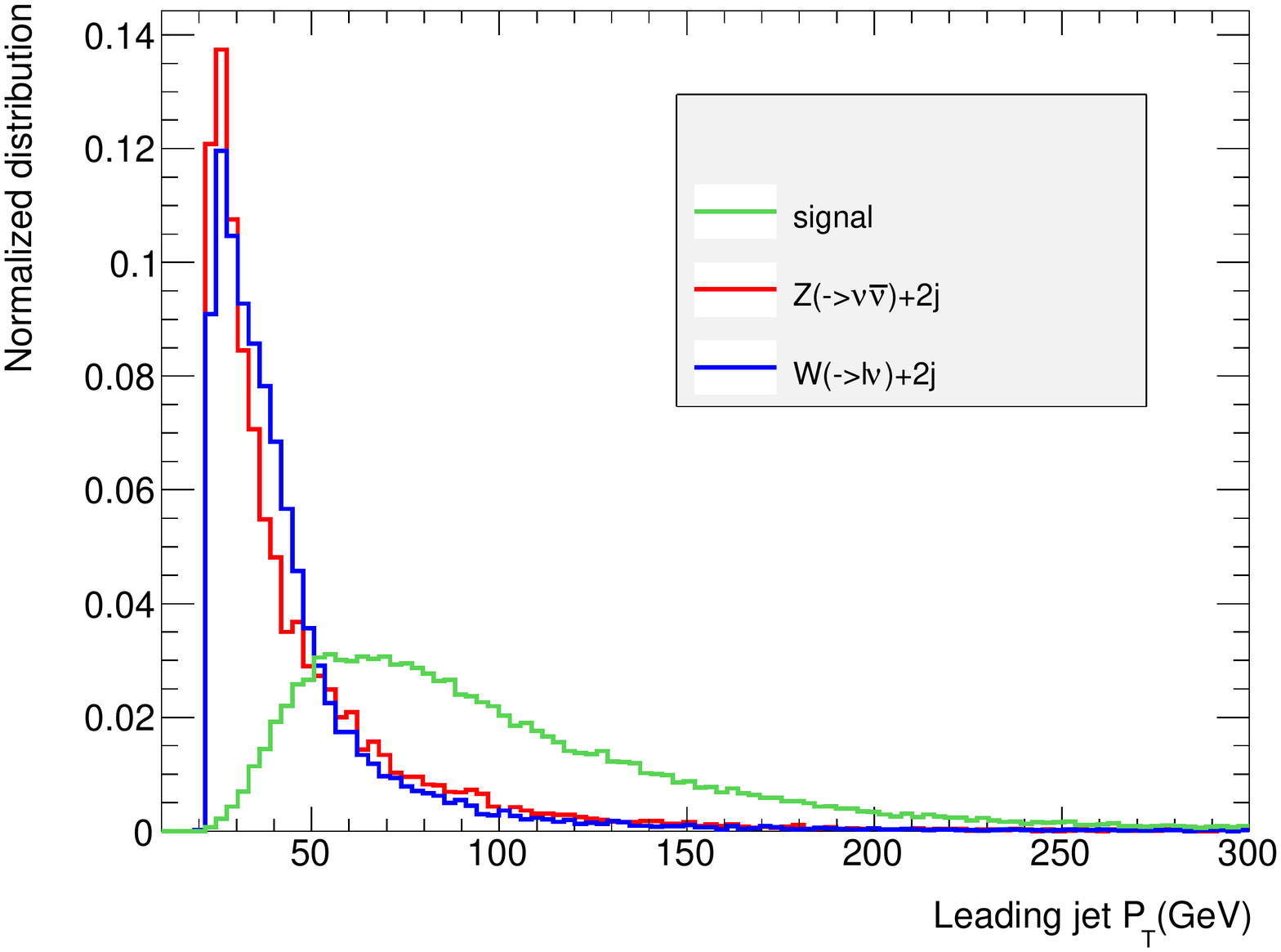}
\hspace{0.02cm}
\includegraphics[width=9.0cm, height=7cm]{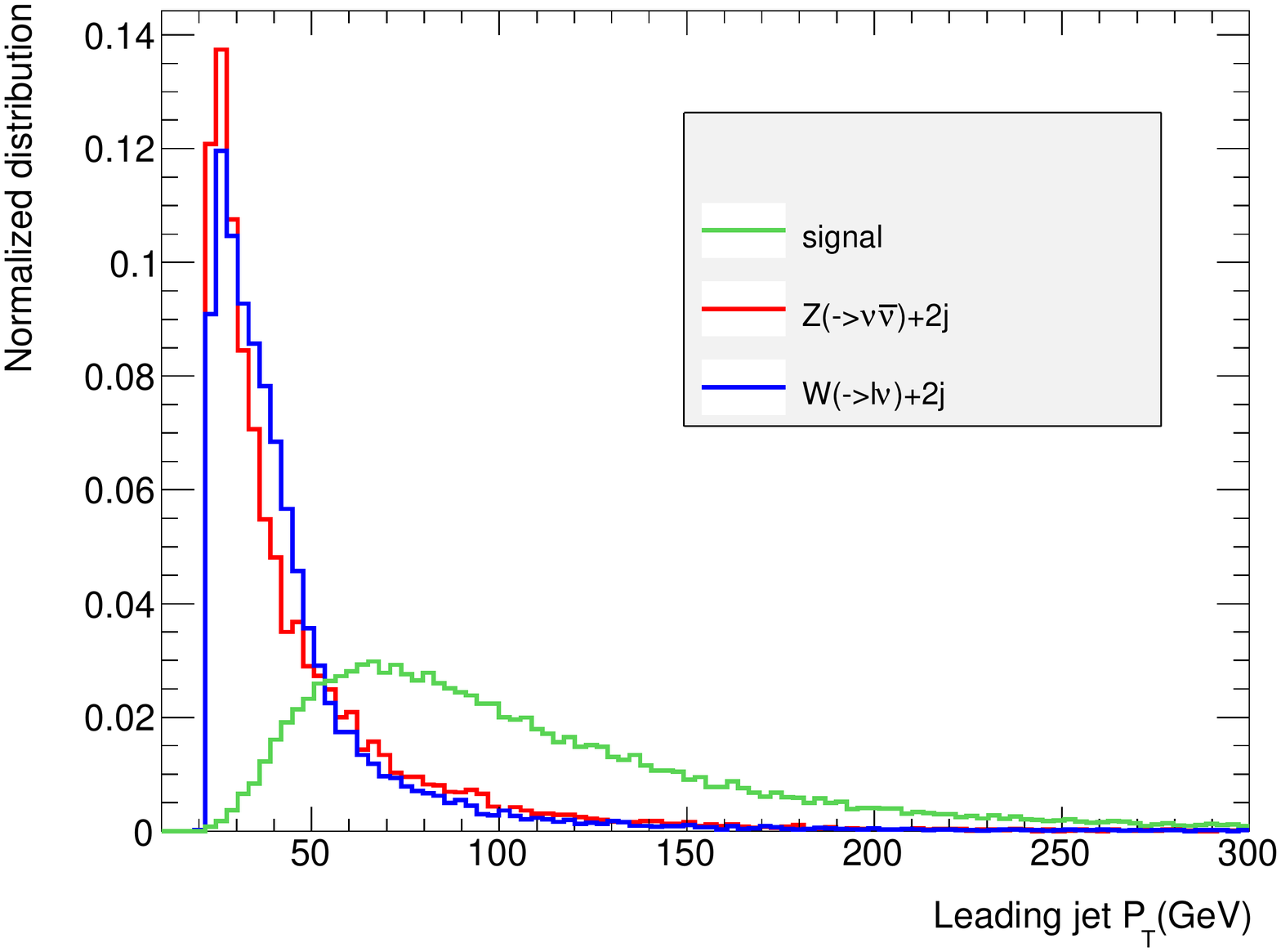} \\
\vspace*{0.02cm}
\includegraphics[width=9.0cm, height=7cm]{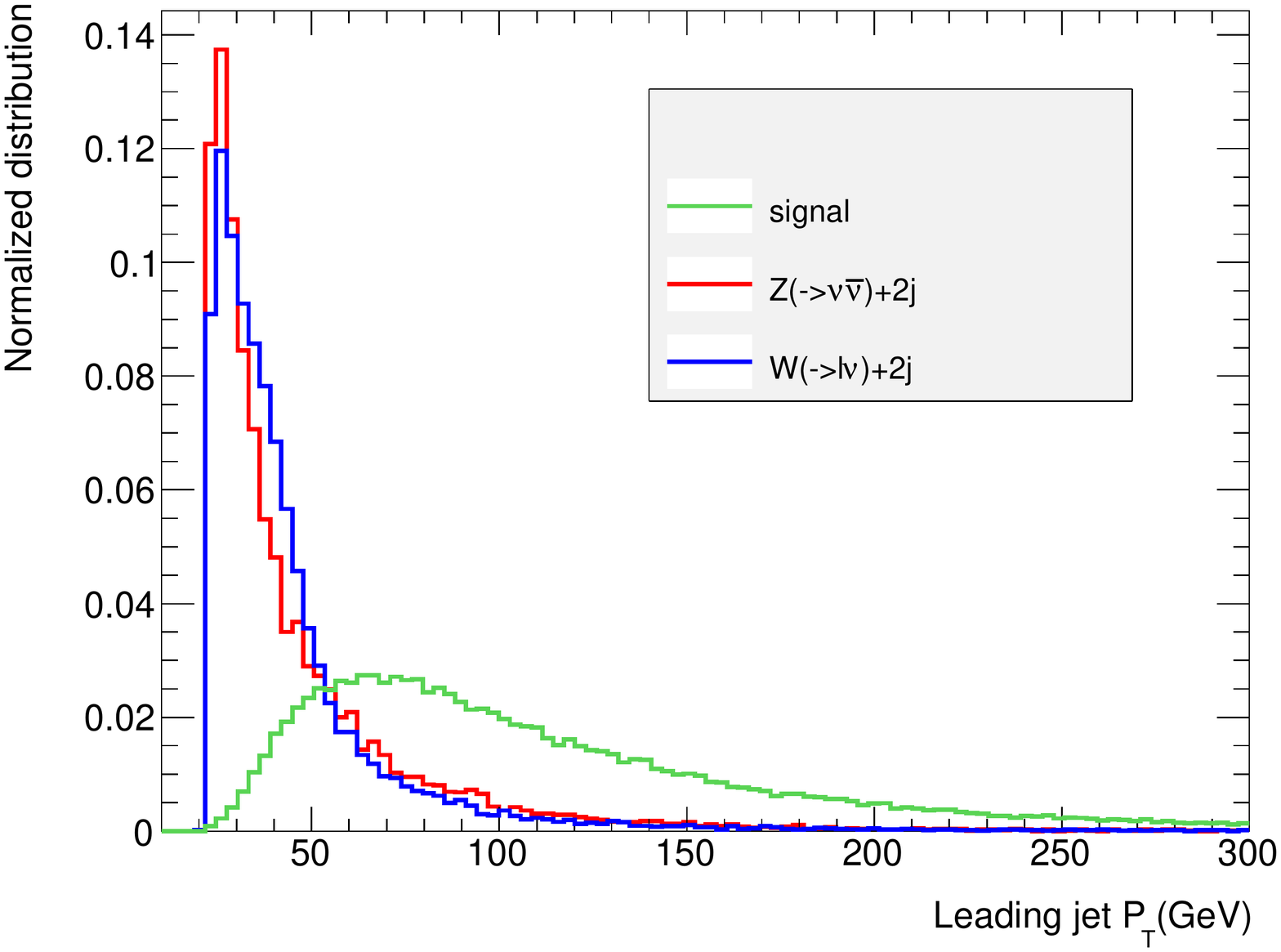}
\hspace{0.02cm}
\includegraphics[width=9.0cm, height=7cm]{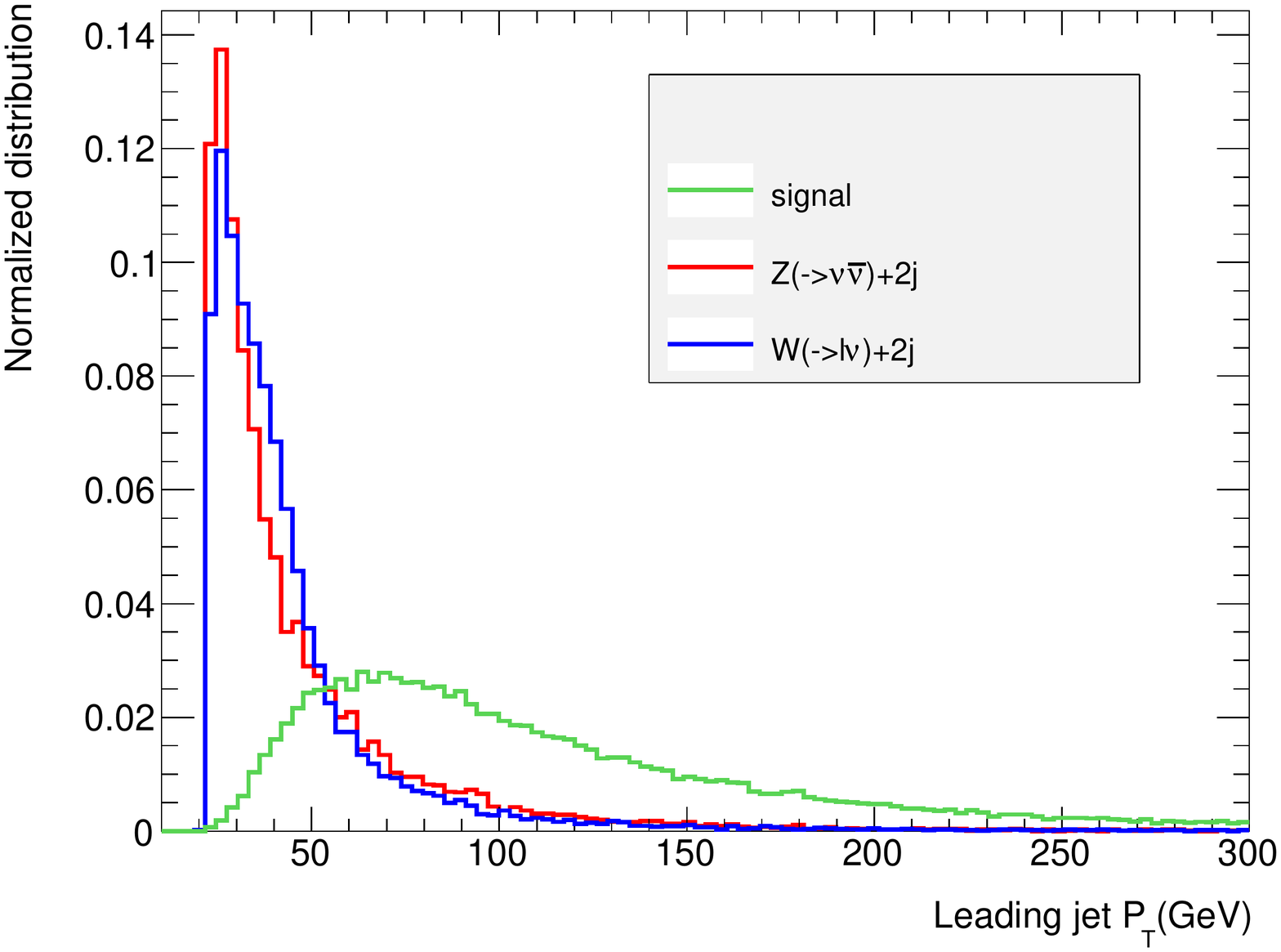} \\
\caption{$P_T$ distribution of the leading jet for vector boson fusion signal and background processes, Type I BP I(top left), Type I BP II(top right), Type II BP I(bottom left) and Type II BP II(bottom right).}
\label{jetpt1vbf}
\end{figure}   

\begin{figure}[!hptb]
\includegraphics[width=9.0cm, height=7cm]{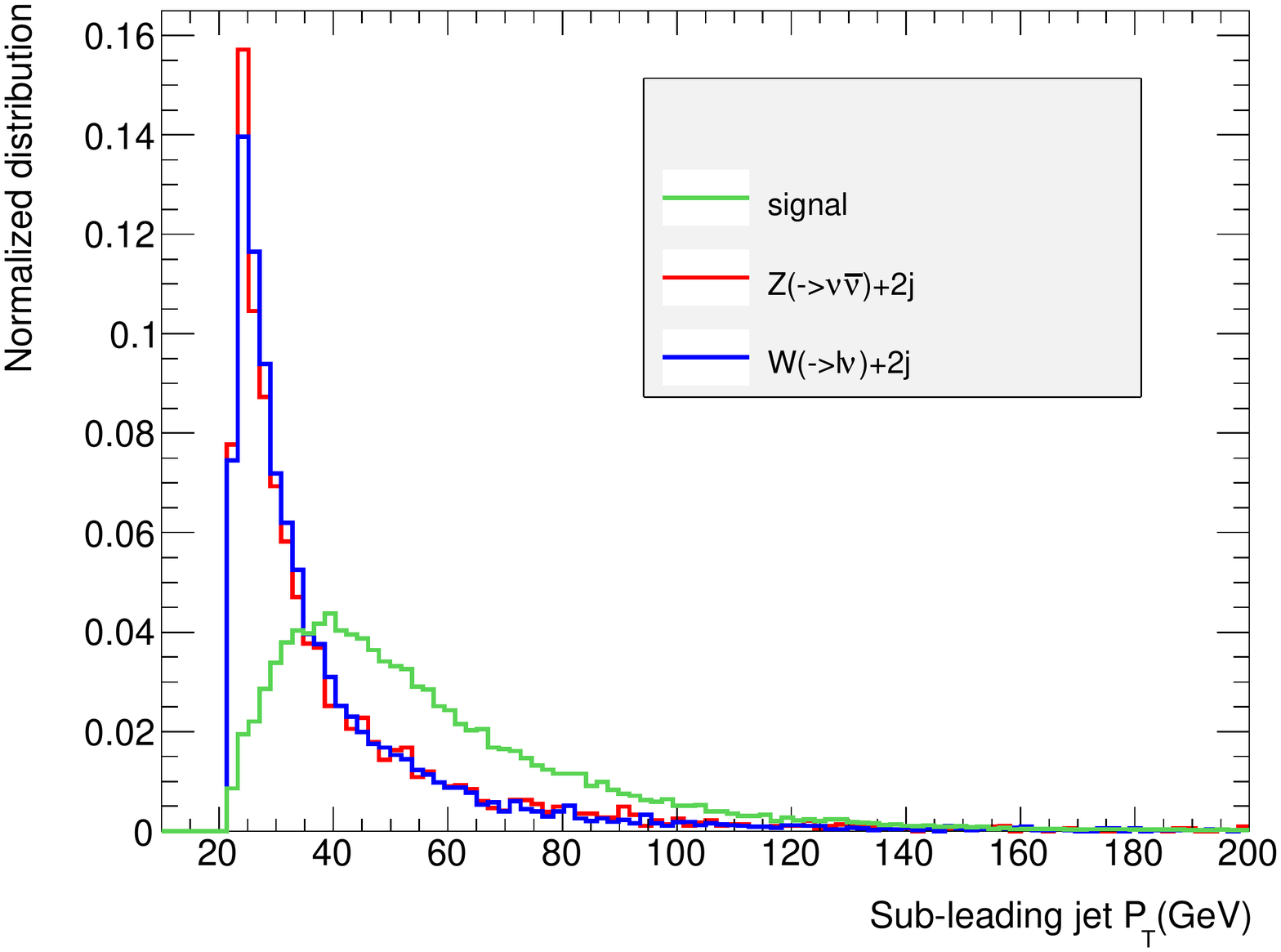}
\hspace{0.02cm}
\includegraphics[width=9.0cm, height=7cm]{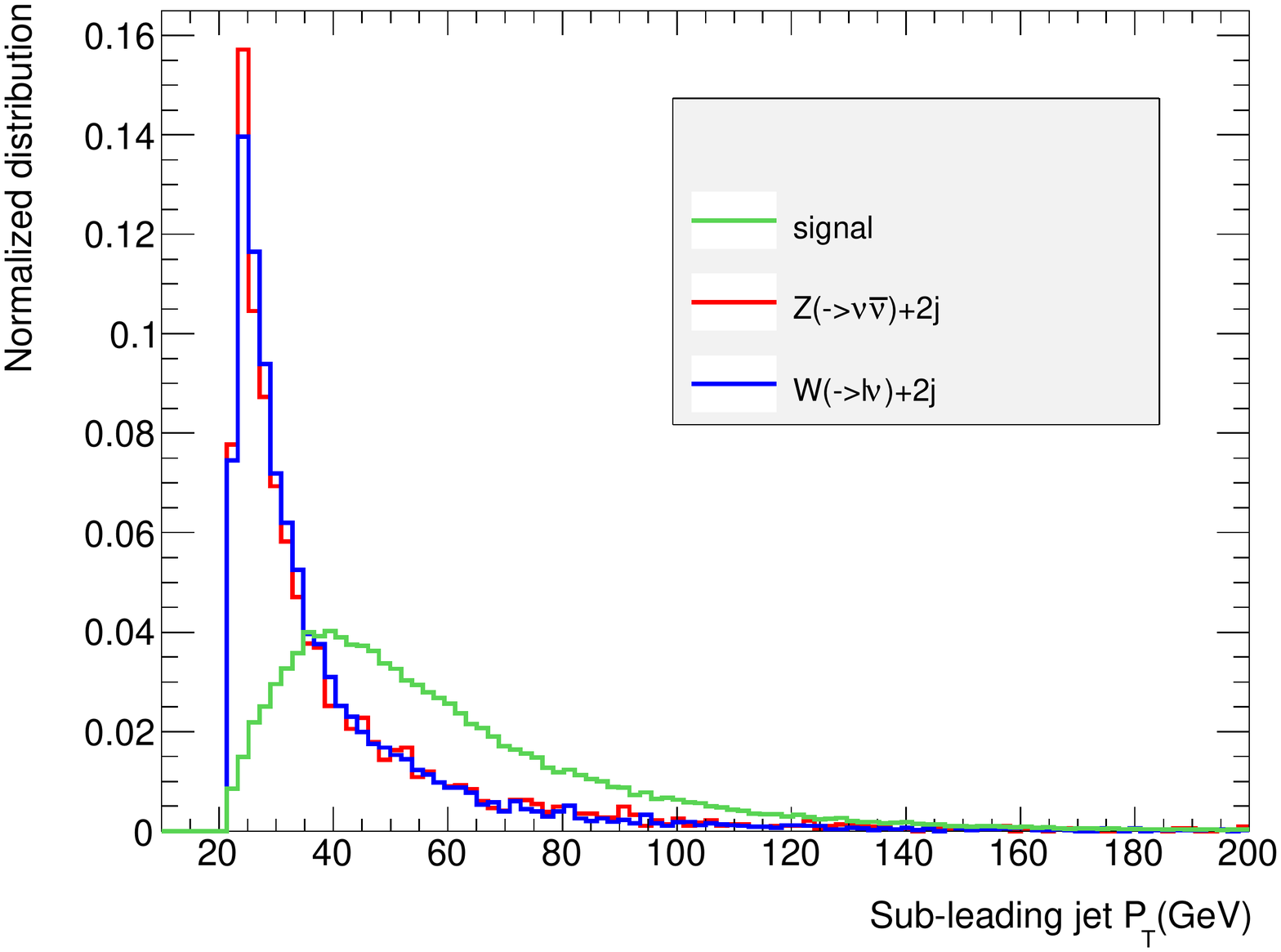} \\
\vspace*{0.02cm}
\includegraphics[width=9.0cm, height=7cm]{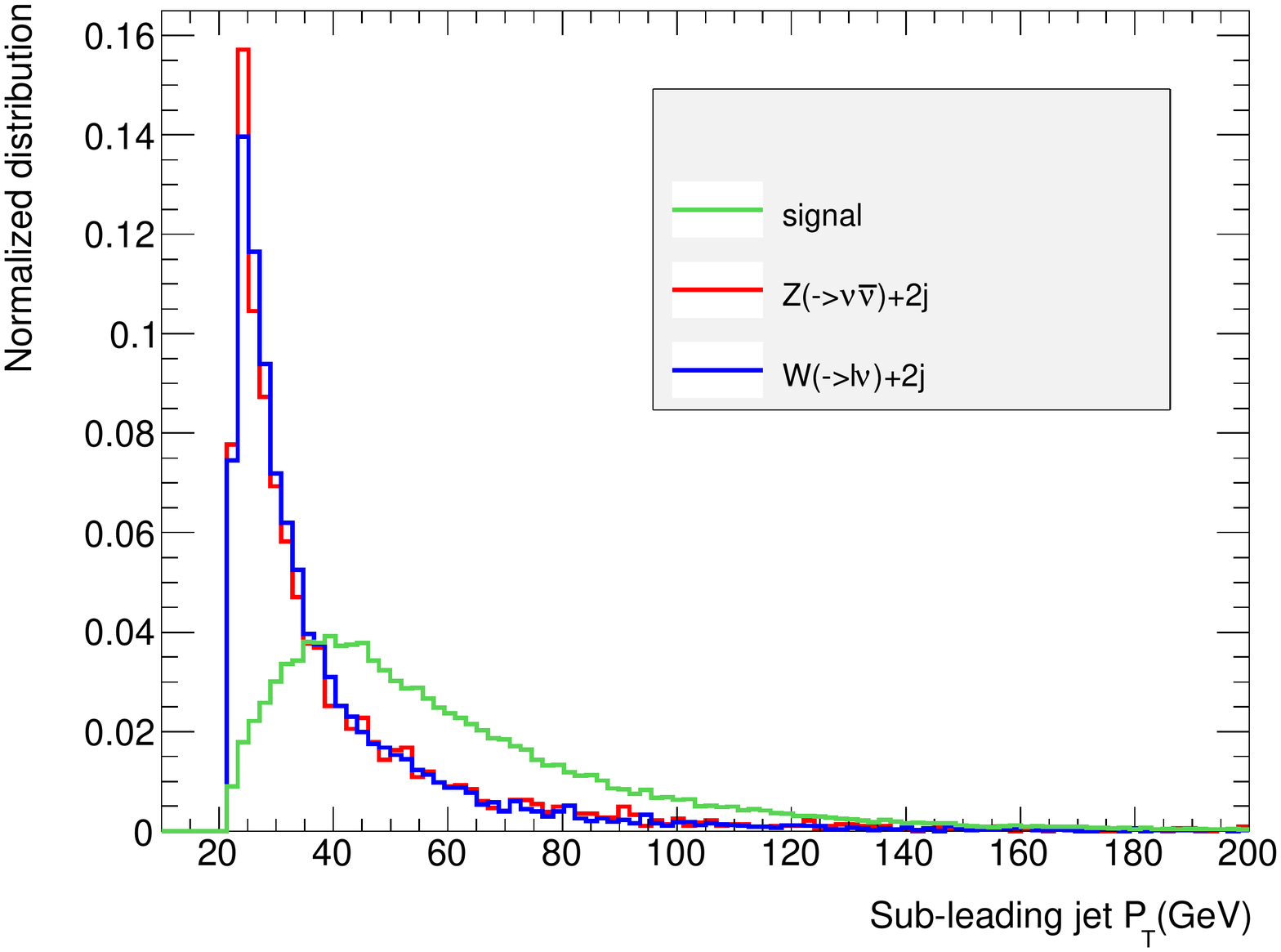}
\hspace{0.02cm}
\includegraphics[width=9.0cm, height=7cm]{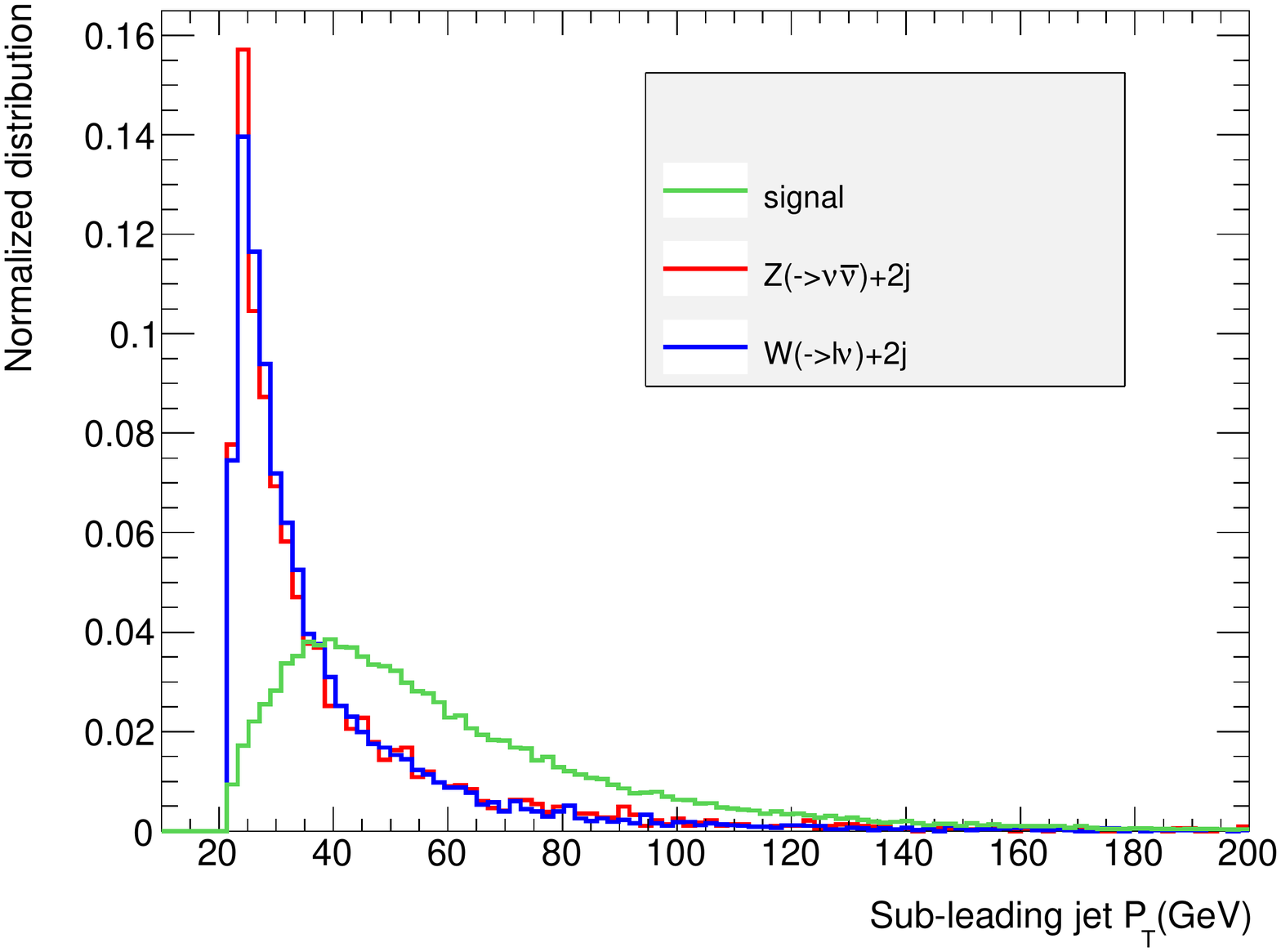} \\
\caption{$P_T$ distribution of the sub-leading jet for vector boson fusion signal and background processes, Type I BP I(top left), Type I BP II(top right), Type II BP I(bottom left) and Type II BP II(bottom right).}
\label{jetpt2vbf}
\end{figure}   
   %


\begin{figure}[!hptb]
\includegraphics[width=9.0cm, height=7cm]{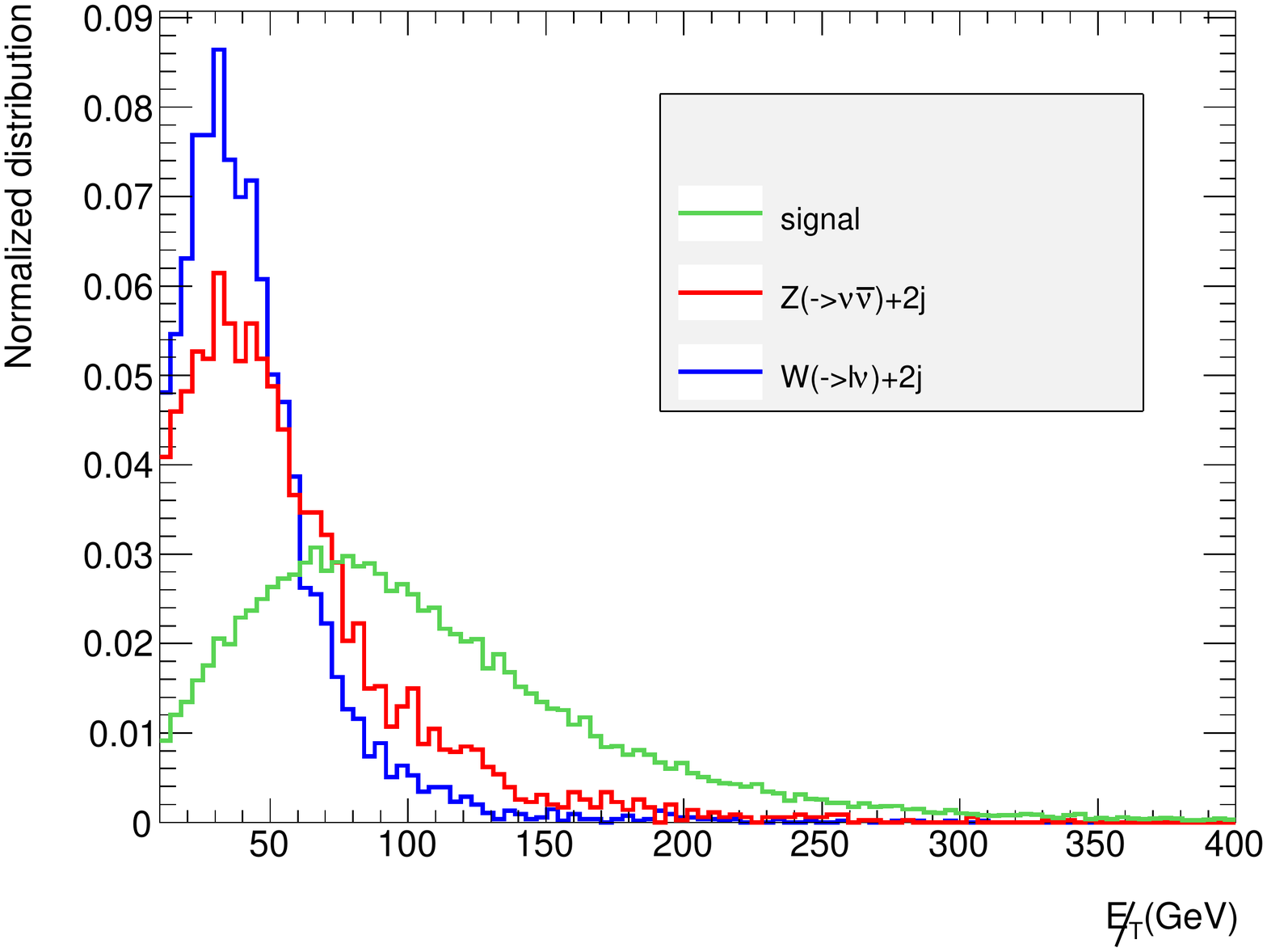}
\hspace{0.02cm}
\includegraphics[width=9.0cm, height=7cm]{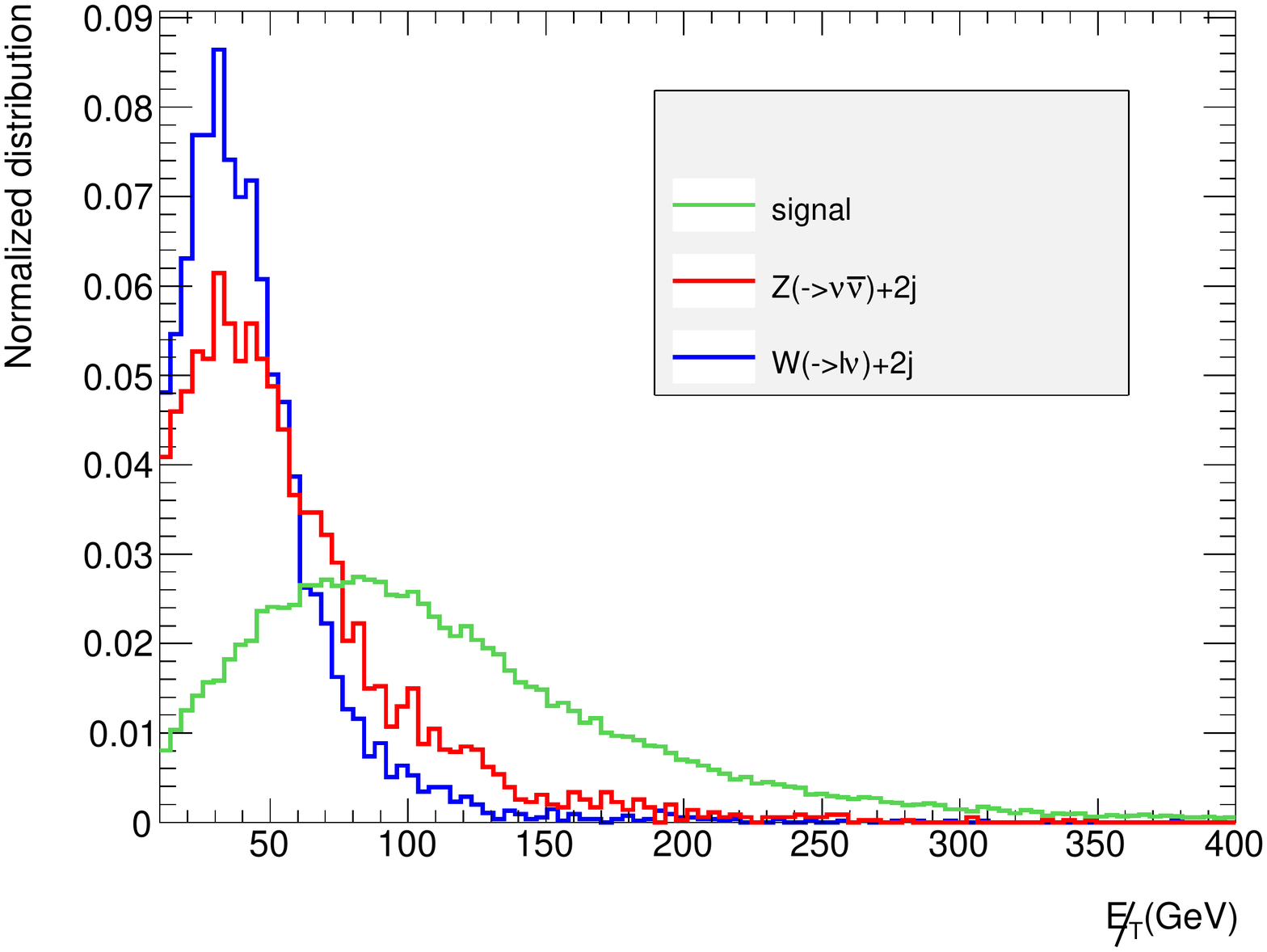} \\
\vspace*{0.02cm}
\includegraphics[width=9.0cm, height=7cm]{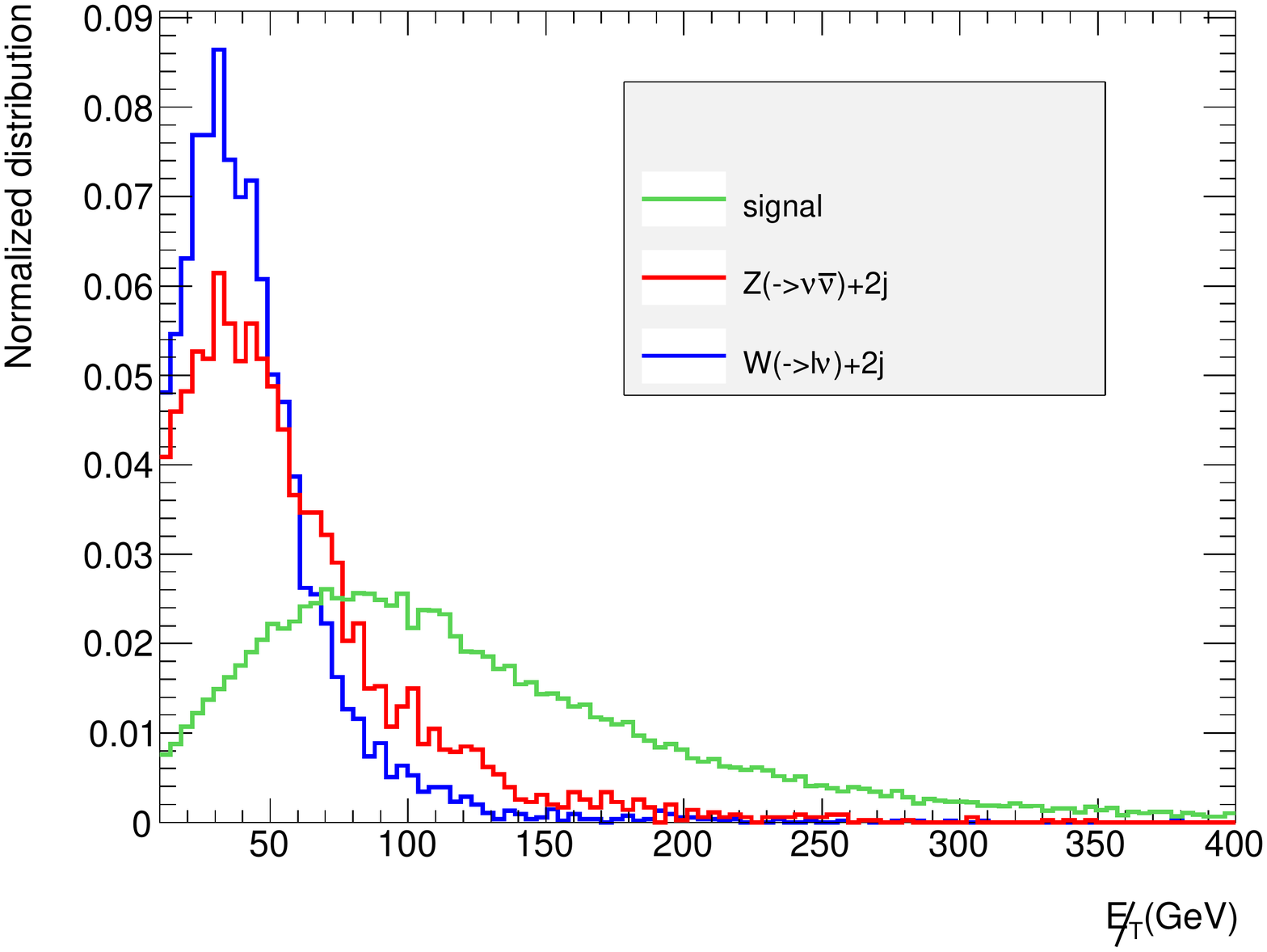}
\hspace{0.02cm}
\includegraphics[width=9.0cm, height=7cm]{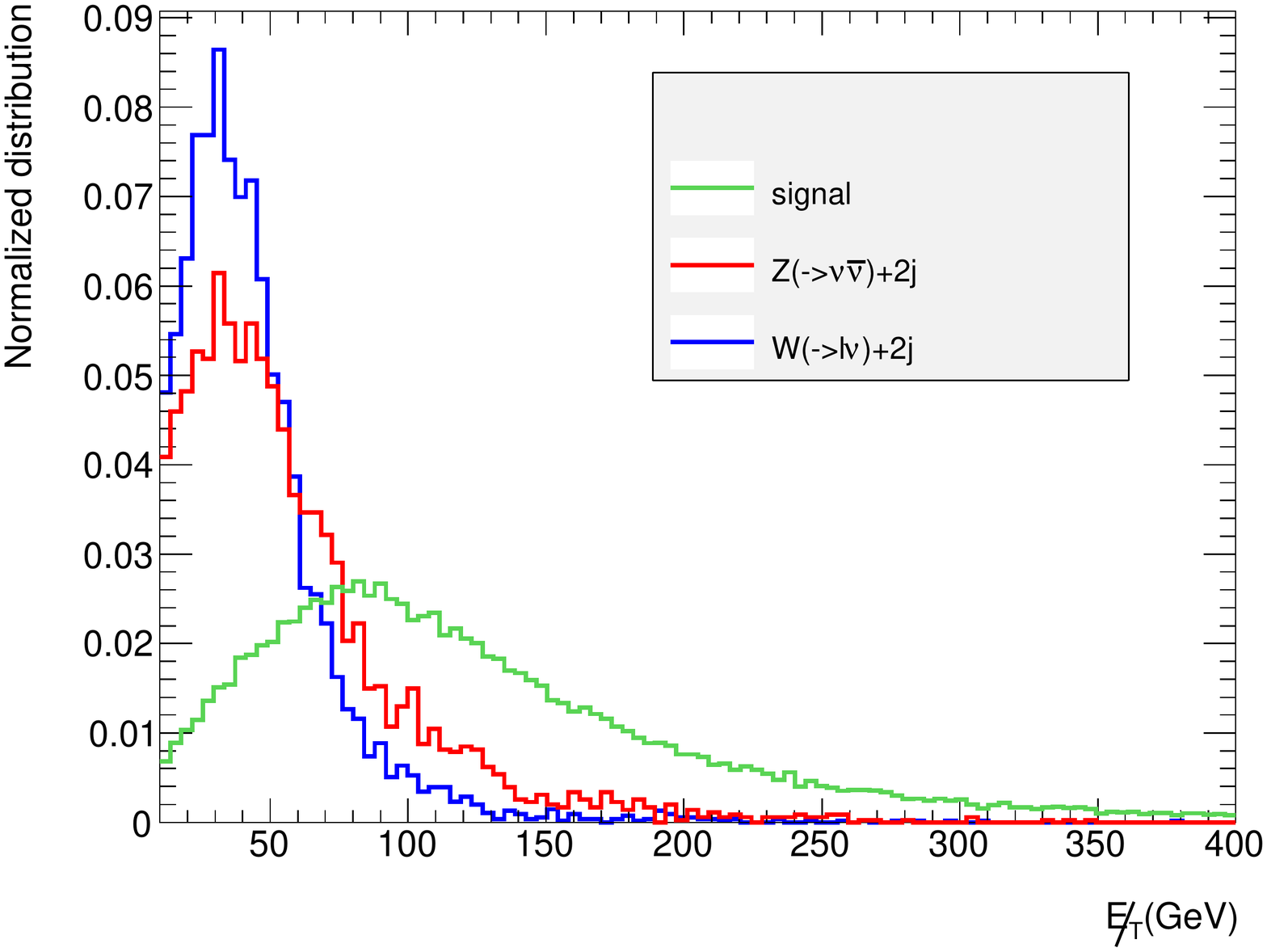} \\
\caption{$\slashed{E_T}$ distribution for vector boson fusion signal and background processes, Type I BP I(top left), Type I BP II(top right), Type II BP I(bottom left) and Type II BP II(bottom right).}
\label{missptvbf}
\end{figure}   

\begin{figure}[!hptb]
\includegraphics[width=9.0cm, height=7cm]{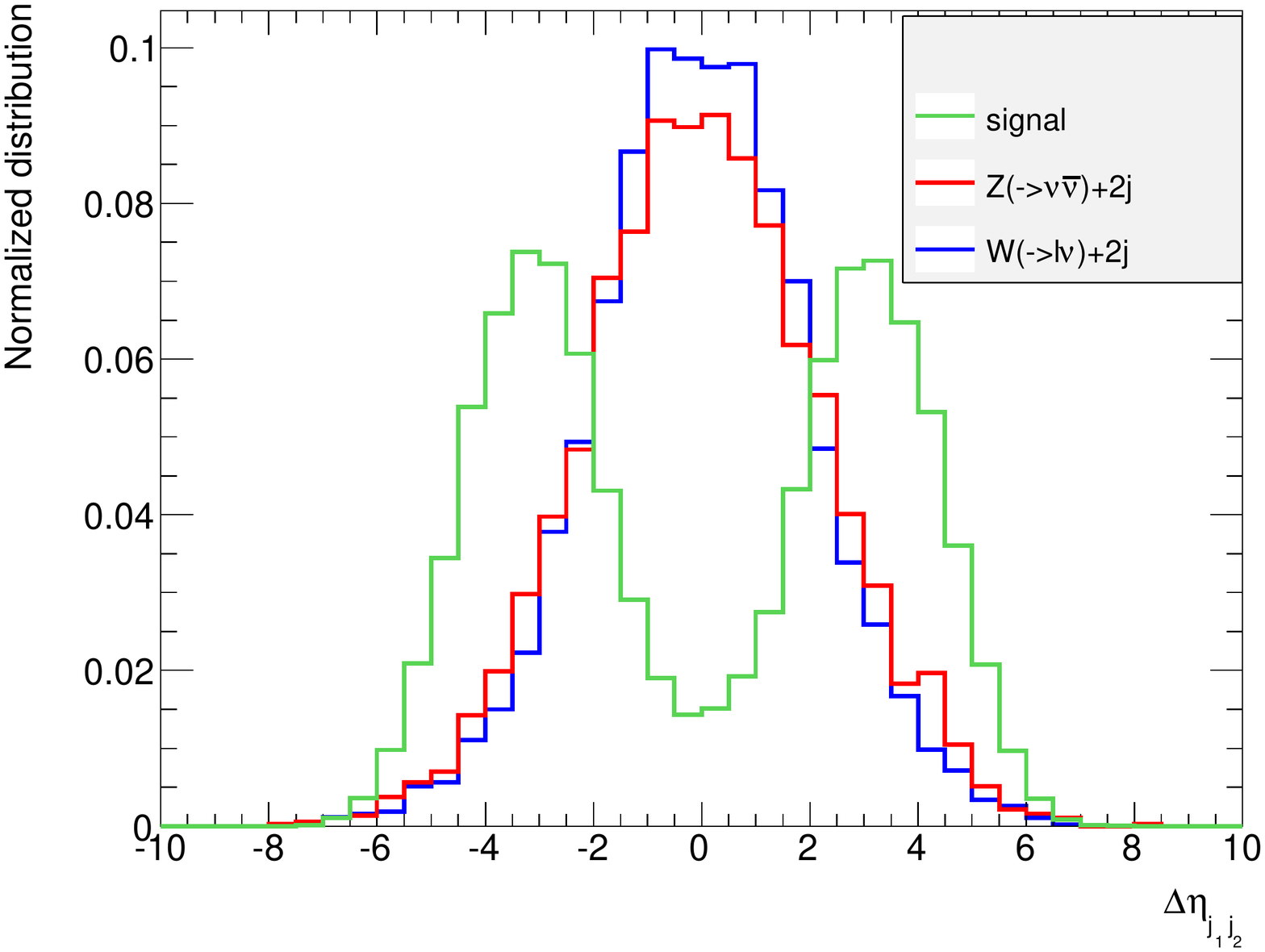}
\hspace{0.02cm}
\includegraphics[width=9.0cm, height=7cm]{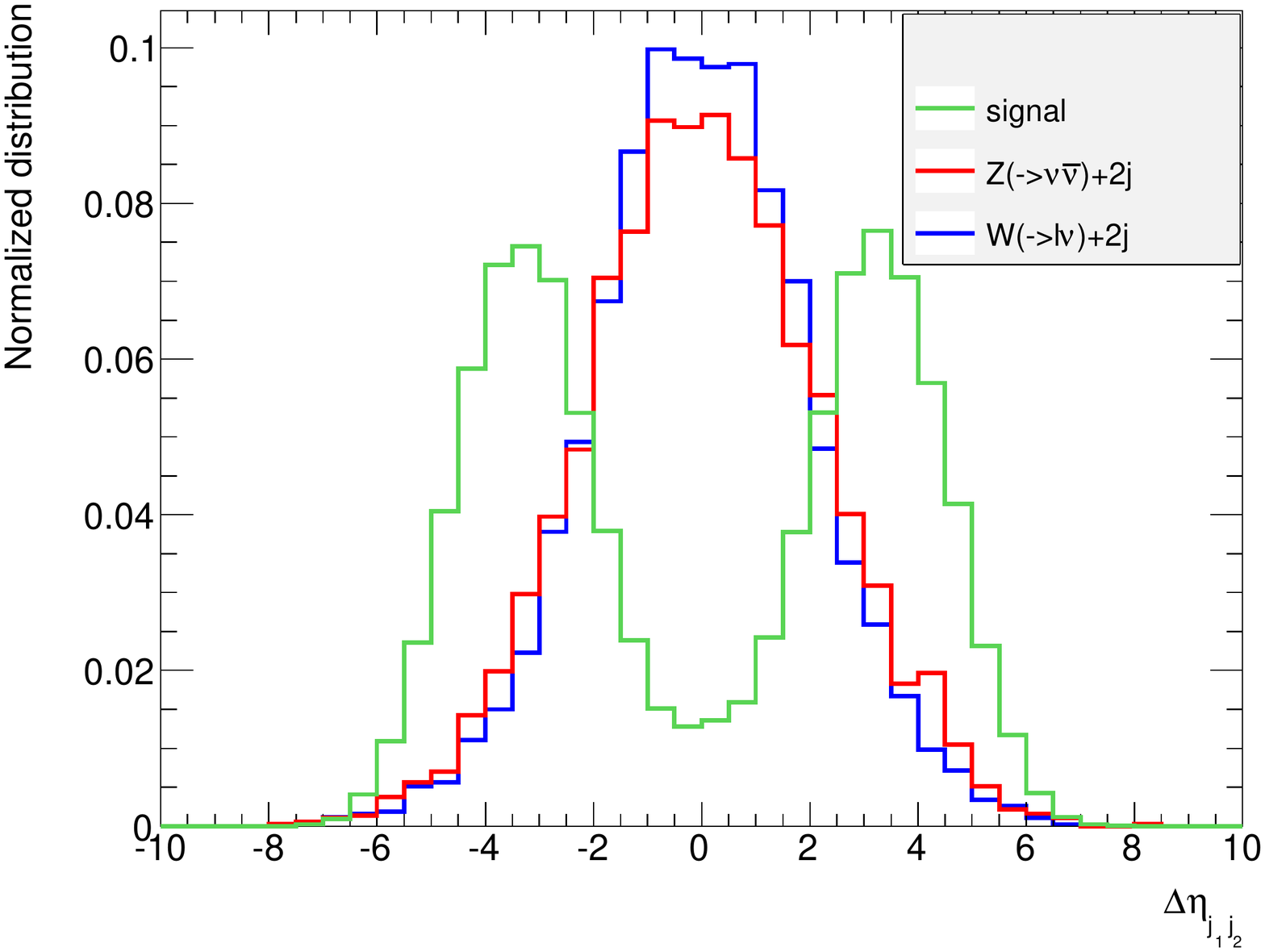} \\
\vspace*{0.02cm}
\includegraphics[width=9.0cm, height=7cm]{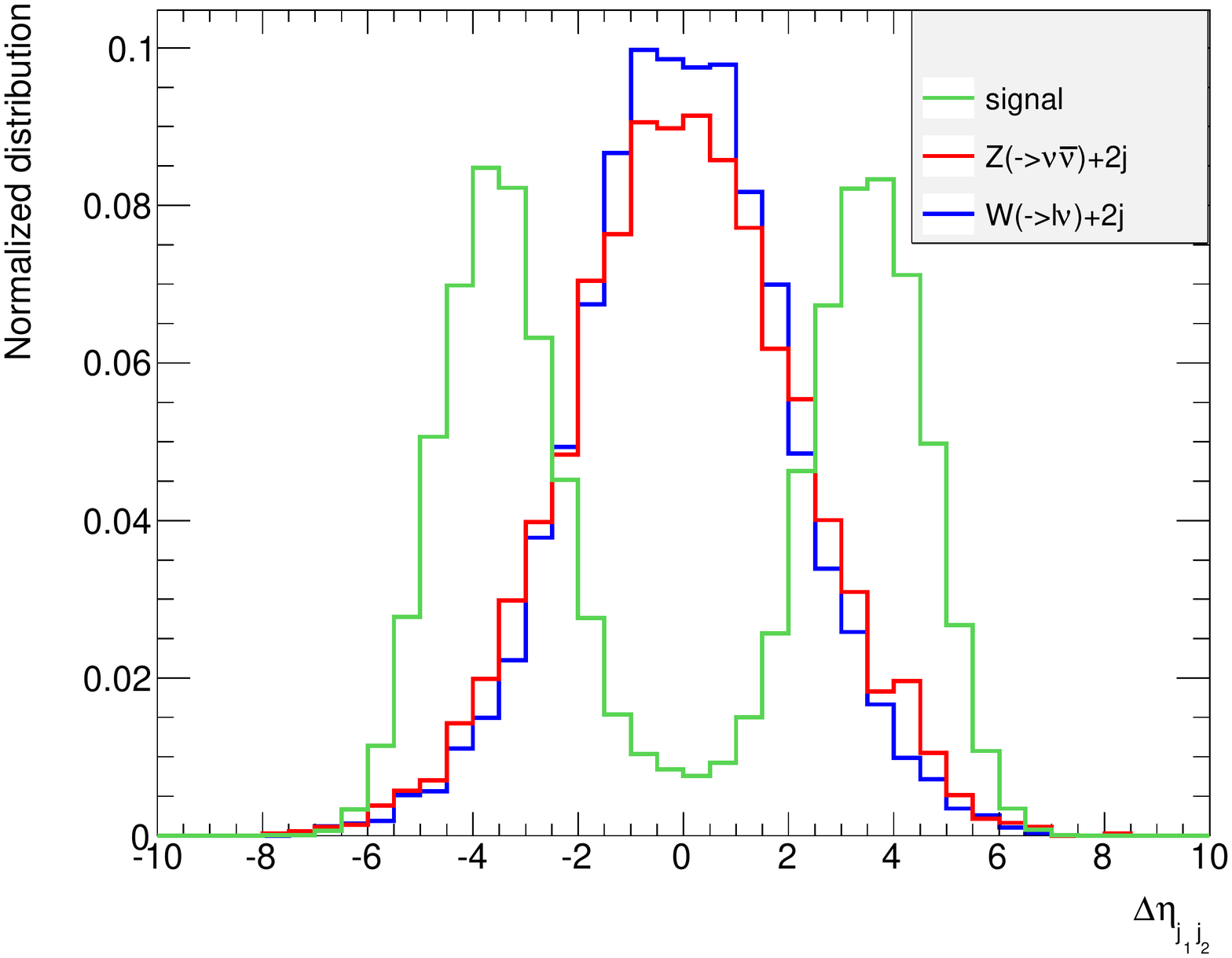}
\hspace{0.02cm}
\includegraphics[width=9.0cm, height=7cm]{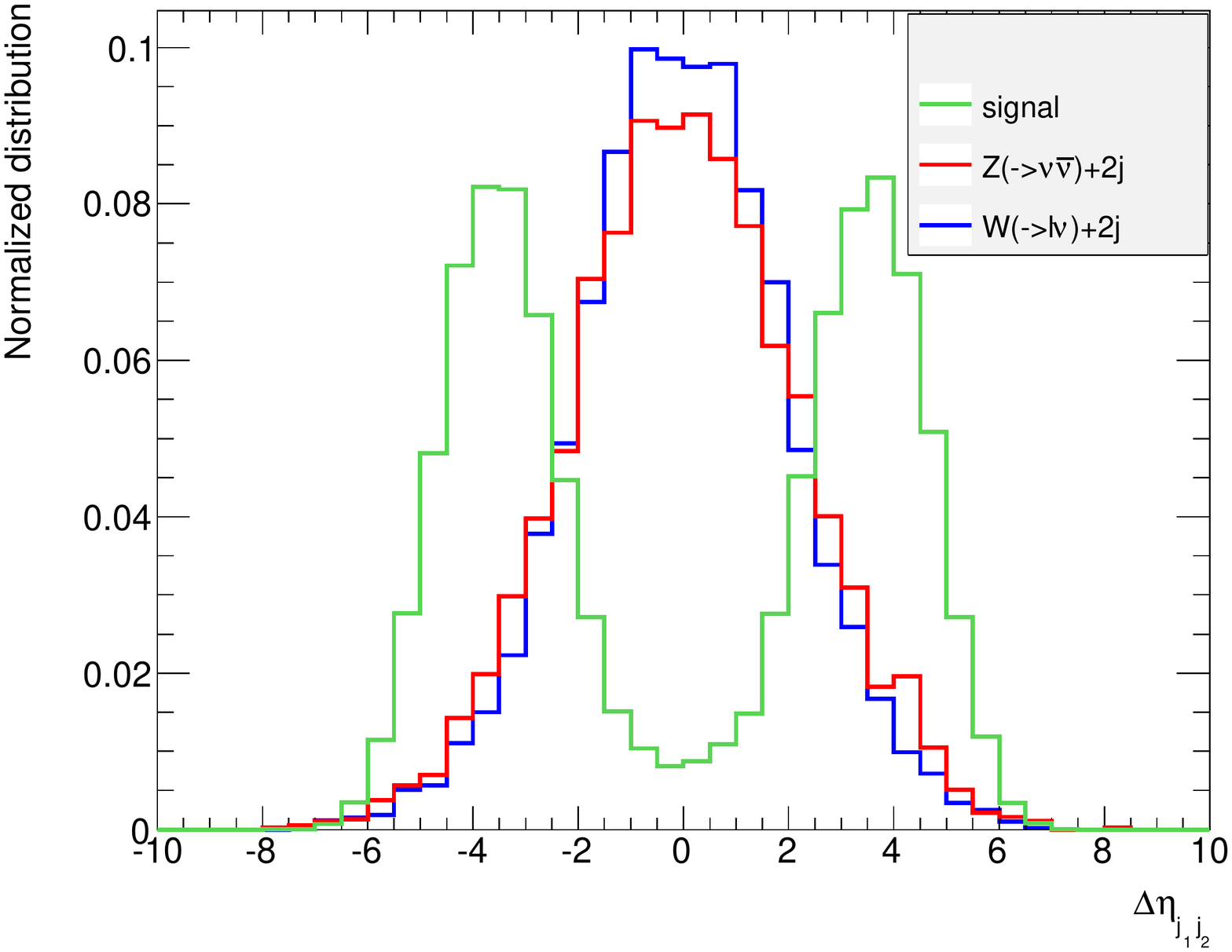} \\
\caption{$\Delta \eta$ distribution between the two forward jets for vector boson fusion signal and background processes, Type I BP I(top left), Type I BP II(top right), Type II BP I(bottom left) and Type II BP II(bottom right).}
\label{deltaetavbf}
\end{figure}

\begin{figure}[!hptb]
\includegraphics[width=9.0cm, height=7cm]{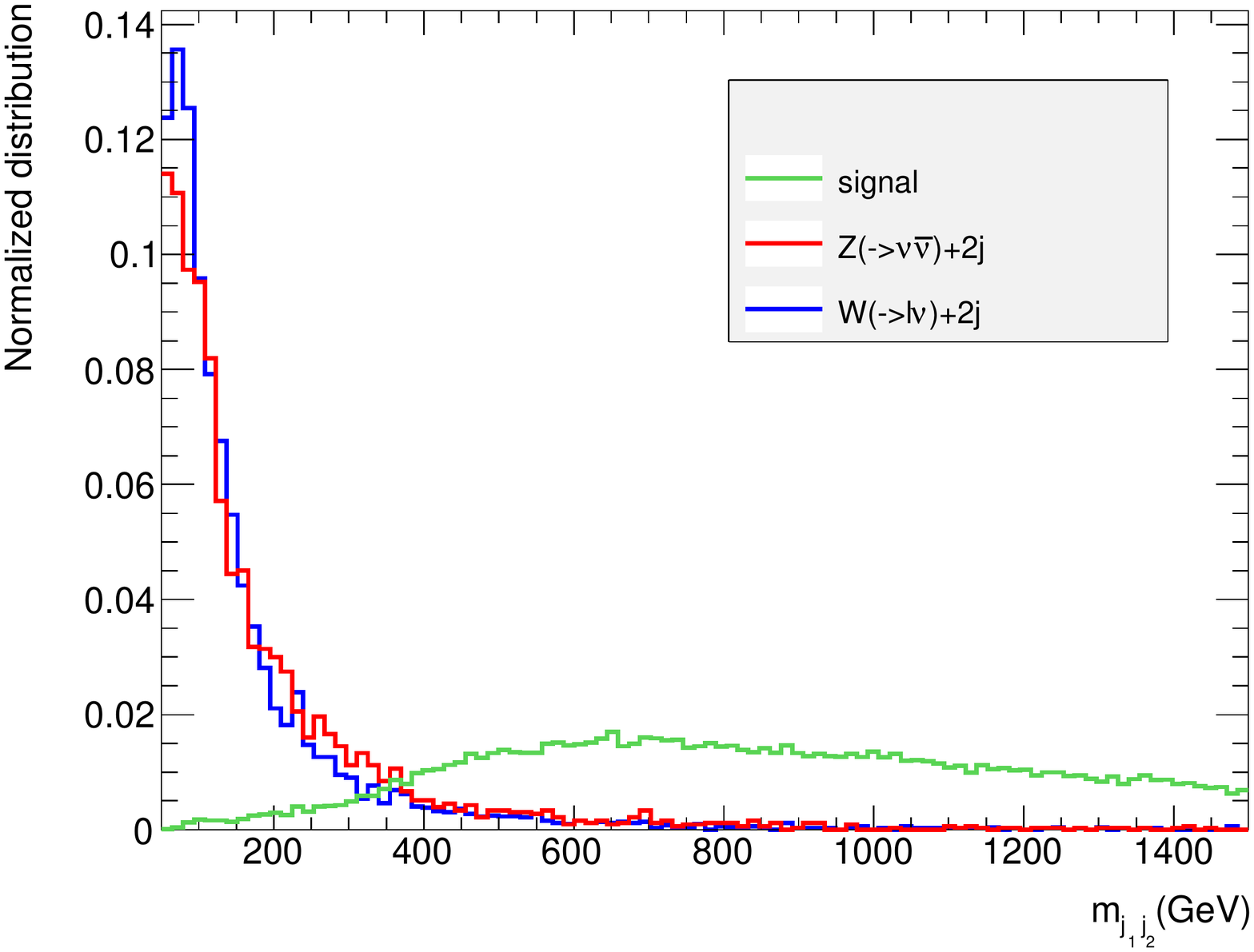}
\hspace{0.02cm}
\includegraphics[width=9.0cm, height=7cm]{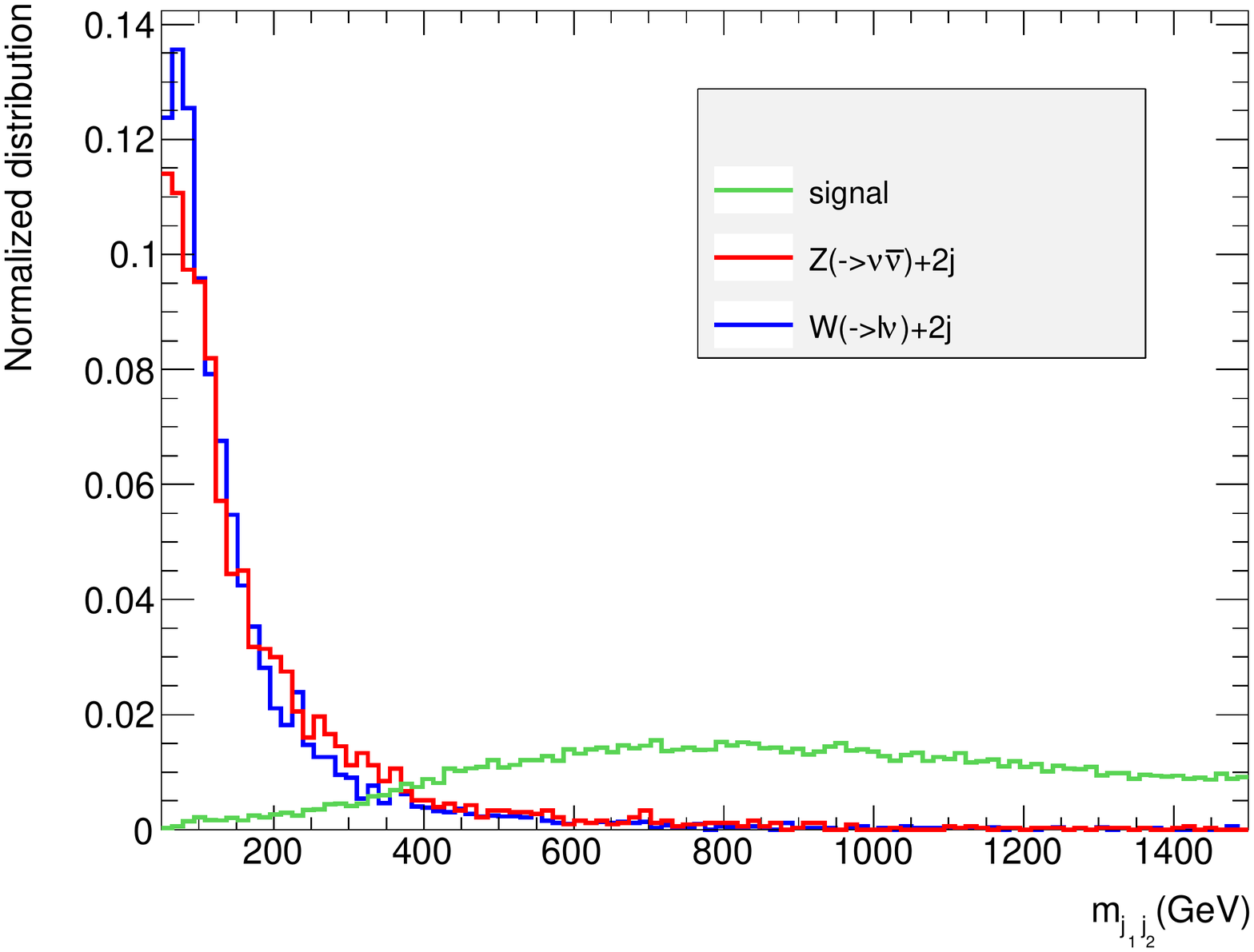} \\
\vspace*{0.02cm}
\includegraphics[width=9.0cm, height=7cm]{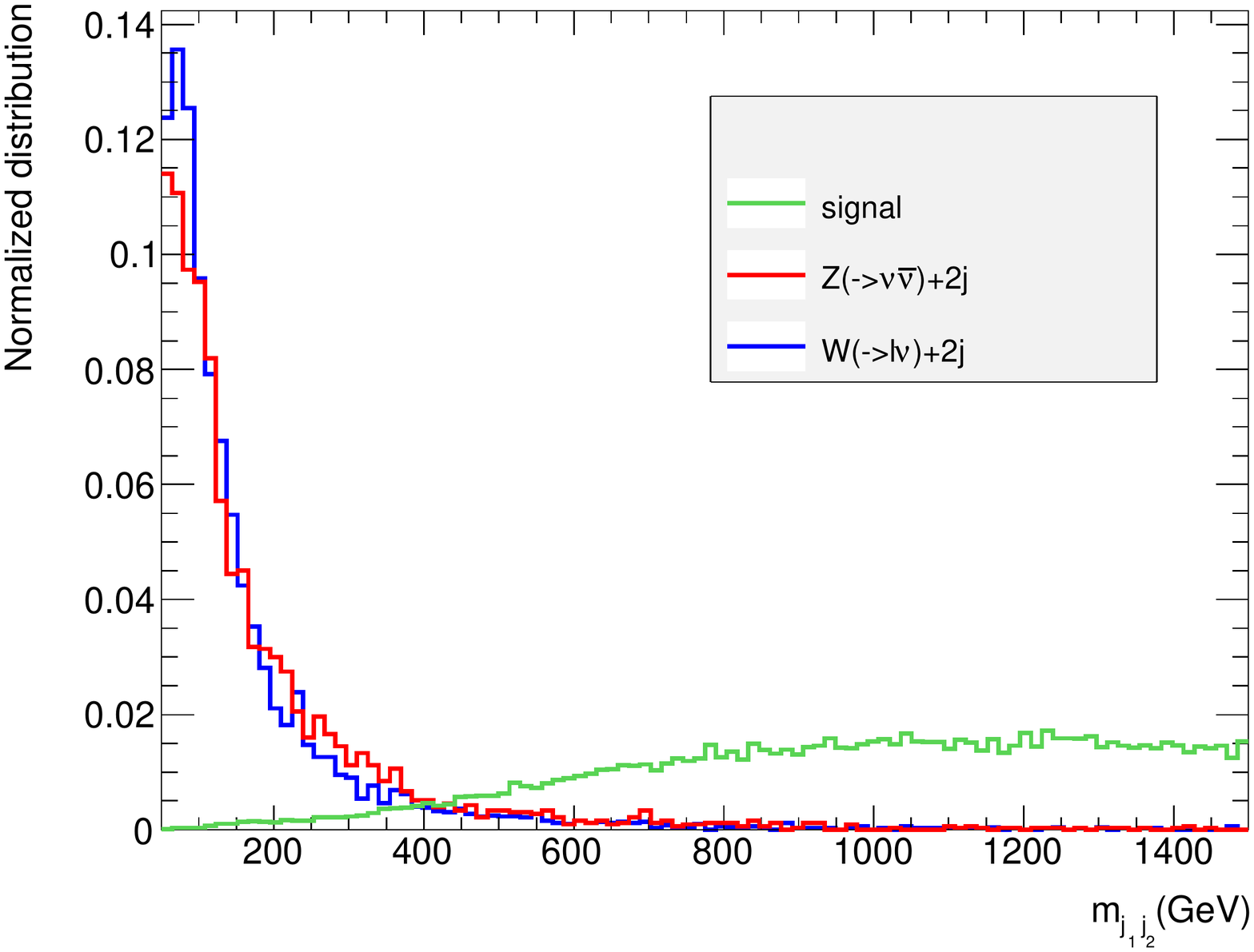}
\hspace{0.02cm}
\includegraphics[width=9.0cm, height=7cm]{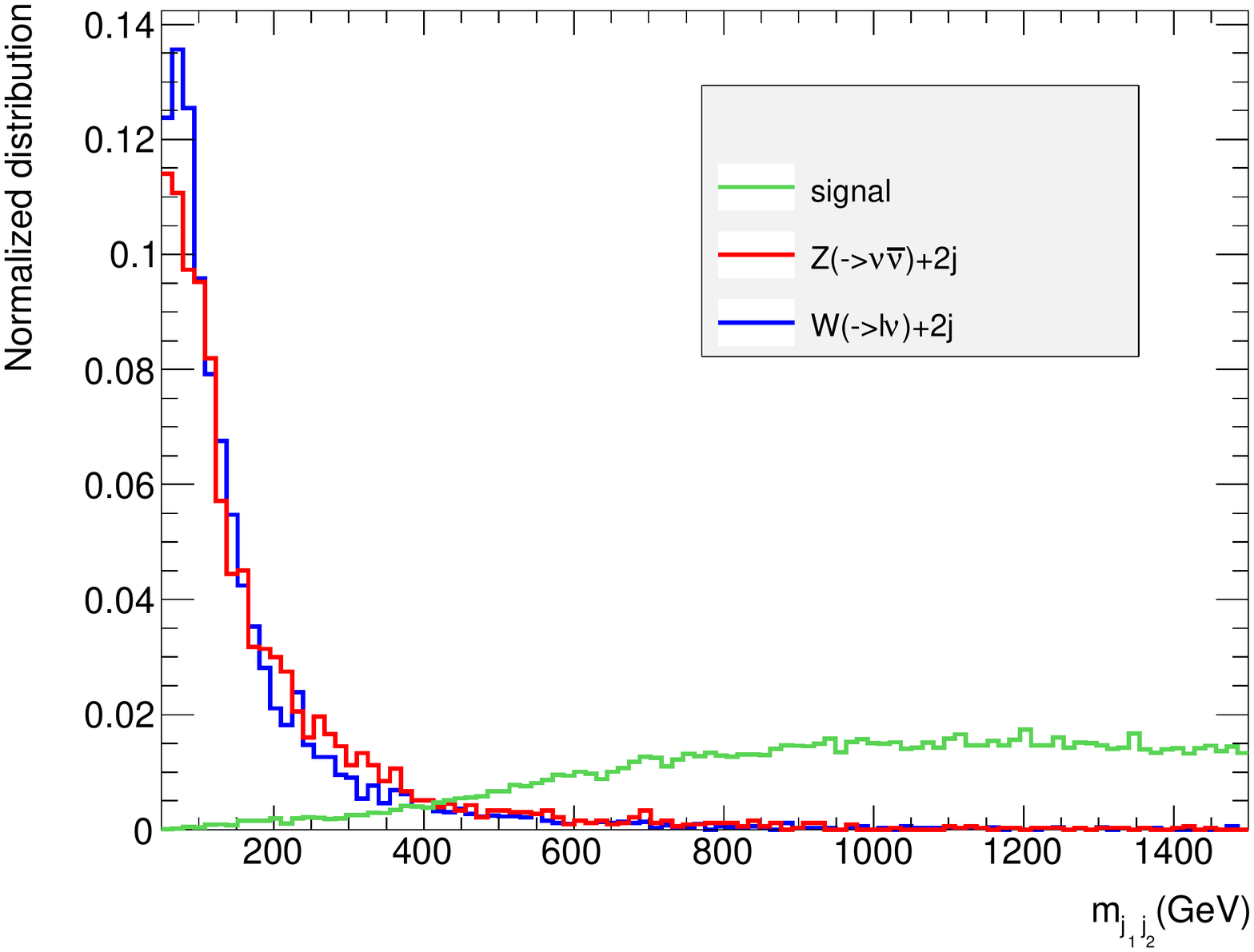} \\
\caption{Invariant mass distribution of the two forward jets for vector boson fusion signal and background processes, Type I BP II(top left), Type I BP I(top right), Type II BP I(bottom left) and Type II BP II(bottom right).}
\label{mjjvbf}
\end{figure}

\begin{figure}[!hptb]
\includegraphics[width=7.5cm, height=6cm]{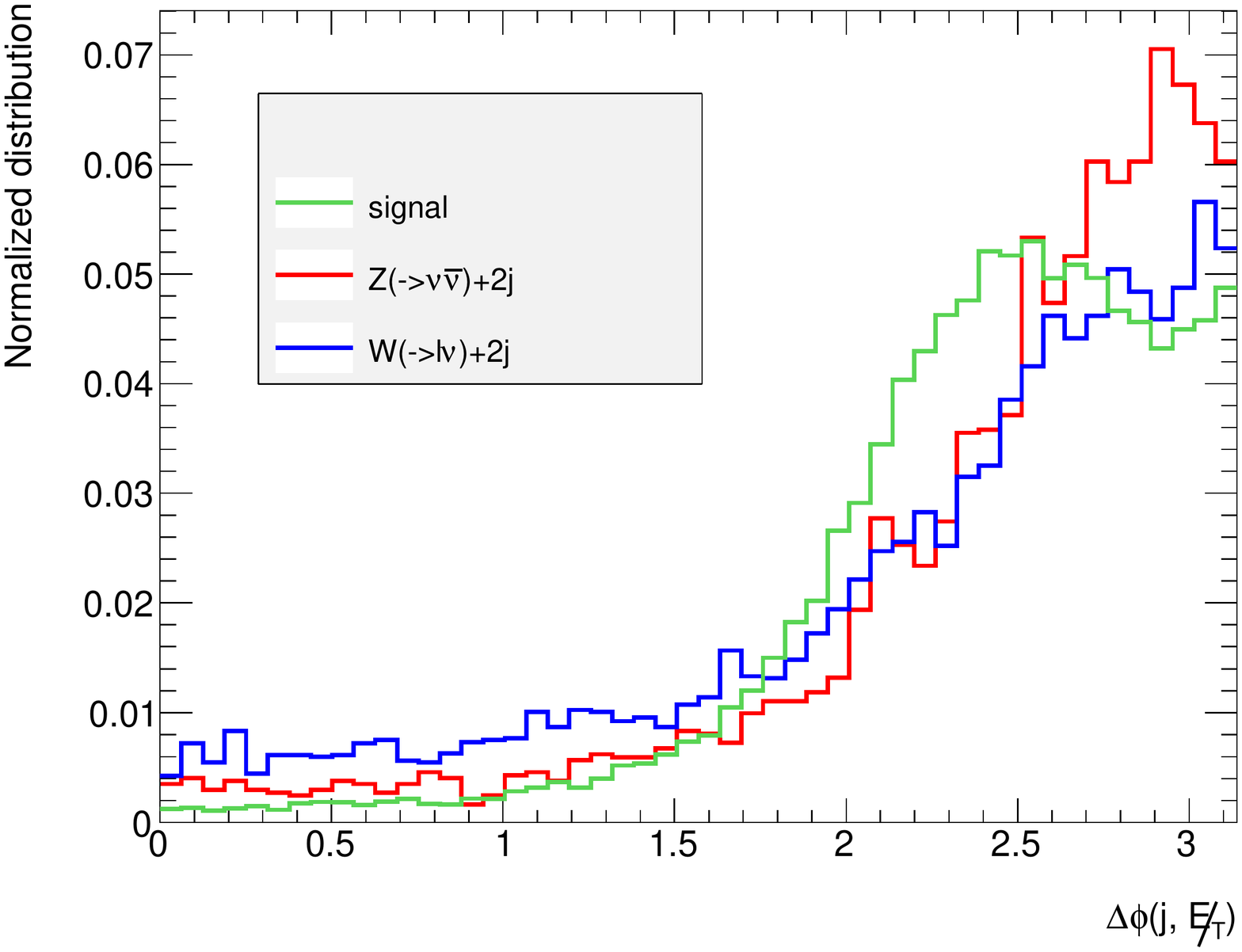}
\hspace{0.9cm}
\includegraphics[width=7.5cm, height=6cm]{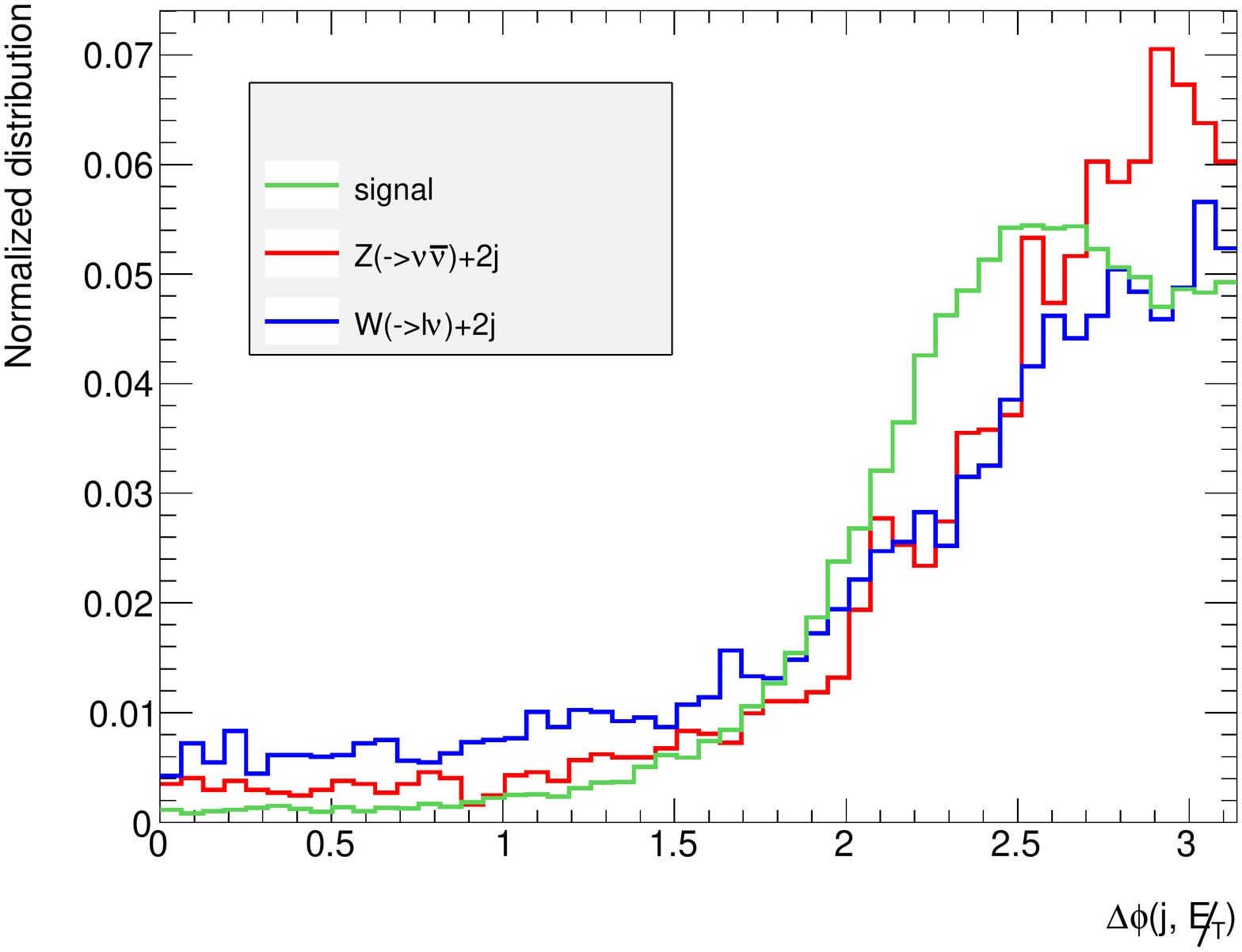} \\
\vspace*{0.02cm}
\includegraphics[width=7.5cm, height=6cm]{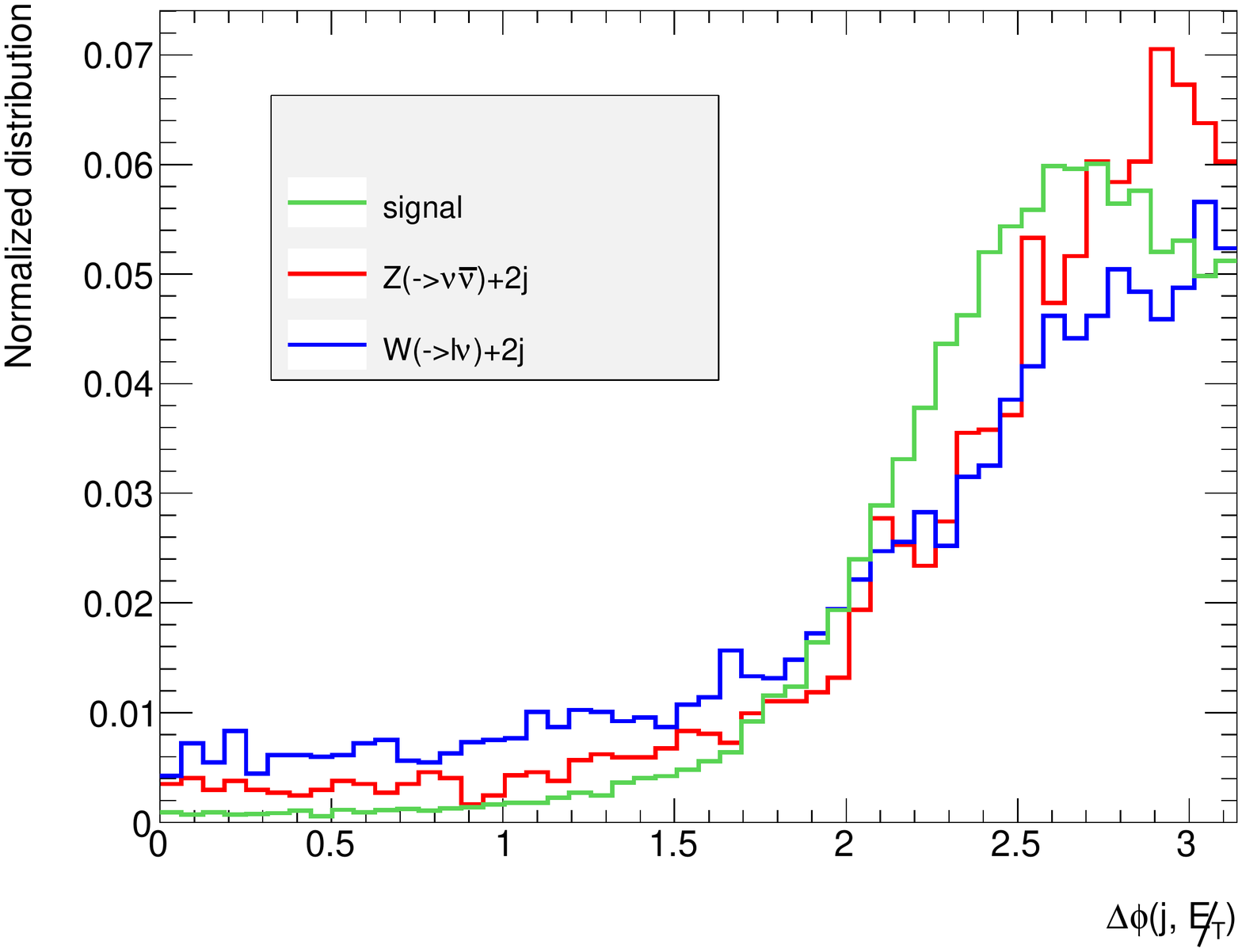}
\hspace{0.9cm}
\includegraphics[width=7.5cm, height=6cm]{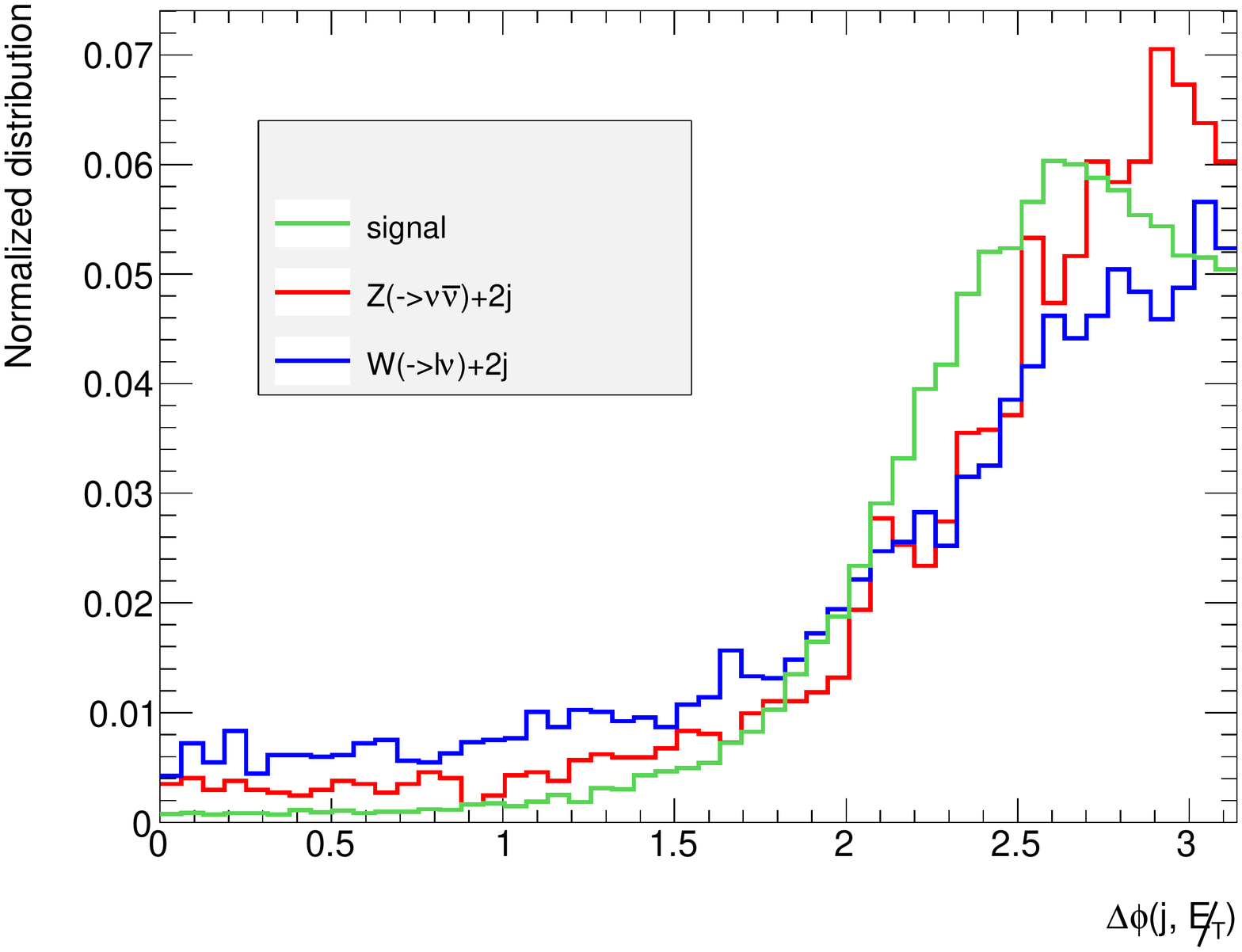} \\
\caption{$\Delta \phi(\slashed{E_T}, p_T^{jet})$ distribution between the two forward jets for vector boson fusion signal and background processes, Type I BP I(top left), Type I BP II(top right), Type II BP I(bottom left) and Type II BP II(bottom right).}
\label{jetmissvbf}
\end{figure}

Having discussed the key feature of VBF process we present distributions of some observables to examine their role to reduce the background contribution and enhance the signal.
We see in Figure.~\ref{jetpt1vbf} and ~\ref{jetpt2vbf}, that the $p_T$ of the leading and sub-leading jet peaks at  much larger values for the signal process than the backgrounds. In Figure.~\ref{missptvbf} we show the $\slashed{E_T}$ distribution for the signal and background, which for the signal peaks at larger values than for the background. The $p_T$ of the jets and $\slashed{E_T}$ are good discriminator between signal and background. 

The absence of hadronic activity in the central region for the signal is evident from Figure.~\ref{deltaetavbf}, where we see the $|\Delta \eta_{j_1 j_2}|$ peaks at larger values for signal compared to backgrounds. The invariant mass of the two forward jets ($m_{j_1 j_2}$ ) related to the variable $|\Delta \eta_{j_1 j_2}|$ is another useful discriminator. The more separated the two jets are, the larger will be their invariant mass. From Figure.~\ref{deltaetavbf} and ~\ref{mjjvbf}, we can see that a cut on the $|\Delta \eta_{j_1 j_2}|$ and $m_{j_1 j_2}$ will help us achieve better signal significance.
We further investigate the distribution of $\Delta \phi(\slashed{E_T}, p_T^{jet})$ for signal and backgrounds. In case of signal the $\slashed{E_T}$ has to balance against the system of two forward jets whereas in case of QCD background the $\slashed{E_T}$ is expected to be along the jet momenta for the reason described earlier. This observable helps us get rid of QCD background to a large extent. 
The background coming from VBF production of SM Higgs, with its invisible decay is an irreducible background in our case. The aforementioned cuts do not perform very well in reducing this background, especially for benchmark points of Type I, where the favoured regions have $m_H$ not too far from 125-GeV. However rate is considerably smaller than the signal. The $Z+$ jets(EW) and $W+$ jets (EW) also constitute a set of irreducible backgrounds, as in these cases $Z$ and $W$ are produced via VBF. We found that these background turn out to be 6-8\% of the $Z+$ jets(QCD) and $W+$ jets (QCD) background. Naturally these backgrounds do not play any important role in the analysis. These backgrounds have still been included in our calculations for the sake of completeness. 

In principle, such signal events can also be faked by strong processes such as color-singlet exchanges in the form of hard or soft positrons~\cite{Khoze:2001ft,Khoze:2000vr}. In the absence of clear predictions on these, the hard $\slashed{E_T}$-cut (as detailed below) has been expected to take care of such fakes.

\subsubsection{Results}

Having discussed the kinematic observables in case of VBF process, we proceed to apply certain set of cuts on them. The cut-flows are shown in Table.~\ref{vbfcuts}.

\subsection{Event selection criteria}

The following cuts are applied to select the events over and above the basic selection cuts~\cite{Sirunyan:2018owy}.
\begin{itemize}

\item Cut 1 =  $|\Delta \eta_{ij}| >$ 3.0.
\item Cut 2 = $m_{jj} >$ 600 GeV.
\item Cut 3 = $\slashed{E_T} >$ 200 GeV.
\item Cut 4 = Lepton veto: Events with electrons with $p_T > 20$ GeV or muons with $p_T > 10$ GeV are not selected.


\end{itemize}

\begin{table}[!hptb]
\begin{center}
\begin{footnotesize}
\begin{tabular}{| c | c | c | c | c | c | c |}
\hline
 & TypeI BP I & TypeI BP II & TypeII BP I & TypeII BP II & $Z(\nu \bar{\nu})+ 2j$ & $W(l \nu) + 2j$  \\
\hline
$\sigma$(pb) & 0.25 & 0.22 & 0.042 & 0.035 & $1.4\times 10^4$  & $5.1 \times 10^4$   \\
\hline
Cut 1 & 53.4\% & 58.2\% & 67.9\% & 67.5\% & 14.3\% & 11.8\% \\
\hline
Cut 2 & 53.2\% & 58.0\% & 67.8\% & 67.4\% & 0.02\% & 0.01\% \\
\hline
Cut 3 & 5.0\% & 7.3\% & 11.5\% & 11.4\% & 0.008\% & 0.002\% \\
\hline
Cut 4 & 5.0\% & 7.3\% & 11.5\% & 11.4\% & 0.008\% & $10^{-6}$\% \\
\hline
\end{tabular}
\end{footnotesize}
\caption{Signal and background efficiencies after applying various cuts for the vector boson fusion production channel at 14 TeV. The cross sections are calculated at NLO. }
\label{vbfcuts}
\end{center}
\end{table}

\begin{table}[!hptb]
\begin{center}
\begin{footnotesize}
\begin{tabular}{| c | c |}
\hline
BP & ${\cal S}$  \\
\hline
Type I BP I  & 3.8 $\sigma$ (600 fb$^{-1}$)  \\
\hline
Type I BP II  & 2.3 $\sigma$ (3 ab$^{-1}$)\\
\hline
Type II BP I  & 4.5 $\sigma$ (3 ab$^{-1}$)\\
\hline
Type II BP II  & 3.3 $\sigma$ (3 ab$^{-1}$) \\
\hline
\end{tabular}
\end{footnotesize}
\caption{Signal significance for the benchmark points at 14 TeV in the VBF channel. }
\label{significance_vbf}
\end{center}
\end{table}

We find out that although VBF production channel has much lower cross section for the production of heavy Higgs than the gluon fusion channel, which gets further reduced in 2HDM because of the multiplicative factor $\cos (\beta - \alpha)$ close to the alignment limit. Table~\ref{significance_ggf} and ~\ref{significance_vbf} reveal this fact. This channel still fares very well, and in fact wins over gluon fusion process, because of its distinctness of the signal, buttressed with the chosen selection criteria. We can see from Table.~\ref{significance_vbf} that all the benchmarks for both Type I and Type II give promising result in terms of probing invisible decay of heavy Higgs in the future collider. The Type I BP I performs exceedingly well among the four benchmark points we considered. Like gluon fusion here also because of large cross section and branching ratio the signal significance in this channel is very large. Therefore this particular benchmark point can be probed an
 d tested with even lower luminosity. We have given projection for 300 fb$^{-1}$ in this case. Comparing Tables~\ref{significance_ggf} and ~\ref{significance_vbf} we can see that for all the benchmark points that we considered VBF supersedes the gluon fusion channel. Although gluon fusion cross section is larger than the VBF cross section, the VBF channel has better separation between signal and background. Consequently the VBF channel seems to have better prospect than gluon fusion channel while looking for decay of heavy Higgs in the invisible mode.

\section{Multivariate analysis and Neural Network techniques}\label{sec6}

Having performed a cut-based analysis for the signal of invisible decay of Higgs in association with a single energetic jet(gluon fusion) and two energetic forward jets(VBF) at the LHC, we further explore the possibility of improvement in the analysis with some recently developed techniques like {\bf Gradient Boosted Decision Trees}~\cite{Chen:2016btl}  and {\bf Artificial Neural Network (ANN)}~\cite{Teodorescu:1100521}. These methods have been used extensively in the literature in the recent past\cite{Baldi:2014kfa,Woodruff:2017geg,Oyulmaz:2019jqr,Bhattacherjee:2019fpt} and have been shown to provide better separation between the signal and background as compared to the rectangular cut-based analysis. Although considerable work has been done in the context of Higgs sector, with these new techniques~\cite{Hultqvist:1995ibm,Bakhet:2015uca,Field:1996rw}, the collider searches for dark matter through Higgs portal scenarios have not been explored in detail with these advanced methods. We have examined and computed the maximum signal significance for the specific signal processes we considered, that can be achieved using these techniques. The toolkit used for Gradient boosting is XGBoost~\cite{Chen:2016btl} and for ANN we used a Python-based deep-learning library Keras~\cite{keras}.

\begin{table}[htpb!]
\centering
 \begin{tabular}{||c | c||} 
 \hline
 Variable & Definition \\ [0.5ex] 
 \hline\hline
 $P^{j}_{T}$ & Transverse momentum of the leading jet \\ 
 $P^{j}_{Z}$ & Longitudinal momentum of the leading jet \\
 $E^{miss}_{T}$ & Missing transverse energy \\
 $N_{j}$ & No of jets in the event \\
 $\phi_{j}$ &   Azimuthal angle of the leading jet \\
 $\phi_{miss}$ & Azimuthal angle of the $\slashed{E_T}$\\
 $\Delta \phi (j, \slashed{E_T} )$ & Angular separation between the leading jet $p_T$ and $\slashed{E_T}$ in the azimuthal plane\\
 $|\eta_{miss}|$ & Pseudorapidity of the missing energy\\
 $|\eta_{j}|$ & Pseudorapidity of the leading jet \\ [1ex] 
 \hline
 \end{tabular}
 \caption{Input variables for XGBoost and ANN analysis for gluon fusion. Azimuthal angles are measured with reference to some arbitrary $x$-axis.}
  \label{ggftab}
\end{table}

\begin{table}[htpb!]
\centering
 \begin{tabular}{||c | c||} 
 \hline
 Variable & Definition \\ [0.5ex] 
 \hline\hline
 $P^{j_1}_{T}$ & Transverse momentum of the leading jet \\
  $P^{j_2}_{T}$ & Transverse momentum of the sub-leading jet \\
 $P^{j_1}_{Z}$ & Longitudinal momentum of the leading jet \\
 $P^{j_2}_{Z}$ & Longitudinal momentum of the sub-leading jet \\
 $E^{miss}_{T}$ & Missing transverse energy \\
 $N_{j}$ & No of jets in the event \\
 $\phi_{j_1}$ &   Azimuthal angle of the leading jet \\
 $\phi_{j_2}$ &   Azimuthal angle of the sub-leading jet \\
 $\phi_{miss}$ & Azimuthal angle of the $\slashed{E_T}$\\
 $\Delta \phi (j, \slashed{E_T} )$ & Angular separation between the leading jet $p_T$ and $\slashed{E_T}$ in the azimuthal plane\\
$\Delta \phi (j_1, j_2)$ & Angular separation between the leading and sub-leading jets in the azimuthal plane\\
 $|\eta_{miss}|$ & Pseudorapidity of the missing energy\\
 $|\eta_{j_1}|$ & Pseudorapidity of the leading jet \\
 $|\eta_{j_2}|$ & Pseudorapidity of the sub-leading jet \\
 $\Delta \eta_{j_1 j_2}$ & Difference of pseudorapidity between the leading and sub-leading jets \\
 $m_{j_1 j_2}$ & Invariant mass of the two leading jets\\ [1ex] 
 \hline
 \end{tabular}
 \caption{Input variables for XGBoost and ANN analysis for vector boson fusion. Azimuthal angles are measured with reference to some arbitrary $x$-axis.}
  \label{vbftab}
\end{table}

We perform the analyses for both signals (gluon fusion and vector boson fusion). We also do a comparative study of the two different techniques mentioned above. From our knowledge of the cut-based analysis done in the previous section of this work, we identified the input feature variables that play important role in separating signal from backgrounds. In both the techniques used here, the choice of input variables play a crucial role. We present in Table.~\ref{ggftab} and Table.~\ref{vbftab} the input variables used for training and validation of our data sample resulting from gluon fusion and vector boson fusion respectively. We have used 9 input variables for gluon fusion and 16 input variables for vector boson fusion channel.

For gradient boosted Decision Tree method of separation, we have taken $\sim 1000$ estimators and maximum depth 4 with learning rate 0.01. For ANN we have chosen a network with 2 hidden layers with activation curve relu at both of them. The batch-size is taken to be 200 with 120 epochs for each batch.  For both XGBoost and ANN analyses we have used 80\% of the total dataset for training purpose and 20\% for validation. One of the possible demerits of these techniques is over-training of the data sample. In case of over-training the training sample gives extremely good accuracy but the test sample fails to achieve that. We have explicitly checked that with our choice of parameters the algorithm does not over-train.

\begin{figure}[!hptb]
\includegraphics[width=7.5cm, height=6cm]{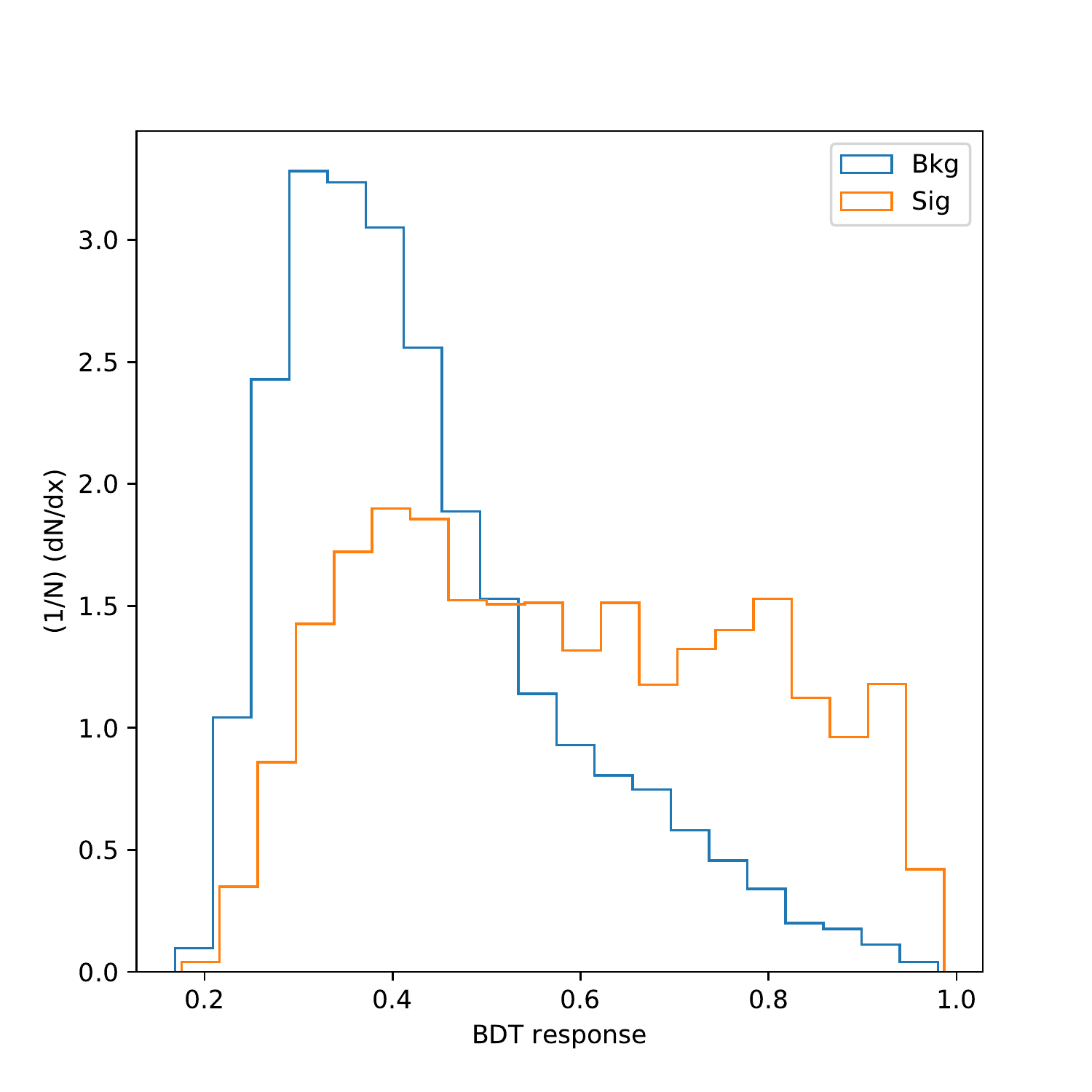}
\hspace{0.9cm}
\includegraphics[width=7.5cm, height=6cm]{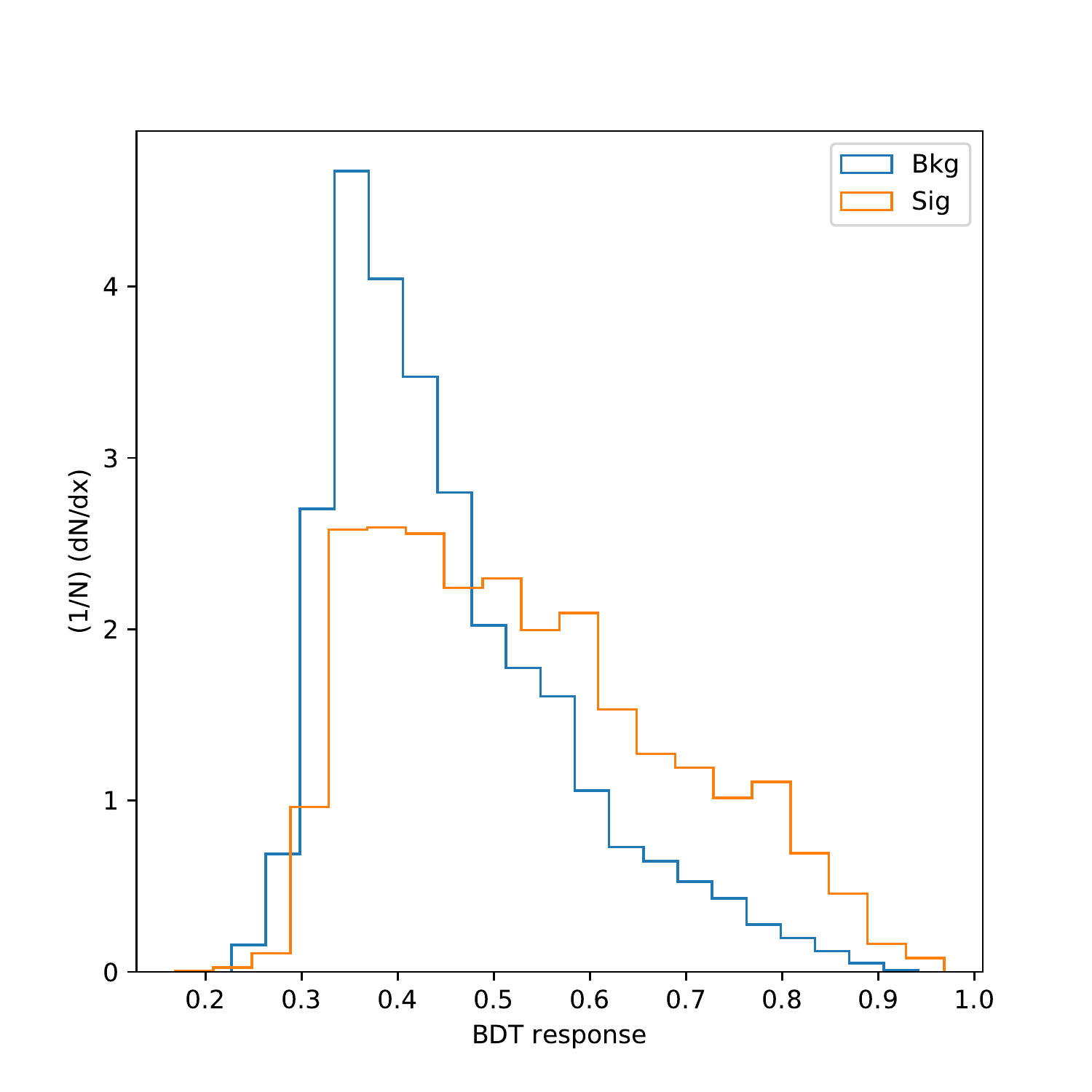} \\
\vspace*{0.02cm}
\includegraphics[width=7.5cm, height=6cm]{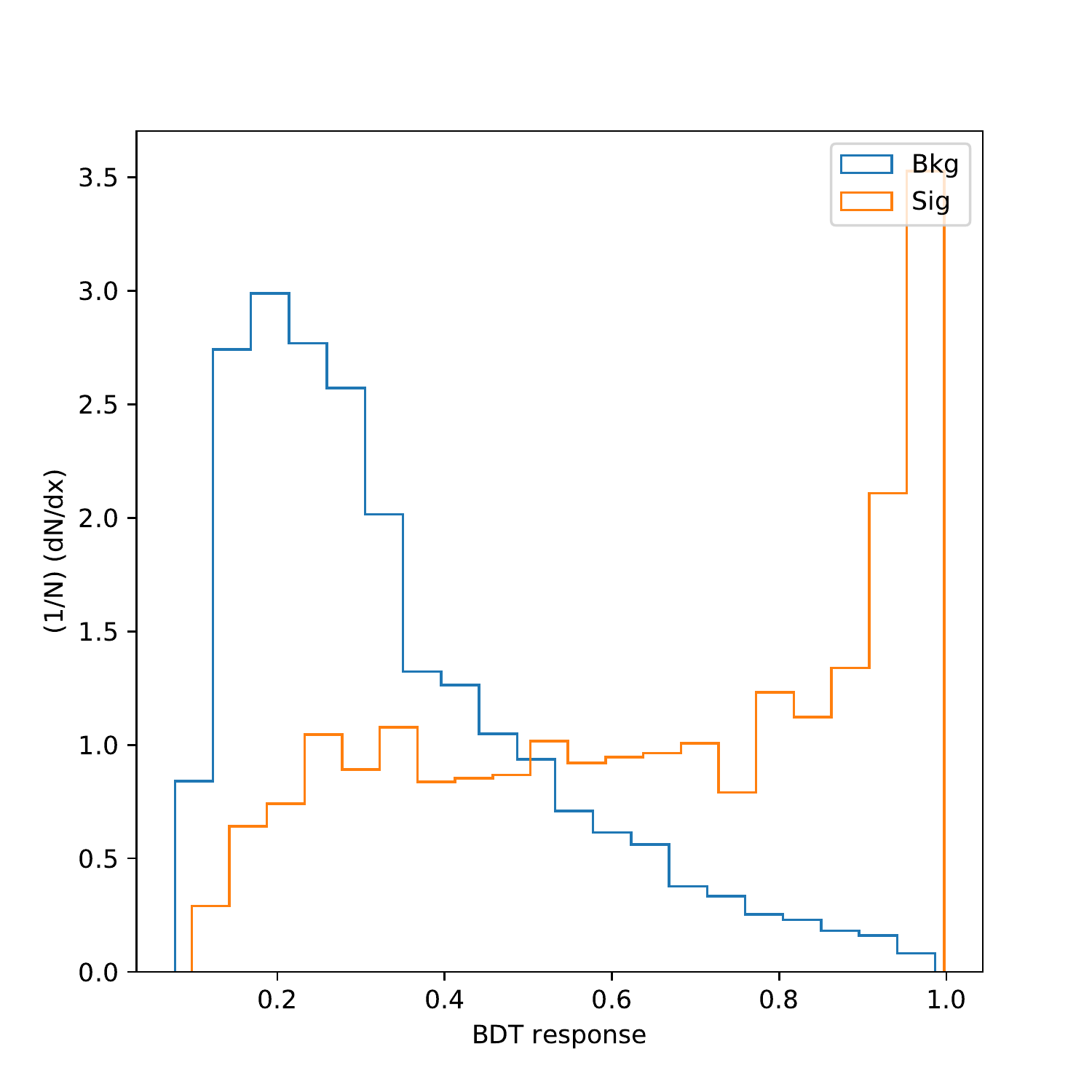}
\hspace{0.9cm}
\includegraphics[width=7.5cm, height=6cm]{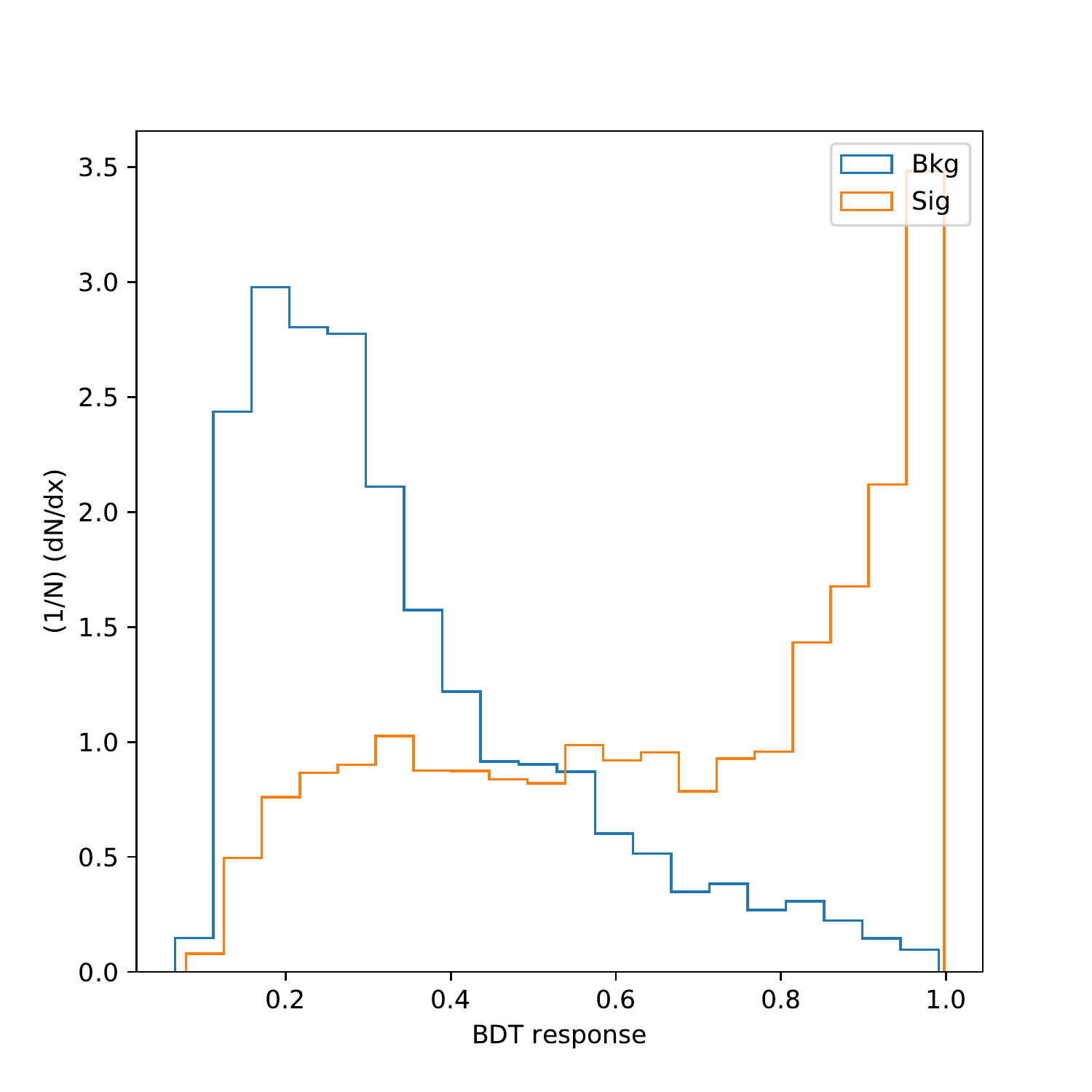} \\
\caption{BDT classifier response for different benchmark points of gluon fusion for Type I BP I (top left), Type I BP II (top right), Type II BP I (bottom left) and Type II BP II (bottom right).}
\label{bdt_classifier}
\end{figure}   

\newpage
\noindent
{\bf Gluon fusion:}

\begin{figure}[!hptb]
\includegraphics[width=7.5cm, height=6cm]{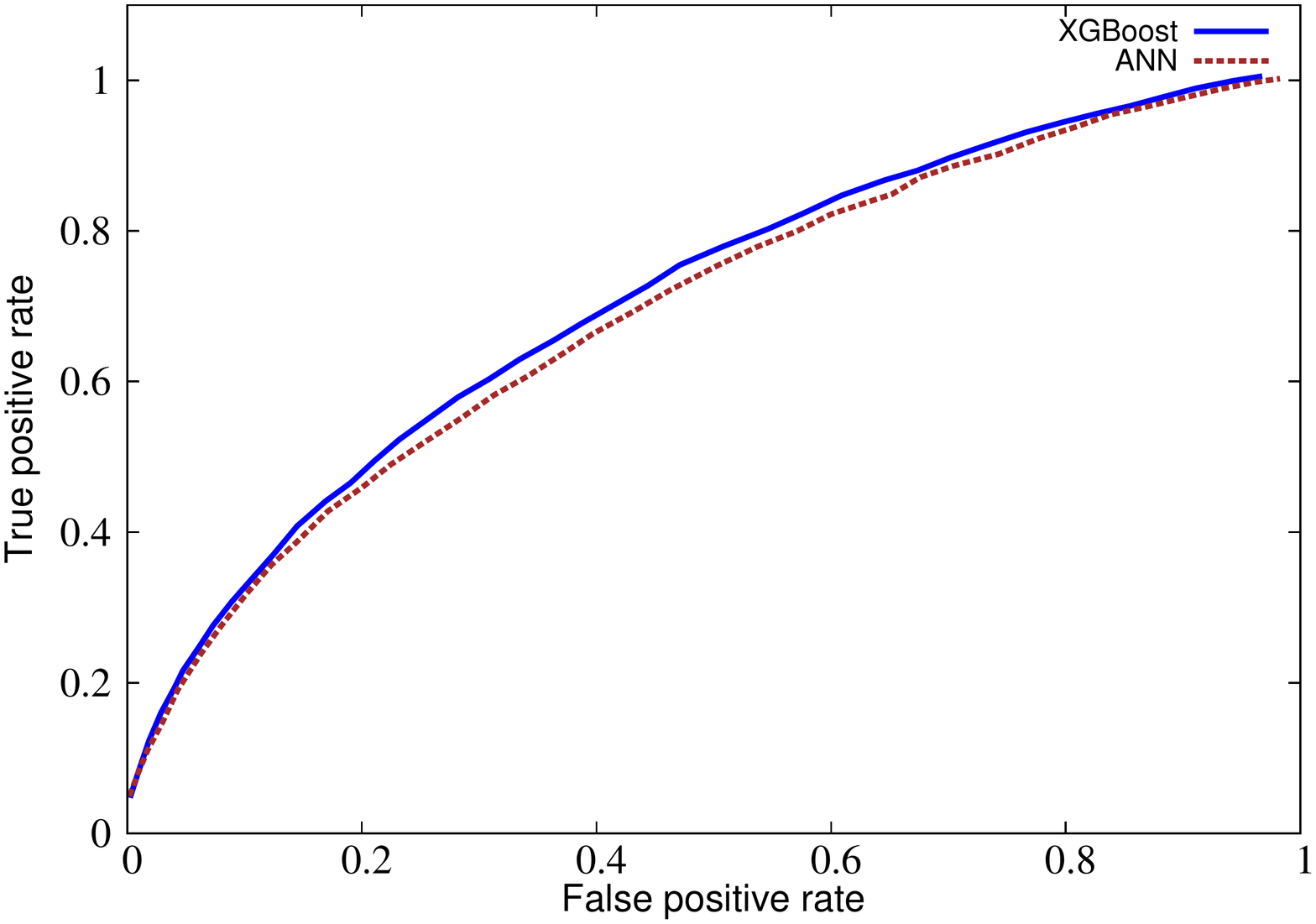}
\hspace{0.9cm}
\includegraphics[width=7.5cm, height=6cm]{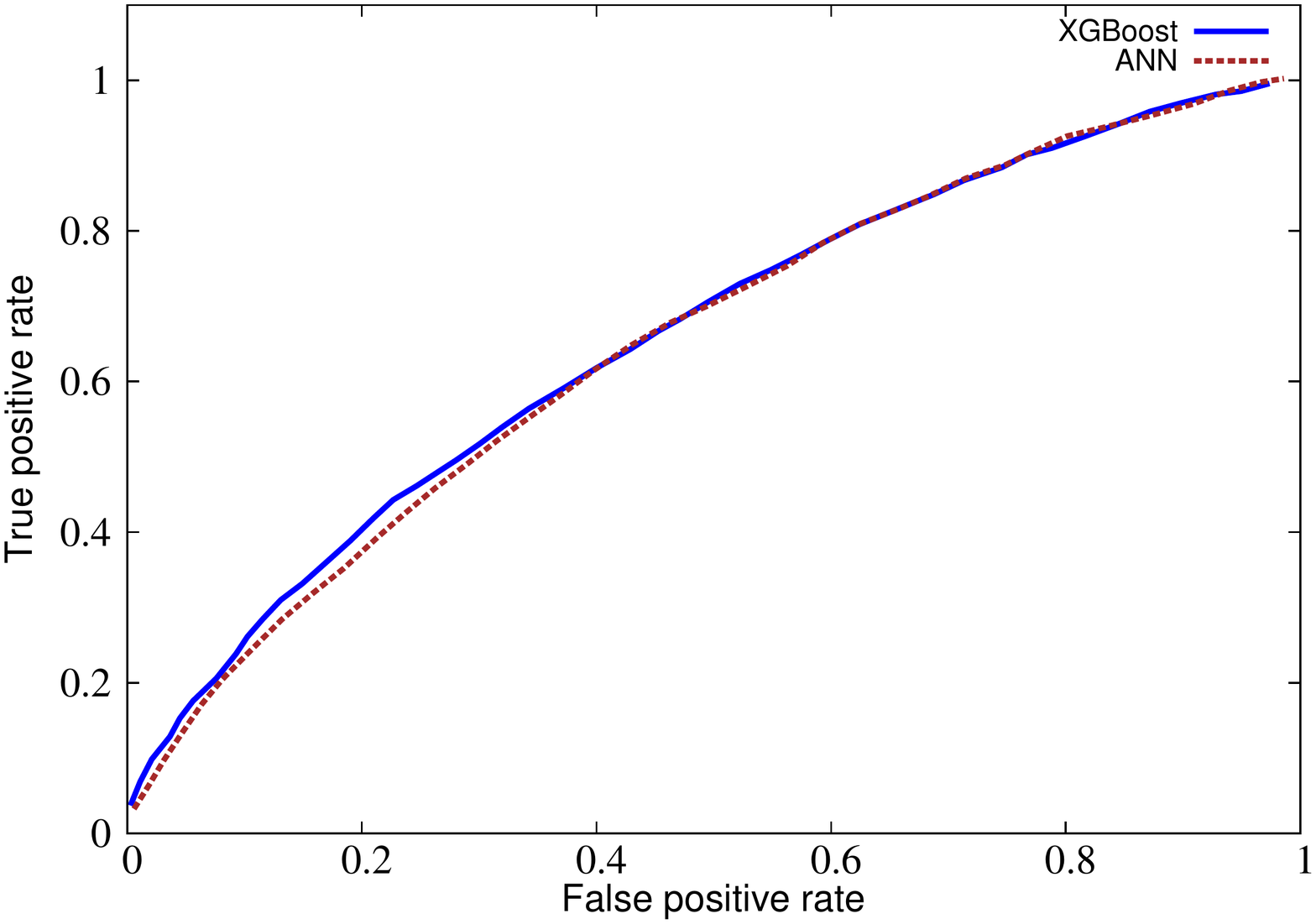} \\
\vspace*{0.02cm}
\includegraphics[width=7.5cm, height=6cm]{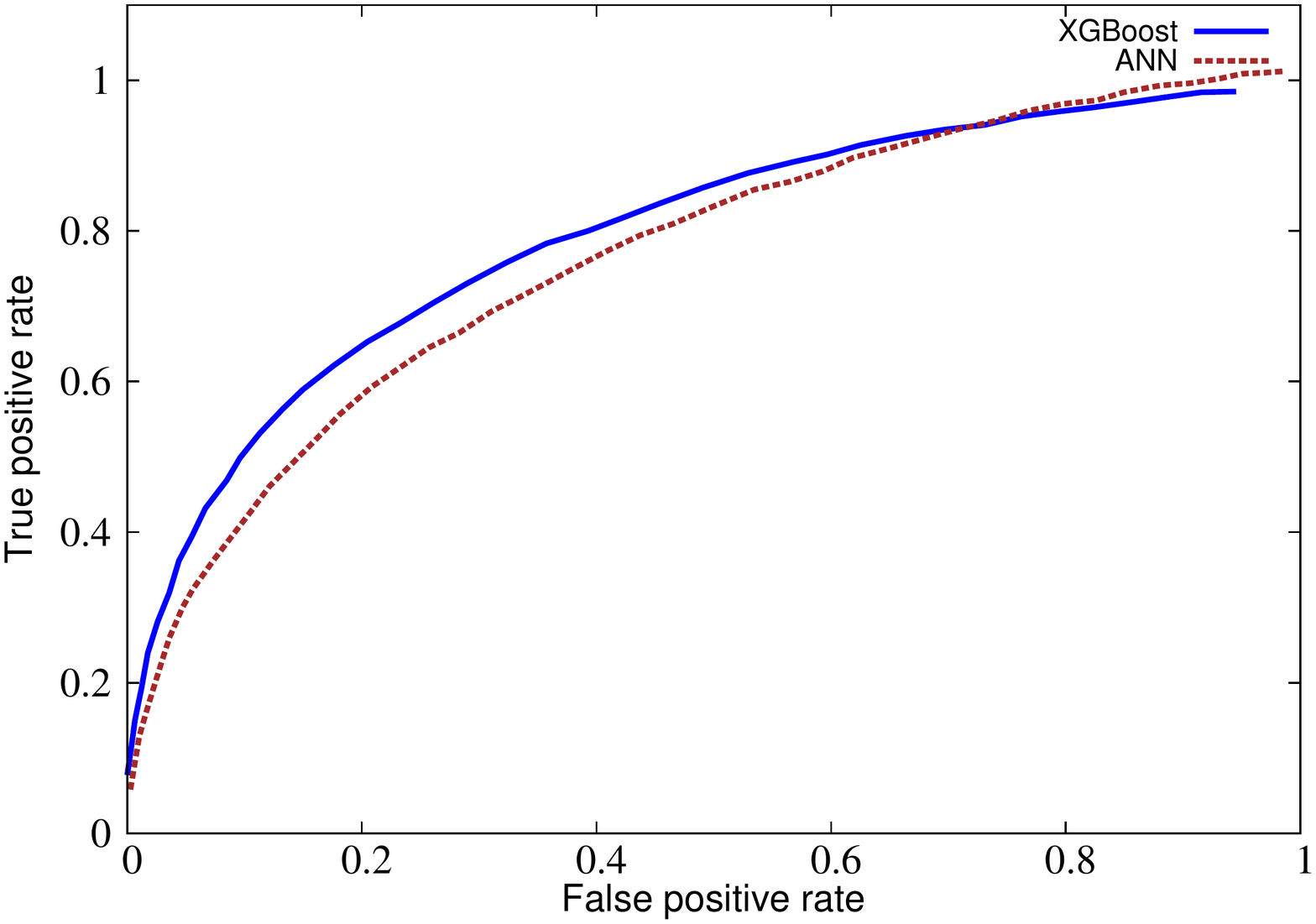}
\hspace{0.9cm}
\includegraphics[width=7.5cm, height=6cm]{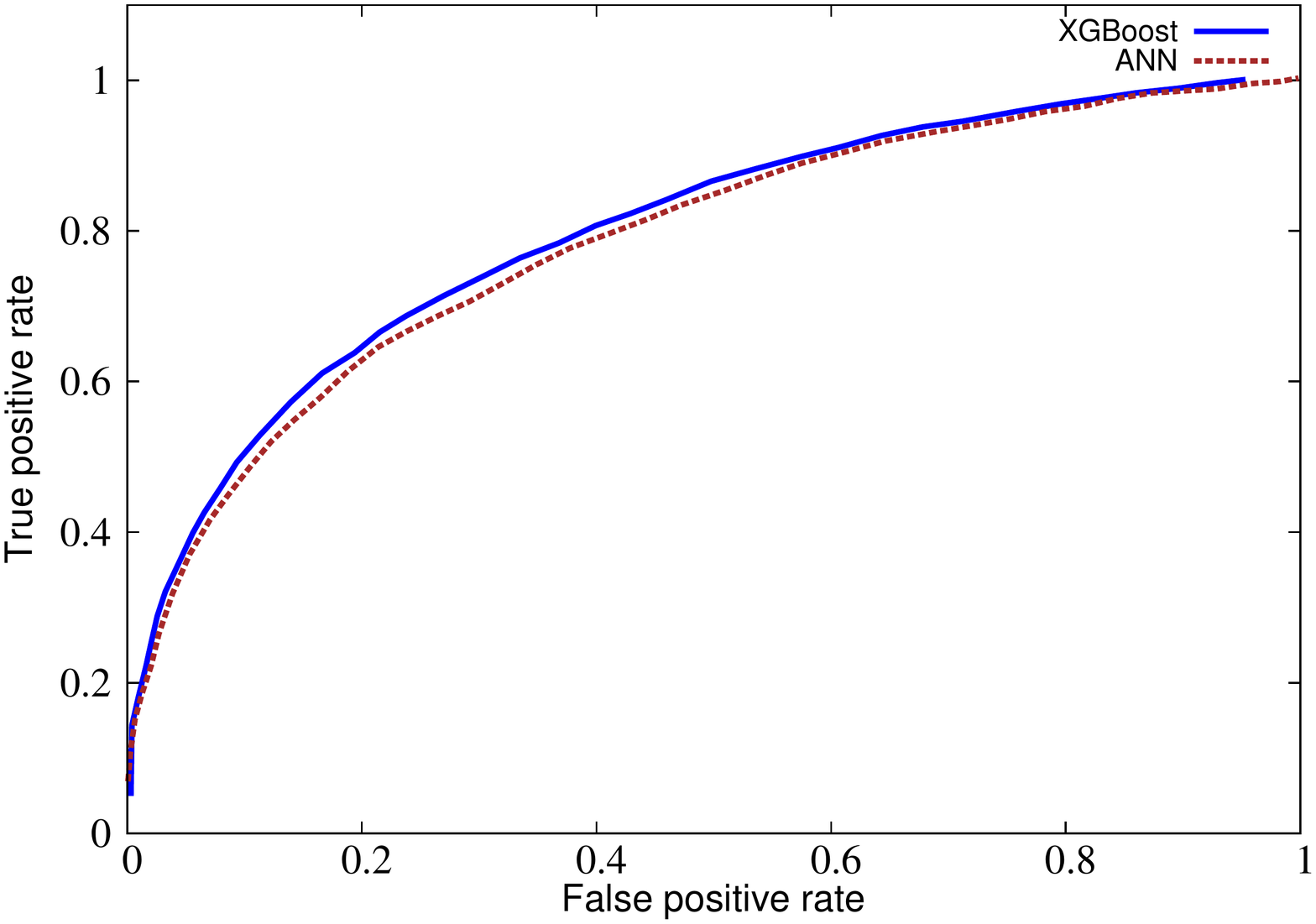} \\
\caption{ROC curves of gluon fusion for Type I BP I (top left), Type I BP II (top right), Type II BP I (bottom left) and Type II BP II (bottom right).}
\label{roc}
\end{figure}

Distribution of the BDT classifier response for the signal and total background events for gluon fusion process have been shown in Figure~\ref{bdt_classifier}. We can see that in case of Type I BP I and II, the classifier does not perform very well. However it does slightly improve our cut-based analysis. BDT classifier performs much better in case of Type II signals. The reason behind this is the separation between signal and background in type II is much more prominent than Type I case as we have already seen from the distributions in the gluon fusion process. We have checked that in the gluon fusion process $\slashed{E_T}$ plays the role of the most important input variable in the BDT analysis and jet multiplicity and $p_T$ distribution of the leading jet are the second and third best discriminator in this case. We have plotted the Receiver Operating Characteristic (ROC) curve for gluon fusion for the four benchmark signal processes in Figure.~\ref{roc}. For comparison we 
 show the ROC curve from XGBoost and ANN in the same plot. However in all the cases XGBoost performs better than ANN, although the difference between them is not much. The area under the ROC curve is 0.69 for Type I BP I, 0.65 for Type I BP II, 0.80 for Type II BP I and 0.79 for Type II BP II. 

\begin{table}[!hptb]
\begin{center}
\begin{footnotesize}
\begin{tabular}{| c | c | c |}
\hline
BP & $
{\cal S}$ (Cuts + XGboost) &  ${\cal S}$ (Cuts + ANN) \\
\hline
Type I BP I  & 5.7 $\sigma$ & 5.6 $\sigma$ \\
\hline
Type I BP II  & 2.1 $\sigma$ & 2.0 $\sigma$ \\
\hline
Type II BP I  & 5.5 $\sigma$ & 4.3 $\sigma$ \\
\hline
Type II BP II  & 4.9 $\sigma$ & 4.8 $\sigma$ \\
\hline
\end{tabular}
\end{footnotesize}
\caption{Signal significance for the benchmark points at 14 TeV with ${\cal L}$ = 3000 fb$^{-1}$ in the gluon fusion channel with XGboost and ANN method with some initial cuts. }
\label{sigmlearn}
\end{center}
\end{table}

In Table~\ref{sigmlearn} we present the modified signal significance (${\cal S}$) after applying basic cuts and then performing XGBoost and ANN algorithm. We compare these results with those obtained in Table~\ref{significance_ggf}. It is evident that Type I results from the cut-based analysis are only slightly modified after applying both ANN and XGboost techniques. This is the consequence of the lack of separation between signal and background in Type I in the space of all variables one would try out. On the other hand, the Type II results show significant improvement from the rectangular cut-based analysis, because of better separation between signal and background distributions. We can see that the XGboost gives better signal significance for all the benchmark points in case of gluon fusion.

One can see from Table~\ref{sigmlearn} that even though the Type I benchmark points do not have the advantage of a clear separation between the signal and background, the larger signal cross section in Type I benchmarks (which is a direct consequence of low enough $m_H$), enables us to achieve signal significance comparable with that of the Type II cases.  

\begin{figure}[!hptb]
\includegraphics[width=7.5cm, height=6cm]{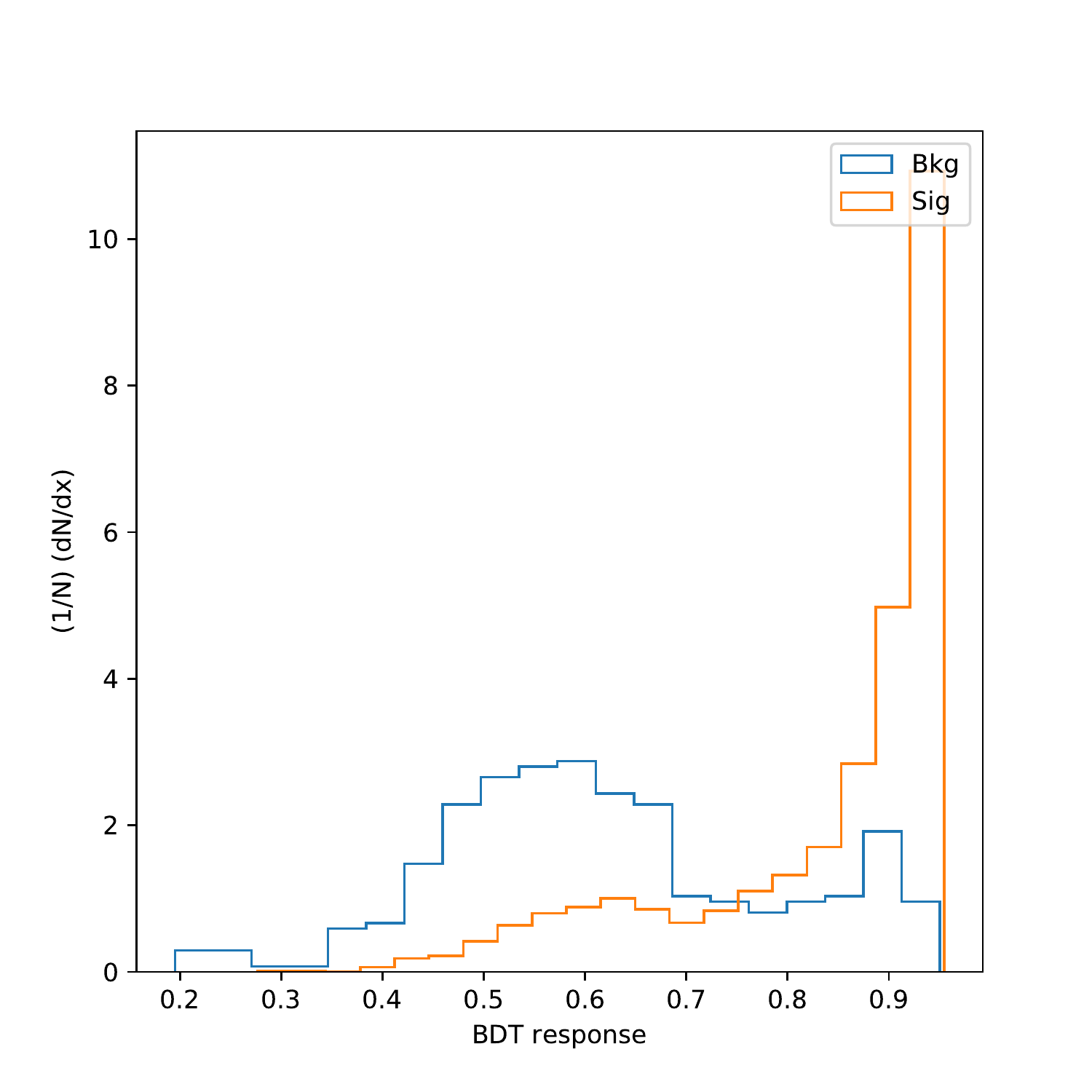}
\hspace{0.9cm}
\includegraphics[width=7.5cm, height=6cm]{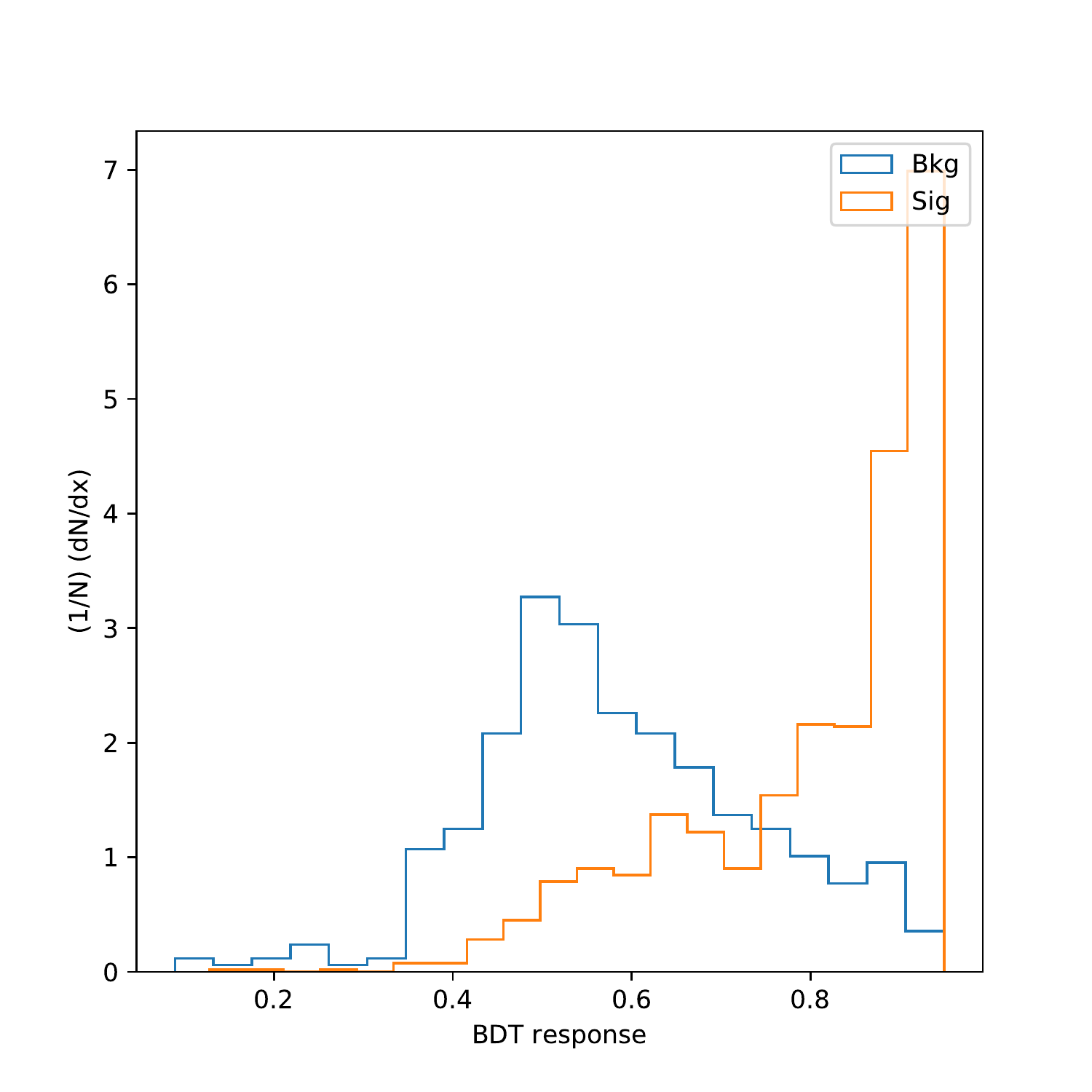} \\
\vspace*{0.02cm}
\includegraphics[width=7.5cm, height=6cm]{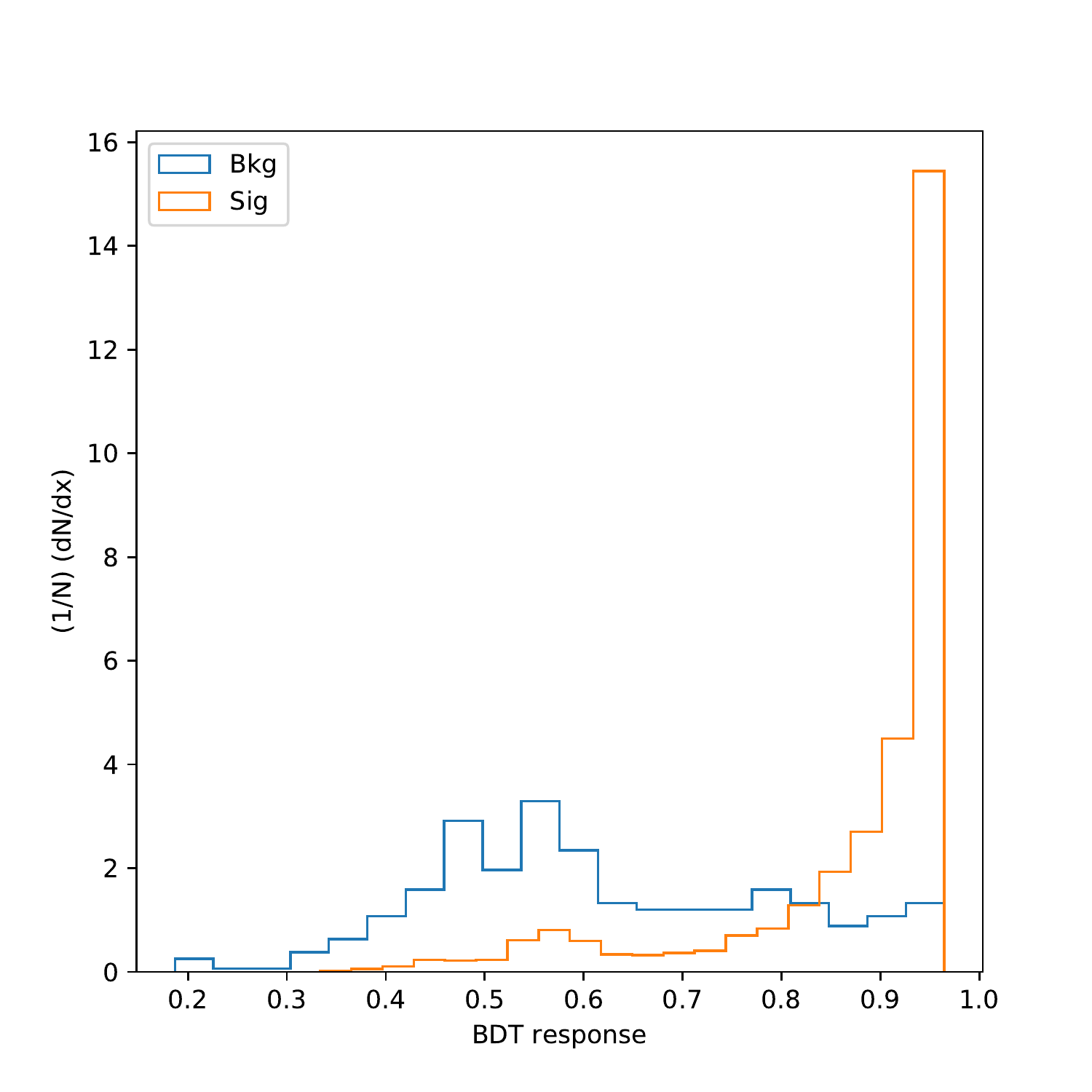}
\hspace{0.9cm}
\includegraphics[width=7.5cm, height=6cm]{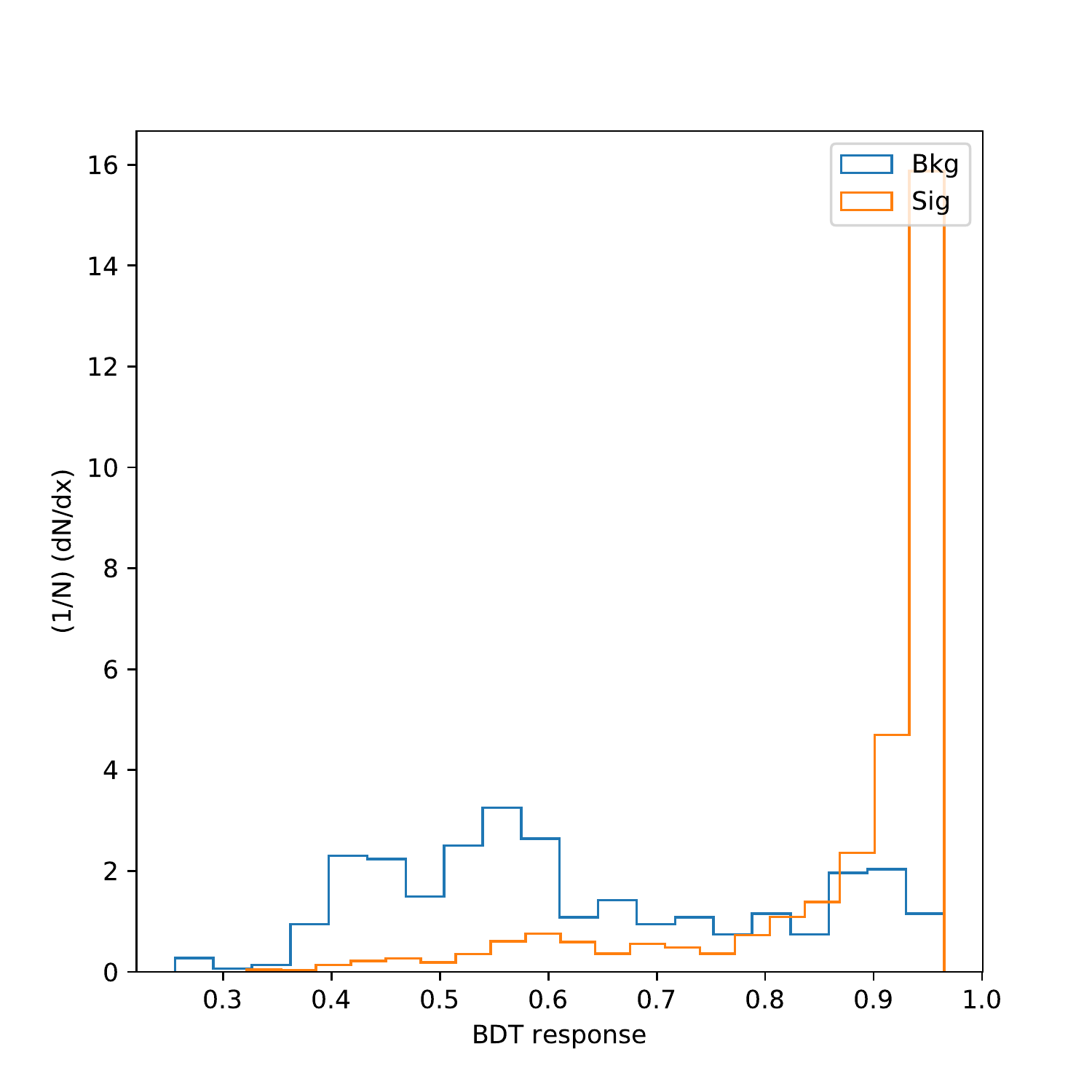} \\
\caption{BDT classifier response for different benchmark points of vector boson fusion for Type I BP I (top left), Type I BP II (top right), Type II BP I (bottom left) and Type II BP II (bottom right).}
\label{bdt_classifier_vbf}
\end{figure}

\begin{figure}[!hptb]
\includegraphics[width=7.5cm, height=6cm]{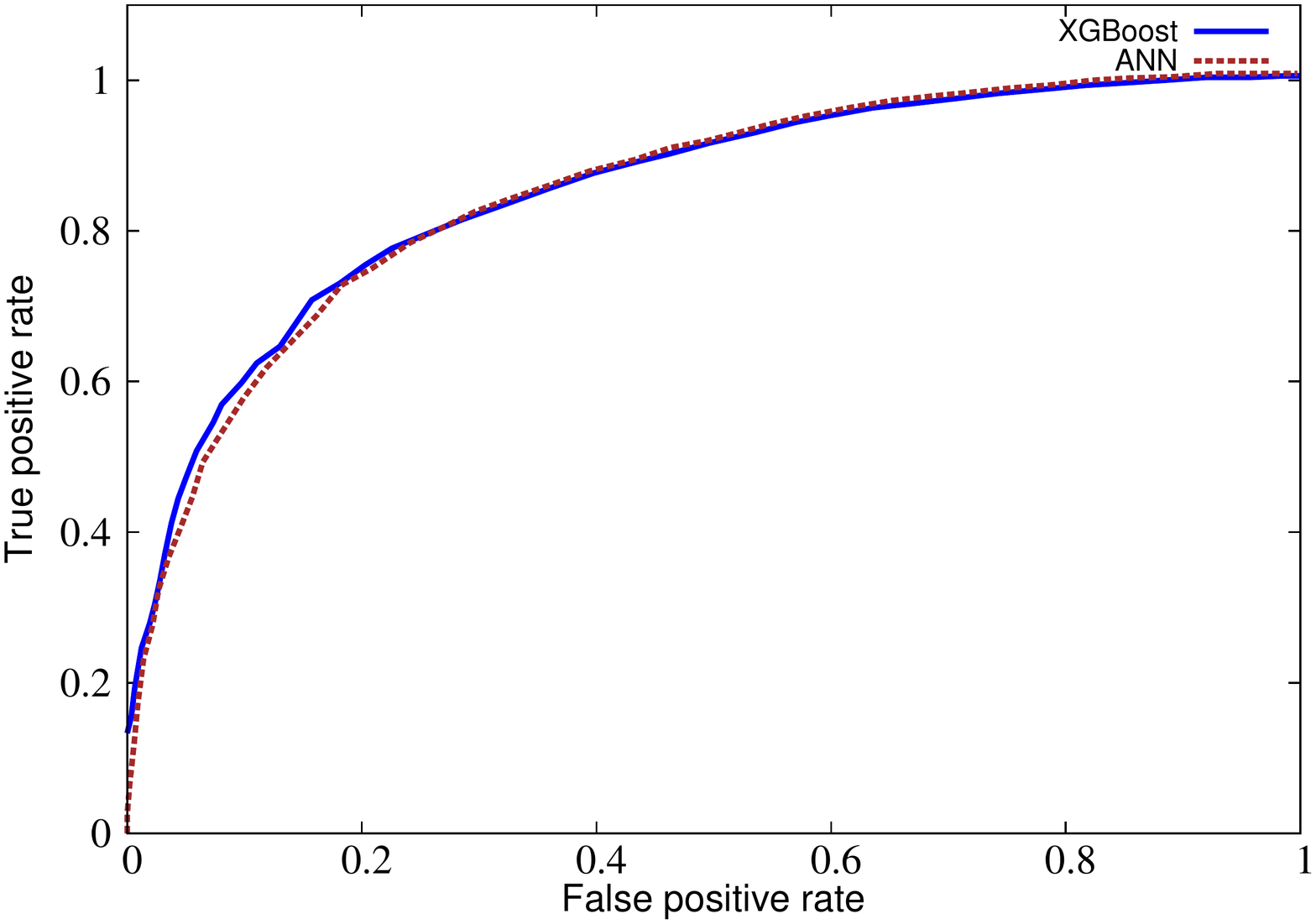}
\hspace{0.9cm}
\includegraphics[width=7.5cm, height=6cm]{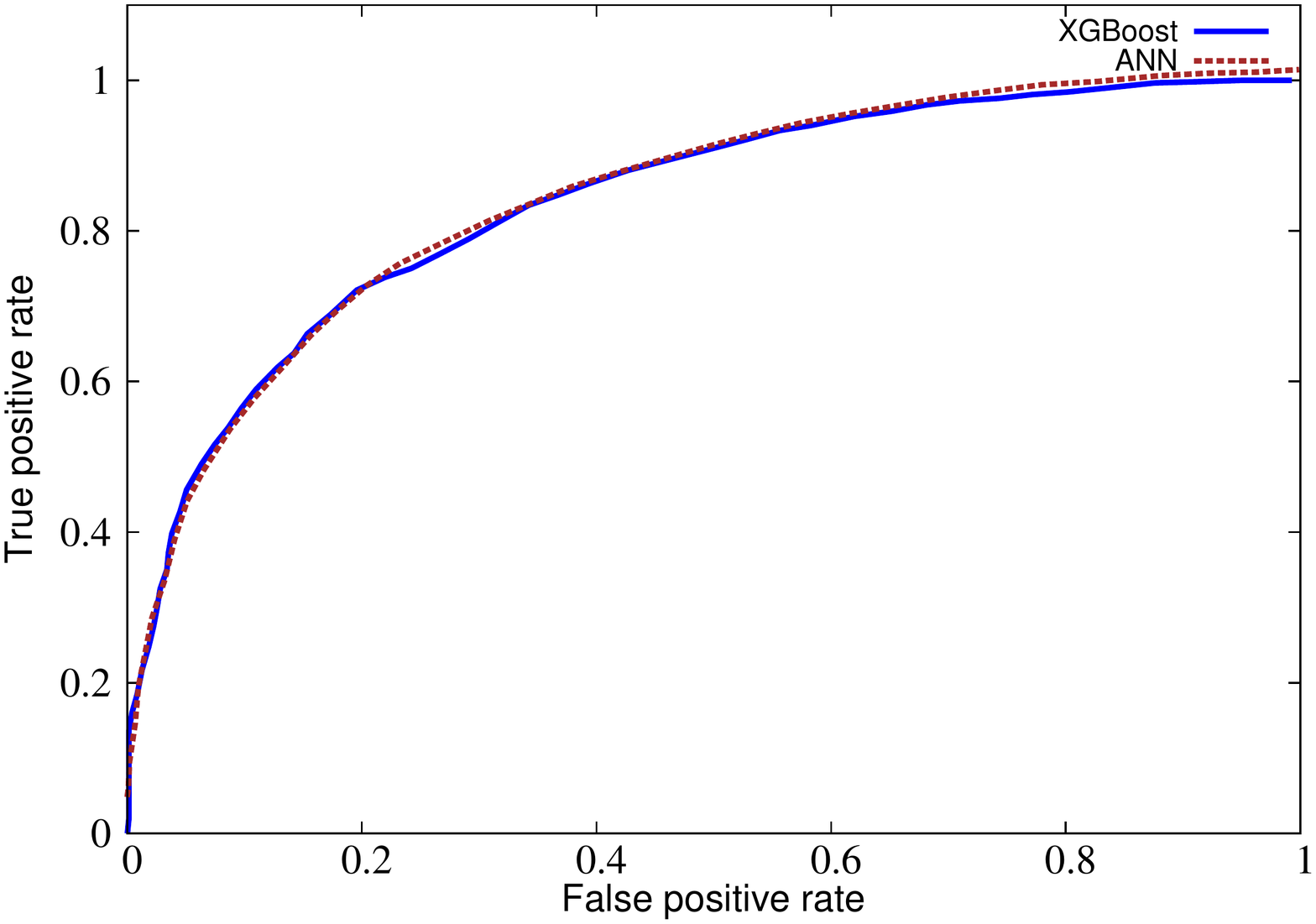} \\
\vspace*{0.02cm}
\includegraphics[width=7.5cm, height=6cm]{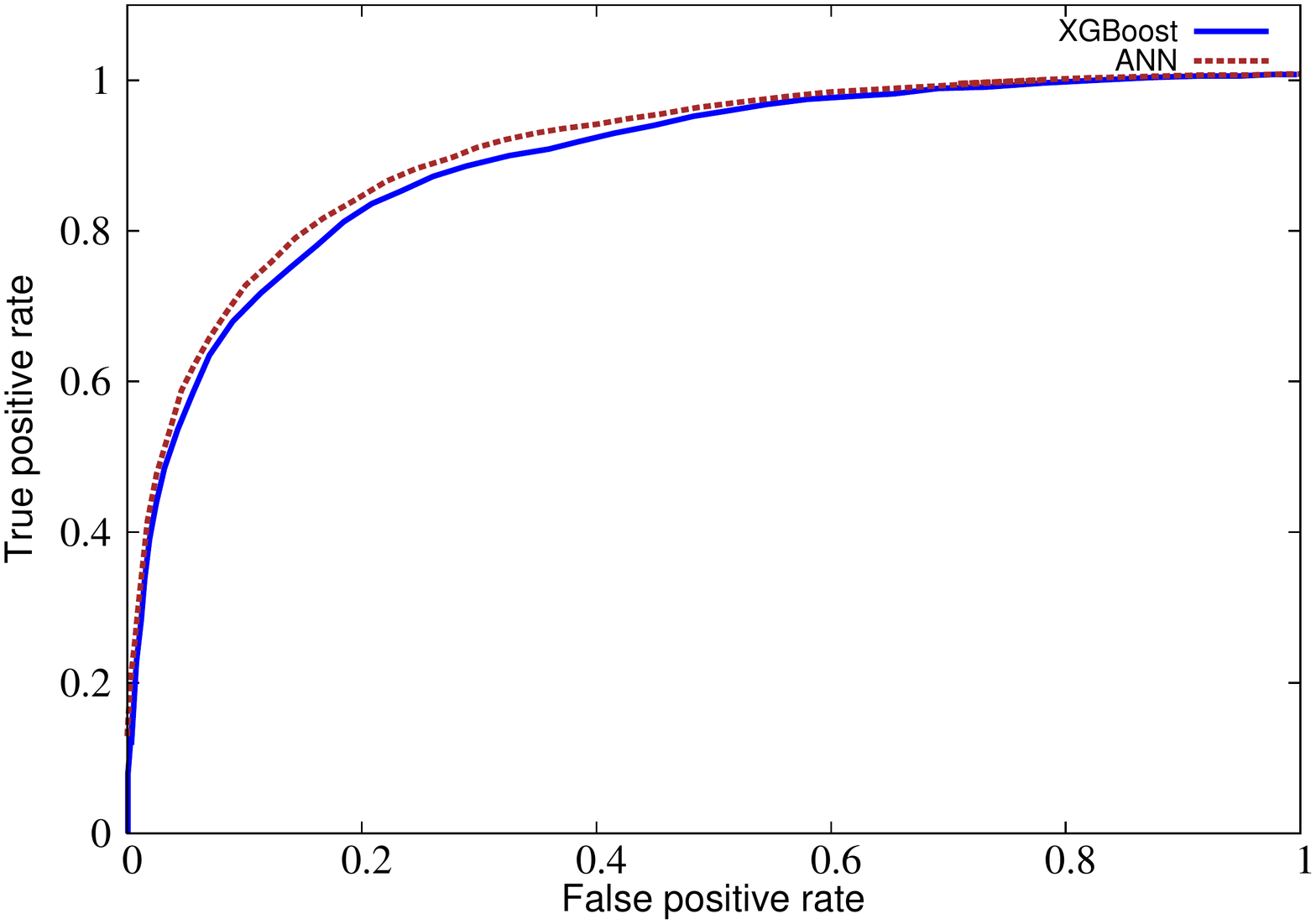}
\hspace{0.9cm}
\includegraphics[width=7.5cm, height=6cm]{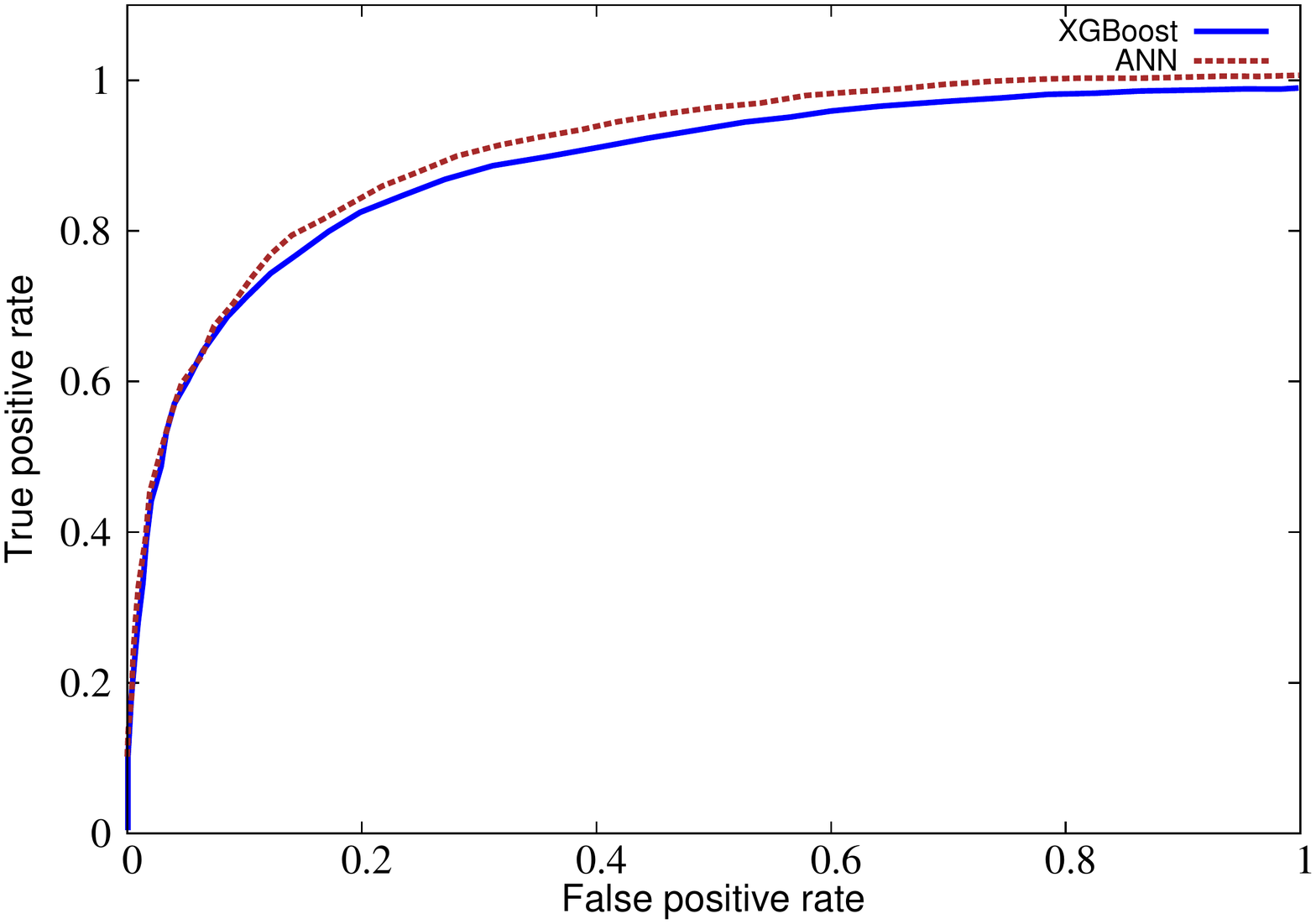} \\
\caption{ROC curves of vector boson fusion channel for Type I BP I (top left), Type I BP II (top right), Type II BP I (bottom left) and Type II BP II (bottom right).}
\label{roc_vbf}
\end{figure}   

\medskip
\noindent
{\bf Vector boson fusion:}

We perform a similar study for vector boson fusion process. Here the number of feature variables are more and therefore considerable separation between signal and backgrounds is possible to achieve in this case. This is true even for Type I, as can be seen in Figure.~\ref{bdt_classifier_vbf}. The enhancement of signal over background here is much more significant as compared to the gluon fusion case discussed earlier. $m_{j_1 j_2}$ (invariant mass of the two forward jets), $\Delta \eta_{j_1 j_2}$, $\slashed{E_T}$ and transverse momenta of the two forward jets play the key role in separating the signal from backgrounds. In Figure.~\ref{roc_vbf} we show the ROC curve in the false positive - true positive plane for both XGboost and ANN methods. We see that both the algorithms perform almost equally well for any specific true positive rate. However, in Type II ANN performs marginally better than the XGboost method. The area under the ROC curve is 0.85 for Type I BP I, 0.84 for Ty
 pe I BP II, 0.89 for Type II BP I and 0.88 for Type II BP II.

After choosing suitable points from the ROC curve, we compute the maximum signal significance ${\cal S}$ which can be achieved at a given integrated luminosity for both XGboost and ANN methods. The results presented in Table.~\ref{sigmlearnvbf} should be compared with those in Table.~\ref{significance_vbf} to understand the improvement in the analysis that can be attained by these newly developed techniques. We can thus see that both Type I and Type II benchmark points undergo significant improvement in case of vector boson fusion.

One should notice that even for Type I BP II, one can obtain $\sim 4.7\sigma$ significance in vector boson fusion after applying ANN or XGboost method whereas in gluon fusion, this benchmark was of hardly any significance($\sim 2 \sigma$). Therefore  it is clear from the discussion that with the application of the new analysis based on the neural network and boosted decision tree algorithm, it is possible to probe the parameter space of our model with considerable significance even at not-so-high integrated luminosity. The observation we made from this analysis is that the benchmark points for which the rectangular cut-based method performs well, the ANN and XGboost method also perform significantly well. Likewise, it is difficult to achieve further improvement through these techniques, from cut-based analysis when the separation between signal and background is extremely poor. 

\begin{table}[!hptb]
\begin{center}
\begin{footnotesize}
\begin{tabular}{| c | c | c |}
\hline
BP & $
{\cal S}$ (Cuts + XGboost) &  ${\cal S}$ (Cuts + ANN) \\
\hline
Type I BP I  & $8.0\sigma$(600 fb$^{-1}$) & 7.4$\sigma$(600 fb$^{-1}$) \\
\hline
Type I BP II  & 4.7$\sigma$ (3000 fb$^{-1}$) & 4.6 $\sigma$(3000 fb$^{-1}$) \\
\hline
Type II BP I  & 11.0 $\sigma$(3000 fb$^{-1}$) & 12.1 $\sigma$(3000 fb$^{-1}$) \\
\hline
Type II BP II  & 9.9 $\sigma$(3000 fb$^{-1}$) & 10.0 $\sigma$(3000 fb$^{-1}$) \\
\hline
\end{tabular}
\end{footnotesize}
\caption{Signal significance for the benchmark points at 14 TeV in the vector fusion channel with XGboost and ANN techniques after applying some initial cuts. }
\label{sigmlearnvbf}
\end{center}
\end{table}

\section{Conclusions}\label{sec7}

For quite some time the Higgs boson of SM has been speculated to be a portal to the dark sector. But this scenario is highly constrained from the consideration of relic density as well as direct search experiments of DM. Therefore one can take a step forward and explore the possibility of the heavy CP-even Higgs boson of 2HDM to be the dark matter portal. We find in our analysis that in various kinds of 2HDM, there is considerable parameter space which is allowed by all the theoretical and experimental constraints. The signals to look for at the LHC in such situations emerge as  monojet + missing energy and two forward jet + missing energy search. The former process can take place when the heavy Higgs is produced through gluon fusion and the latter can take place when the heavy Higgs is produced through vector boson fusion, in association with two forward jets. 

We first perform a complete cut-based analysis for both gluon fusion and vector boson fusion process. We identify the observables which can be used to separate the signal and background with a desired efficiency and put optimum cuts on these input variables. Then we calculate the projected signal significance that can be achieved at the future high-luminosity LHC.

Next we have employed two recently developed techniques namely XGBoost and artificial neural network. These methods have obvious advantage over rectangular cut-based analyses. We have used optimal parameters for each method and obtained improved significance for all our benchmark points. We see that in case of Type I benchmark points in gluon fusion process the new algorithms do not improve the cut-based results much, but in case of Type II the results are significantly improved. As $m_H$ can be light in Type I 2HDM in contrast to the Type II case, with the large production cross section of heavy Higgs in Type I, one ends up getting comparable signal significance in both Type I and Type II scenarios. In vector boson fusion Type I BP I scenario can be probed even at a lower luminosity (300 fb$^{-1}$) at $\sim 5 \sigma$  CL. Thus a heavy Higgs portal scenario based on 2HDM is testable at the upcoming phase of the LHC, with the promise of sizeable statistical significance especially through multivariate/neural network techniques.

\section{Acknowledgement}

We thank Akanksha Bharadwaj, Satyaki Bhattacharya, Debabrata Bhowmik, Asesh K. Datta, Samik Ghosh, Arun Nayak and Soumya Sadhukhan for useful discussions and also for helping with codes. This  work  was  supported  by  funding  available  from  the  Department of Atomic Energy, Government of India, for the Regional Centre for Accelerator-based Particle Physics (RECAPP), Harish-Chandra Research Institute.  

\bibliographystyle{JHEP}
\bibliography{paperbib}

\providecommand{\href}[2]{#2}\begingroup\raggedright\begin{thebibliography}{10}

\bibitem{Djouadi:2012zc}
A.~Djouadi, A.~Falkowski, Y.~Mambrini and J.~Quevillon, \emph{{Direct Detection
  of Higgs-Portal Dark Matter at the LHC}},
  \href{https://doi.org/10.1140/epjc/s10052-013-2455-1}{\emph{Eur. Phys. J.}
  {\bfseries C73} (2013) 2455},
  [\href{https://arxiv.org/abs/1205.3169}{{\ttfamily 1205.3169}}].

\bibitem{Han:2016gyy}
H.~Han, J.~M. Yang, Y.~Zhang and S.~Zheng, \emph{{Collider Signatures of
  Higgs-portal Scalar Dark Matter}},
  \href{https://doi.org/10.1016/j.physletb.2016.03.010}{\emph{Phys. Lett.}
  {\bfseries B756} (2016) 109--112},
  [\href{https://arxiv.org/abs/1601.06232}{{\ttfamily 1601.06232}}].

\bibitem{LopezHonorez:2012kv}
L.~Lopez-Honorez, T.~Schwetz and J.~Zupan, \emph{{Higgs portal, fermionic dark
  matter, and a Standard Model like Higgs at 125 GeV}},
  \href{https://doi.org/10.1016/j.physletb.2012.07.017}{\emph{Phys. Lett.}
  {\bfseries B716} (2012) 179--185},
  [\href{https://arxiv.org/abs/1203.2064}{{\ttfamily 1203.2064}}].

\bibitem{Greljo:2013wja}
A.~Greljo, J.~Julio, J.~F. Kamenik, C.~Smith and J.~Zupan, \emph{{Constraining
  Higgs mediated dark matter interactions}},
  \href{https://doi.org/10.1007/JHEP11(2013)190}{\emph{JHEP} {\bfseries 11}
  (2013) 190}, [\href{https://arxiv.org/abs/1309.3561}{{\ttfamily 1309.3561}}].

\bibitem{Fedderke:2014wda}
M.~A. Fedderke, J.-Y. Chen, E.~W. Kolb and L.-T. Wang, \emph{{The Fermionic
  Dark Matter Higgs Portal: an effective field theory approach}},
  \href{https://doi.org/10.1007/JHEP08(2014)122}{\emph{JHEP} {\bfseries 08}
  (2014) 122}, [\href{https://arxiv.org/abs/1404.2283}{{\ttfamily 1404.2283}}].

\bibitem{Ade:2013zuv}
{\scshape Planck} collaboration, P.~A.~R. Ade et~al., \emph{{Planck 2013
  results. XVI. Cosmological parameters}},
  \href{https://doi.org/10.1051/0004-6361/201321591}{\emph{Astron. Astrophys.}
  {\bfseries 571} (2014) A16},
  [\href{https://arxiv.org/abs/1303.5076}{{\ttfamily 1303.5076}}].

\bibitem{Gunion:1989we}
J.~F. Gunion, H.~E. Haber, G.~L. Kane and S.~Dawson, \emph{{The Higgs Hunter's
  Guide}}, {\emph{Front. Phys.} {\bfseries 80} (2000) 1--404}.

\bibitem{Branco:2011iw}
G.~C. Branco, P.~M. Ferreira, L.~Lavoura, M.~N. Rebelo, M.~Sher and J.~P.
  Silva, \emph{{Theory and phenomenology of two-Higgs-doublet models}},
  \href{https://doi.org/10.1016/j.physrep.2012.02.002}{\emph{Phys. Rept.}
  {\bfseries 516} (2012) 1--102},
  [\href{https://arxiv.org/abs/1106.0034}{{\ttfamily 1106.0034}}].

\bibitem{Han:2017etg}
L.~Wang, R.~Shi and X.-F. Han, \emph{{Wrong sign Yukawa coupling of the 2HDM
  with a singlet scalar as dark matter confronted with dark matter and Higgs
  data}}, \href{https://doi.org/10.1103/PhysRevD.96.115025}{\emph{Phys. Rev.}
  {\bfseries D96} (2017) 115025},
  [\href{https://arxiv.org/abs/1708.06882}{{\ttfamily 1708.06882}}].

\bibitem{Bandyopadhyay:2017tlq}
P.~Bandyopadhyay, E.~J. Chun and R.~Mandal, \emph{{Scalar Dark Matter in
  Leptophilic Two-Higgs-Doublet Model}},
  \href{https://doi.org/10.1016/j.physletb.2018.01.071}{\emph{Phys. Lett.}
  {\bfseries B779} (2018) 201--205},
  [\href{https://arxiv.org/abs/1709.08581}{{\ttfamily 1709.08581}}].

\bibitem{Boucenna:2011hy}
M.~S. Boucenna and S.~Profumo, \emph{{Direct and Indirect Singlet Scalar Dark
  Matter Detection in the Lepton-Specific two-Higgs-doublet Model}},
  \href{https://doi.org/10.1103/PhysRevD.84.055011}{\emph{Phys. Rev.}
  {\bfseries D84} (2011) 055011},
  [\href{https://arxiv.org/abs/1106.3368}{{\ttfamily 1106.3368}}].

\bibitem{Han:2018bni}
L.~Wang, X.-F. Han and B.~Zhu, \emph{{Light scalar dark matter extension of the
  type-II two-Higgs-doublet model}},
  \href{https://doi.org/10.1103/PhysRevD.98.035024}{\emph{Phys. Rev.}
  {\bfseries D98} (2018) 035024},
  [\href{https://arxiv.org/abs/1801.08317}{{\ttfamily 1801.08317}}].

\bibitem{Arhrib:2018qmw}
A.~Arhrib, R.~Benbrik, M.~El~Kacimi, L.~Rahili and S.~Semlali, \emph{{Extended
  Higgs sector of 2HDM with real singlet facing LHC data}},
  \href{https://arxiv.org/abs/1811.12431}{{\ttfamily 1811.12431}}.

\bibitem{Berlin:2015wwa}
A.~Berlin, S.~Gori, T.~Lin and L.-T. Wang, \emph{{Pseudoscalar Portal Dark
  Matter}}, \href{https://doi.org/10.1103/PhysRevD.92.015005}{\emph{Phys. Rev.}
  {\bfseries D92} (2015) 015005},
  [\href{https://arxiv.org/abs/1502.06000}{{\ttfamily 1502.06000}}].

\bibitem{Arcadi:2019lka}
G.~Arcadi, A.~Djouadi and M.~Raidal, \emph{{Dark Matter through the Higgs
  portal}},  \href{https://arxiv.org/abs/1903.03616}{{\ttfamily 1903.03616}}.

\bibitem{Drozd:2014yla}
A.~Drozd, B.~Grzadkowski, J.~F. Gunion and Y.~Jiang, \emph{{Extending
  two-Higgs-doublet models by a singlet scalar field - the Case for Dark
  Matter}}, \href{https://doi.org/10.1007/JHEP11(2014)105}{\emph{JHEP}
  {\bfseries 11} (2014) 105},
  [\href{https://arxiv.org/abs/1408.2106}{{\ttfamily 1408.2106}}].

\bibitem{Deshpande:1977rw}
N.~G. Deshpande and E.~Ma, \emph{{Pattern of Symmetry Breaking with Two Higgs
  Doublets}}, \href{https://doi.org/10.1103/PhysRevD.18.2574}{\emph{Phys. Rev.}
  {\bfseries D18} (1978) 2574}.

\bibitem{Nie:1998yn}
S.~Nie and M.~Sher, \emph{{Vacuum stability bounds in the two Higgs doublet
  model}}, \href{https://doi.org/10.1016/S0370-2693(99)00019-2}{\emph{Phys.
  Lett.} {\bfseries B449} (1999) 89--92},
  [\href{https://arxiv.org/abs/hep-ph/9811234}{{\ttfamily hep-ph/9811234}}].

\bibitem{Arhrib:2000is}
A.~Arhrib, \emph{{Unitarity constraints on scalar parameters of the standard
  and two Higgs doublets model}},  in \emph{{Workshop on Noncommutative
  Geometry, Superstrings and Particle Physics Rabat, Morocco, June 16-17,
  2000}}, 2000, \href{https://arxiv.org/abs/hep-ph/0012353}{{\ttfamily
  hep-ph/0012353}}.

\bibitem{Kanemura:1993hm}
S.~Kanemura, T.~Kubota and E.~Takasugi, \emph{{Lee-Quigg-Thacker bounds for
  Higgs boson masses in a two doublet model}},
  \href{https://doi.org/10.1016/0370-2693(93)91205-2}{\emph{Phys. Lett.}
  {\bfseries B313} (1993) 155--160},
  [\href{https://arxiv.org/abs/hep-ph/9303263}{{\ttfamily hep-ph/9303263}}].

\bibitem{Lee:1977yc}
B.~W. Lee, C.~Quigg and H.~B. Thacker, \emph{{The Strength of Weak Interactions
  at Very High-Energies and the Higgs Boson Mass}},
  \href{https://doi.org/10.1103/PhysRevLett.38.883}{\emph{Phys. Rev. Lett.}
  {\bfseries 38} (1977) 883--885}.

\bibitem{Lee:1977eg}
B.~W. Lee, C.~Quigg and H.~B. Thacker, \emph{{Weak Interactions at Very
  High-Energies: The Role of the Higgs Boson Mass}},
  \href{https://doi.org/10.1103/PhysRevD.16.1519}{\emph{Phys. Rev.} {\bfseries
  D16} (1977) 1519}.

\bibitem{Ginzburg:2005dt}
I.~F. Ginzburg and I.~P. Ivanov, \emph{{Tree-level unitarity constraints in the
  most general 2HDM}},
  \href{https://doi.org/10.1103/PhysRevD.72.115010}{\emph{Phys. Rev.}
  {\bfseries D72} (2005) 115010},
  [\href{https://arxiv.org/abs/hep-ph/0508020}{{\ttfamily hep-ph/0508020}}].

\bibitem{Peskin:1991sw}
M.~E. Peskin and T.~Takeuchi, \emph{{Estimation of oblique electroweak
  corrections}}, \href{https://doi.org/10.1103/PhysRevD.46.381}{\emph{Phys.
  Rev.} {\bfseries D46} (1992) 381--409}.

\bibitem{Erler:2019hds}
J.~Erler and M.~Schott, \emph{{Electroweak Precision Tests of the Standard
  Model after the Discovery of the Higgs Boson}},
  \href{https://arxiv.org/abs/1902.05142}{{\ttfamily 1902.05142}}.

\bibitem{Bennett:2006fi}
{\scshape Muon g-2} collaboration, G.~W. Bennett et~al., \emph{{Final Report of
  the Muon E821 Anomalous Magnetic Moment Measurement at BNL}},
  \href{https://doi.org/10.1103/PhysRevD.73.072003}{\emph{Phys. Rev.}
  {\bfseries D73} (2006) 072003},
  [\href{https://arxiv.org/abs/hep-ex/0602035}{{\ttfamily hep-ex/0602035}}].

\bibitem{Aoyama:2012wk}
T.~Aoyama, M.~Hayakawa, T.~Kinoshita and M.~Nio, \emph{{Complete Tenth-Order
  QED Contribution to the Muon g-2}},
  \href{https://doi.org/10.1103/PhysRevLett.109.111808}{\emph{Phys. Rev. Lett.}
  {\bfseries 109} (2012) 111808},
  [\href{https://arxiv.org/abs/1205.5370}{{\ttfamily 1205.5370}}].

\bibitem{Czarnecki:2002nt}
A.~Czarnecki, W.~J. Marciano and A.~Vainshtein, \emph{{Refinements in
  electroweak contributions to the muon anomalous magnetic moment}},
  \href{https://doi.org/10.1103/PhysRevD.67.073006,
  10.1103/PhysRevD.73.119901}{\emph{Phys. Rev.} {\bfseries D67} (2003) 073006},
  [\href{https://arxiv.org/abs/hep-ph/0212229}{{\ttfamily hep-ph/0212229}}].

\bibitem{Cherchiglia:2017uwv}
A.~Cherchiglia, D.~Stöckinger and H.~Stöckinger-Kim, \emph{{Muon g-2 in the
  2HDM: maximum results and detailed phenomenology}},
  \href{https://doi.org/10.1103/PhysRevD.98.035001}{\emph{Phys. Rev.}
  {\bfseries D98} (2018) 035001},
  [\href{https://arxiv.org/abs/1711.11567}{{\ttfamily 1711.11567}}].

\bibitem{Czakon:2015exa}
M.~Czakon, P.~Fiedler, T.~Huber, M.~Misiak, T.~Schutzmeier and M.~Steinhauser,
  \emph{{The $(Q_{7}, Q_{1,2})$ contribution to $ \overline{B}\to {X}_s\gamma $
  at $ \mathcal{O}\left({\alpha}_{\mathrm{s}}^2\right) $}},
  \href{https://doi.org/10.1007/JHEP04(2015)168}{\emph{JHEP} {\bfseries 04}
  (2015) 168}, [\href{https://arxiv.org/abs/1503.01791}{{\ttfamily
  1503.01791}}].

\bibitem{Misiak:2006zs}
M.~Misiak et~al., \emph{{Estimate of $\mathcal{B} (\bar B \to X_s \gamma)$ at
  $O(\alpha_s^2)$}},
  \href{https://doi.org/10.1103/PhysRevLett.98.022002}{\emph{Phys. Rev. Lett.}
  {\bfseries 98} (2007) 022002},
  [\href{https://arxiv.org/abs/hep-ph/0609232}{{\ttfamily hep-ph/0609232}}].

\bibitem{Amhis:2016xyh}
{\scshape HFLAV} collaboration, Y.~Amhis et~al., \emph{{Averages of $b$-hadron,
  $c$-hadron, and $\tau$-lepton properties as of summer 2016}},
  \href{https://doi.org/10.1140/epjc/s10052-017-5058-4}{\emph{Eur. Phys. J.}
  {\bfseries C77} (2017) 895},
  [\href{https://arxiv.org/abs/1612.07233}{{\ttfamily 1612.07233}}].

\bibitem{CMS:2014xfa}
{\scshape CMS, LHCb} collaboration, V.~Khachatryan et~al., \emph{{Observation
  of the rare $B^0_s\to\mu^+\mu^-$ decay from the combined analysis of CMS and
  LHCb data}}, \href{https://doi.org/10.1038/nature14474}{\emph{Nature}
  {\bfseries 522} (2015) 68--72},
  [\href{https://arxiv.org/abs/1411.4413}{{\ttfamily 1411.4413}}].

\bibitem{Adachi:2012mm}
{\scshape Belle} collaboration, I.~Adachi et~al., \emph{{Evidence for $B^- \to
  \tau^- \bar{\nu}_\tau$ with a Hadronic Tagging Method Using the Full Data
  Sample of Belle}},
  \href{https://doi.org/10.1103/PhysRevLett.110.131801}{\emph{Phys. Rev. Lett.}
  {\bfseries 110} (2013) 131801},
  [\href{https://arxiv.org/abs/1208.4678}{{\ttfamily 1208.4678}}].

\bibitem{Kronenbitter:2015kls}
{\scshape Belle} collaboration, B.~Kronenbitter et~al., \emph{{Measurement of
  the branching fraction of $B^+ -> \tau^+ \nu_{\tau}$ decays with the
  semileptonic tagging method}},
  \href{https://doi.org/10.1103/PhysRevD.92.051102}{\emph{Phys. Rev.}
  {\bfseries D92} (2015) 051102},
  [\href{https://arxiv.org/abs/1503.05613}{{\ttfamily 1503.05613}}].

\bibitem{Khachatryan:2016vau}
{\scshape ATLAS, CMS} collaboration, G.~Aad et~al., \emph{{Measurements of the
  Higgs boson production and decay rates and constraints on its couplings from
  a combined ATLAS and CMS analysis of the LHC pp collision data at $
  \sqrt{s}=7 $ and 8 TeV}},
  \href{https://doi.org/10.1007/JHEP08(2016)045}{\emph{JHEP} {\bfseries 08}
  (2016) 045}, [\href{https://arxiv.org/abs/1606.02266}{{\ttfamily
  1606.02266}}].

\bibitem{CMS:2017rli}
{\scshape CMS} collaboration, C.~Collaboration, \emph{{Measurements of
  properties of the Higgs boson in the diphoton decay channel with the full
  2016 data set}}, .

\bibitem{CMS:2017jkd}
{\scshape CMS} collaboration, C.~Collaboration, \emph{{Measurements of
  properties of the Higgs boson decaying into four leptons in pp collisions at
  sqrt{s} = 13 TeV}}, .

\bibitem{CMS:2017pzi}
{\scshape CMS} collaboration, C.~Collaboration, \emph{{Higgs to WW measurements
  with $15.2~\mathrm{fb}^{-1}$ of 13 TeV proton-proton collisions}}, .

\bibitem{Sirunyan:2017khh}
{\scshape CMS} collaboration, A.~M. Sirunyan et~al., \emph{{Observation of the
  Higgs boson decay to a pair of $\tau$ leptons with the CMS detector}},
  \href{https://doi.org/10.1016/j.physletb.2018.02.004}{\emph{Phys. Lett.}
  {\bfseries B779} (2018) 283--316},
  [\href{https://arxiv.org/abs/1708.00373}{{\ttfamily 1708.00373}}].

\bibitem{ATLAS-CONF-2017-045}
{\scshape ATLAS Collaboration} collaboration, \emph{{Measurements of Higgs
  boson properties in the diphoton decay channel with 36.1 fb$^{−1}$ $pp$
  collision data at the center-of-mass energy of 13 TeV with the ATLAS
  detector}},  Tech. Rep. ATLAS-CONF-2017-045, CERN, Geneva, Jul, 2017.

\bibitem{ATLAS-CONF-2017-043}
{\scshape ATLAS Collaboration} collaboration, \emph{{Measurement of the Higgs
  boson coupling properties in the $H\rightarrow ZZ^{*} \rightarrow 4\ell$
  decay channel at $\sqrt{s}$ = 13 TeV with the ATLAS detector}},  Tech. Rep.
  ATLAS-CONF-2017-043, CERN, Geneva, Jul, 2017.

\bibitem{Sirunyan:2018owy}
{\scshape CMS} collaboration, A.~M. Sirunyan et~al., \emph{{Search for
  invisible decays of a Higgs boson produced through vector boson fusion in
  proton-proton collisions at $\sqrt{s} =$ 13 TeV}},
  \href{https://arxiv.org/abs/1809.05937}{{\ttfamily 1809.05937}}.

\bibitem{Chowdhury:2017aav}
D.~Chowdhury and O.~Eberhardt, \emph{{Update of Global Two-Higgs-Doublet Model
  Fits}}, \href{https://doi.org/10.1007/JHEP05(2018)161}{\emph{JHEP} {\bfseries
  05} (2018) 161}, [\href{https://arxiv.org/abs/1711.02095}{{\ttfamily
  1711.02095}}].

\bibitem{Aprile:2018dbl}
{\scshape XENON} collaboration, E.~Aprile et~al., \emph{{Dark Matter Search
  Results from a One Ton-Year Exposure of XENON1T}},
  \href{https://doi.org/10.1103/PhysRevLett.121.111302}{\emph{Phys. Rev. Lett.}
  {\bfseries 121} (2018) 111302},
  [\href{https://arxiv.org/abs/1805.12562}{{\ttfamily 1805.12562}}].

\bibitem{Ackermann:2015zua}
{\scshape Fermi-LAT} collaboration, M.~Ackermann et~al., \emph{{Searching for
  Dark Matter Annihilation from Milky Way Dwarf Spheroidal Galaxies with Six
  Years of Fermi Large Area Telescope Data}},
  \href{https://doi.org/10.1103/PhysRevLett.115.231301}{\emph{Phys. Rev. Lett.}
  {\bfseries 115} (2015) 231301},
  [\href{https://arxiv.org/abs/1503.02641}{{\ttfamily 1503.02641}}].

\bibitem{Khachatryan:2016whc}
{\scshape CMS} collaboration, V.~Khachatryan et~al., \emph{{Searches for
  invisible decays of the Higgs boson in pp collisions at $\sqrt{s}$ = 7, 8,
  and 13 TeV}}, \href{https://doi.org/10.1007/JHEP02(2017)135}{\emph{JHEP}
  {\bfseries 02} (2017) 135},
  [\href{https://arxiv.org/abs/1610.09218}{{\ttfamily 1610.09218}}].

\bibitem{Aaboud:2019rtt}
{\scshape ATLAS} collaboration, M.~Aaboud et~al., \emph{{Combination of
  searches for invisible Higgs boson decays with the ATLAS experiment}},
  \href{https://arxiv.org/abs/1904.05105}{{\ttfamily 1904.05105}}.

\bibitem{Alwall:2014hca}
J.~Alwall, R.~Frederix, S.~Frixione, V.~Hirschi, F.~Maltoni, O.~Mattelaer
  et~al., \emph{{The automated computation of tree-level and next-to-leading
  order differential cross sections, and their matching to parton shower
  simulations}}, \href{https://doi.org/10.1007/JHEP07(2014)079}{\emph{JHEP}
  {\bfseries 07} (2014) 079},
  [\href{https://arxiv.org/abs/1405.0301}{{\ttfamily 1405.0301}}].

\bibitem{Sjostrand:2006za}
T.~Sjostrand, S.~Mrenna and P.~Z. Skands, \emph{{PYTHIA 6.4 Physics and
  Manual}}, \href{https://doi.org/10.1088/1126-6708/2006/05/026}{\emph{JHEP}
  {\bfseries 05} (2006) 026},
  [\href{https://arxiv.org/abs/hep-ph/0603175}{{\ttfamily hep-ph/0603175}}].

\bibitem{deFavereau:2013fsa}
{\scshape DELPHES 3} collaboration, J.~de~Favereau, C.~Delaere, P.~Demin,
  A.~Giammanco, V.~Lemaître, A.~Mertens et~al., \emph{{DELPHES 3, A modular
  framework for fast simulation of a generic collider experiment}},
  \href{https://doi.org/10.1007/JHEP02(2014)057}{\emph{JHEP} {\bfseries 02}
  (2014) 057}, [\href{https://arxiv.org/abs/1307.6346}{{\ttfamily 1307.6346}}].

\bibitem{Aaboud:2017phn}
{\scshape ATLAS} collaboration, M.~Aaboud et~al., \emph{{Search for dark matter
  and other new phenomena in events with an energetic jet and large missing
  transverse momentum using the ATLAS detector}},
  \href{https://doi.org/10.1007/JHEP01(2018)126}{\emph{JHEP} {\bfseries 01}
  (2018) 126}, [\href{https://arxiv.org/abs/1711.03301}{{\ttfamily
  1711.03301}}].

\bibitem{Rainwater:1998kj}
D.~L. Rainwater, D.~Zeppenfeld and K.~Hagiwara, \emph{{Searching for
  $H\to\tau^+\tau^-$ in weak boson fusion at the CERN LHC}},
  \href{https://doi.org/10.1103/PhysRevD.59.014037}{\emph{Phys. Rev.}
  {\bfseries D59} (1998) 014037},
  [\href{https://arxiv.org/abs/hep-ph/9808468}{{\ttfamily hep-ph/9808468}}].

\bibitem{Hankele:2006ma}
V.~Hankele, G.~Klamke, D.~Zeppenfeld and T.~Figy, \emph{{Anomalous Higgs boson
  couplings in vector boson fusion at the CERN LHC}},
  \href{https://doi.org/10.1103/PhysRevD.74.095001}{\emph{Phys. Rev.}
  {\bfseries D74} (2006) 095001},
  [\href{https://arxiv.org/abs/hep-ph/0609075}{{\ttfamily hep-ph/0609075}}].

\bibitem{Datta:2001hv}
A.~Datta, P.~Konar and B.~Mukhopadhyaya, \emph{{Invisible charginos and
  neutralinos from gauge boson fusion: A Way to explore anomaly mediation?}},
  \href{https://doi.org/10.1103/PhysRevLett.88.181802}{\emph{Phys. Rev. Lett.}
  {\bfseries 88} (2002) 181802},
  [\href{https://arxiv.org/abs/hep-ph/0111012}{{\ttfamily hep-ph/0111012}}].

\bibitem{Khoze:2001ft}
V.~A. Khoze, A.~D. Martin and M.~G. Ryskin, \emph{{Double-diffractive Higgs
  production and pomeron pomeron luminometry in proton collisions}},
  \href{https://doi.org/10.1016/S0920-5632(01)01331-7}{\emph{Nucl. Phys. Proc.
  Suppl.} {\bfseries 99B} (2001) 188--191}.

\bibitem{Khoze:2000vr}
V.~A. Khoze, A.~D. Martin and M.~G. Ryskin, \emph{{Soft diffraction at the LHC
  and properties of the pomeron}},
  \href{https://doi.org/10.1016/S0920-5632(01)01336-6}{\emph{Nucl. Phys. Proc.
  Suppl.} {\bfseries 99B} (2001) 213--216},
  [\href{https://arxiv.org/abs/hep-ph/0011319}{{\ttfamily hep-ph/0011319}}].

\bibitem{Chen:2016btl}
T.~Chen and C.~Guestrin, \emph{{XGBoost: A Scalable Tree Boosting System}},
  \href{https://arxiv.org/abs/1603.02754}{{\ttfamily 1603.02754}}.

\bibitem{Teodorescu:1100521}
L.~Teodorescu, \emph{{Artificial neural networks in high-energy physics}}, .

\bibitem{Baldi:2014kfa}
P.~Baldi, P.~Sadowski and D.~Whiteson, \emph{{Searching for Exotic Particles in
  High-Energy Physics with Deep Learning}},
  \href{https://doi.org/10.1038/ncomms5308}{\emph{Nature Commun.} {\bfseries 5}
  (2014) 4308}, [\href{https://arxiv.org/abs/1402.4735}{{\ttfamily
  1402.4735}}].

\bibitem{Woodruff:2017geg}
{\scshape MicroBooNE} collaboration, K.~Woodruff, \emph{{Automated Proton Track
  Identification in MicroBooNE Using Gradient Boosted Decision Trees}},  in
  \emph{{Proceedings, Meeting of the APS Division of Particles and Fields (DPF
  2017): Fermilab, Batavia, Illinois, USA, July 31 - August 4, 2017}}, 2018,
  \href{https://arxiv.org/abs/1710.00898}{{\ttfamily 1710.00898}},
  \href{http://lss.fnal.gov/archive/2017/conf/fermilab-conf-17-440-e.pdf}{http://lss.fnal.gov/archive/2017/conf/fermilab-conf-17-440-e.pdf}.

\bibitem{Oyulmaz:2019jqr}
K.~Y. Oyulmaz, A.~Senol, H.~Denizli and O.~Cakir, \emph{{Top quark anomalous
  FCNC production via $tqg$ couplings at FCC-hh}},
  \href{https://arxiv.org/abs/1902.03037}{{\ttfamily 1902.03037}}.

\bibitem{Bhattacherjee:2019fpt}
B.~Bhattacherjee, S.~Mukherjee and R.~Sengupta, \emph{{Discrimination between
  prompt and long-lived particles using convolutional neural network}},
  \href{https://arxiv.org/abs/1904.04811}{{\ttfamily 1904.04811}}.

\bibitem{Hultqvist:1995ibm}
K.~Hultqvist, R.~Jacobsson and K.~E. Johansson, \emph{{Using a neural network
  in the search for the Higgs boson}}, .

\bibitem{Bakhet:2015uca}
N.~Bakhet, M.~{\relax Yu}. Khlopov and T.~Hussein, \emph{{Neural Networks
  Search for Charged Higgs Boson of Two Doublet Higgs Model at the Hadrons
  Colliders}},  \href{https://arxiv.org/abs/1507.06547}{{\ttfamily
  1507.06547}}.

\bibitem{Field:1996rw}
R.~D. Field, Y.~Kanev, M.~Tayebnejad and P.~A. Griffin, \emph{{Using neural
  networks to enhance the Higgs boson signal at hadron colliders}},
  \href{https://doi.org/10.1103/PhysRevD.53.2296}{\emph{Phys. Rev.} {\bfseries
  D53} (1996) 2296--2308}.

\bibitem{keras}
J.~R. Hermans, \emph{https://github.com/cerndb/dist-keras}, .

\end{thebibliography}\endgroup



\providecommand{\href}[2]{#2}\begingroup\raggedright\endgroup

\end{document}